\documentclass[preprint2]{aastex}

\def\lae{\lower 2pt \hbox{$\, \buildrel {\scriptstyle <}\over {\scriptstyle \sim}\,$}} 

\begin{document}

\title{Relations Between Timing Features and Colors in Accreting Millisecond pulsars} 

\author{Steve van Straaten\altaffilmark{1}, Michiel van der Klis\altaffilmark{1}, Rudy Wijnands\altaffilmark{1}}

\email{straaten@science.uva.nl}
\altaffiltext{1}{Astronomical Institute, ``Anton Pannekoek'',
University of Amsterdam, and Center for High Energy Astrophysics, 
Kruislaan 403, 1098 SJ Amsterdam, The Netherlands.}

\begin{abstract}

We have studied the aperiodic X-ray timing and color behavior of the accreting millisecond pulsars SAX J1808.4--3658, 
XTE J1751--305, XTE J0929--314, and XTE J1814-338 using large data sets obtained with the {\it Rossi X-ray Timing Explorer}.
We find that the accreting millisecond pulsars have very similar timing properties to the atoll sources and the low luminosity
bursters. Based on the correlation of timing and color behavior SAX J1808.4--3658 can be classified as an atoll source,
and XTE J0929--314 and XTE J1814--338 are consistent with being atoll sources, but the color behavior of XTE J1751--305 
is different. Unlike in other atoll sources the hard color in the extreme island state of XTE J1751--305 is strongly 
correlated with soft color and intensity, and is not anti-correlated with any of the characteristic frequencies.
We found previously, that the frequencies of the variability components of atoll sources follow a universal scheme of 
correlations. The frequency correlations of the accreting millisecond pulsars are similar, but in the case of SAX J1808.4--3658 
and XTE J0929--314 shifted compared to those of the other atoll sources in a way that is most easily described as a shift in 
upper and lower kilohertz QPO frequency by a factor close to 1.5. 
Although, we note that the shift in lower kilohertz QPO frequency is based on only one observation for SAX J1808.4--3658. 
XTE J1751--305 and XTE J1814--338, as well as the low luminosity 
bursters show no or maybe small shifts.

\end{abstract}

\section{INTRODUCTION}

Many low mass X-ray binaries containing a neutron star show kilohertz QPOs, quasi periodic oscillations 
with frequencies ranging from a few hundred to more than a 1200 Hz. The low frequency ($<200$ Hz) part of the power 
spectrum taken from their X-ray lightcurves is usually dominated by a broad band-limited noise component. 
In addition to the band-limited noise several Lorentzian components below 200 Hz are 
present (see e.g. \cite{vstr02:apj568}). These Lorentzians become broader as their characteristic frequency 
decreases (\citealp*{psaltis99:apj520}; \citealp{vstr02:apj568}). Therefore, several features can, only at high 
frequencies, be classified as a QPO \citep[traditionally defined by the condition FWHM $<$ 
centroid frequency/2, see][]{vdk95:xrb252}.
In addition to these timing features in the persistent emission,
13 sources show millisecond oscillations (270--619 Hz) during thermonuclear X-ray bursts 
\citep[see][for a review]{muno04:rossi}. Although these burst oscillations show a slight drift, they are thought to be 
connected with the spin of the neutron star and were the first direct indication that these sources harbour neutron 
stars with millisecond spins.

Based on the correlated behavior of their timing properties and the behavior of the sources in color-color diagrams,
that are used to study their X-ray spectral properties, many of the neutron-star low-mass X-ray binaries are traditionally 
classified as either ``atoll'' or ``Z'' sources \citep[][]{hasinger89:aa225}. 
Recently, \citet{muno02:apj568} and independently \citet{gierlinski02:mnras331} used the large 
{\it Rossi X-ray Timing Explorer} data sets now available for several of the atoll sources, to show 
that some of the atoll sources look much more like Z sources when plotted in color-color diagrams 
than was previously thought. Although the appearances in the color-color diagram of the two classes now is 
more similar, the classification still holds, as clear differences between the two classes in both timing and 
color behavior remain \citep{muno02:apj568,barret02:apj576,olive03:apj583,vstr03:apj596,reig04:apj}.
Note, that next to the differences in timing and color behavior there is also a difference in luminosity between 
the two classes; where the Z sources are all very luminous \citep[$\sim 3 \times 10^{38} erg s^{-1}$][]{ford00:apj537}, 
the atoll sources occur over a wide range of luminosities \citep[$10^{36-38} erg s^{-1}$][]{ford00:apj537}.

The sources 1E 1724--3045, GS 1826--24, and SLX 1735--269, are all low-luminosity neutron stars that show 
thermonuclear X-ray bursts. The power spectra of these ``low-luminosity bursters'' resemble those of the atoll 
sources in their hard states (\citealp{olive98:aa333}; \citealp{barret00:apj533}; \citealp*{belloni02:apj572}); they show a broad band-limited noise component and several broad Lorentzian components.
However, where the characteristic frequencies of the low-frequency Lorentzians were found to be similar to those in the atoll 
sources, the characteristic frequency of the highest frequency Lorentzian seemed to be higher \citep[][]{belloni02:apj572}.

In 1998 the first accreting millisecond pulsar was found with the {\it Rossi X-ray Timing Explorer} ({\it RXTE})
during an outburst of SAX J1808.4--3658 \citep[][]{wijnands98:nat394}. During this outburst no thermonuclear X-ray bursts
were observed and the source did not show kilohertz QPOs above 200 Hz \citep[][]{wijnands98:apj507}. The low frequency
power spectrum was very similar to that of the atoll sources \citep[][]{wijnands98:apj507} and both the atoll sources 
and SAX J1808.4--3658 followed one relation when the characteristic frequency of a low-frequency Lorentzian was 
plotted versus that of the band-limited noise \citep[][]{wijnands99:apj514}. During a new outburst of SAX J1808.4--3658
in October 2002, kilohertz QPOs (up to 700 Hz) and burst oscillations were observed in an accreting 
millisecond pulsar for the first time \citep[][]{wijnands03:nat424,chakrabarty03:nat424}. Recently, four additional accreting millisecond pulsars
were discovered; XTE J1751--305, XTE J0929--314, XTE J1807--294, and XTE J1814--338 
\citep[see][for a observational review]{wijnands04:bepposax}. Of these XTE J1807--294 showed kiloherz QPOs 
(Markwardt, private communication) and XTE J1814--338 showed burst oscillations \citep{strohmayer03:apj596}. 

Here we present an extensive analysis of the aperiodic-timing and color behavior of the accreting millisecond pulsars 
SAX J1808.4--3658, XTE J1751--305, 
XTE J0929--314, and XTE J1814--338. We study the correlations between the characteristic frequencies of their various timing 
features, and compare these with those of four well-studied atoll sources and three
low-luminosity bursters. We find, that the accreting millisecond pulsars have very similar timing properties as the atoll sources
and the low luminosity bursters, although discrepancies occur for SAX J1808.4--3658 and XTE J0929--314. 
We also compare the behavior of the accreting millisecond pulsars in the color-color diagrams to that of the atoll sources 
and find that one pulsar, XTE J1751--305, exhibits unusual behavior while the others act like ordinary atoll sources.
In the next section we describe the observations and data analysis. We suggest that readers not
interested in the technical details of observations and analysis skip immediately to the results section.

\section{OBSERVATIONS AND DATA ANALYSIS}
\label{sec.observ}

In this work we study four accreting millisecond pulsars: SAX J1808.4--3658, 
XTE J1751--305, XTE J0929--314, and XTE J1814--338. We also re-analyzed data 
of \citet{belloni02:apj572} for the low luminosity bursters 1E 1724--3045, GS 1826--24, and 
SLX 1735--269 (see \S \ref{sec.compare_timing}). For the accreting millisecond pulsars we analyzed data from 
{\it RXTE}'s proportional counter array \citep[PCA; for instrument 
information see][]{zhang93:spie2006}; for SAX J1808.4--3658 we used data from 
March 11 to May 6, 1998 (the 1998 outburst) and from October 15 to 
November 26, 2002 (the 2002 outburst), for XTE J1751--305  
from April 4 to April 30, 2002, for XTE J0929--314  
from May 2 to May 21, 2002, and for XTE J1814--338  
from June 8 to August 4, 2003. The 
data were divided into observations, identified by a unique {\it RXTE} observation ID, 
that cover one to several satellite 
orbits with $\sim3000$ s of useful data each. 
We excluded data for which the angle of the source above the Earth 
limb is less than 10 degrees or for which the pointing offset is greater 
then 0.02 degrees. We also excluded type I X-ray bursts. We 
found 4 bursts in the 2002 outburst of SAX J1808.4--3658 
\citep[for results on these bursts see][]{chakrabarty03:nat424} for which we excluded 16 s 
before and 384 s after the onset of each burst, 1 burst in XTE J1751--305 
for which we excluded 16 s before and 592 s after the onset of the burst 
(note that this burst occurred during a weak reflare from April 27--30 for which 
it is unclear whether it originated from XTE J1751--305, see \citealt{markwardt02:apj575}, 
the burst itself probably originated from GRS 1747--312, which is however ruled out 
as the source of the reflare, see \citealt{intzand03:aa409})
and 22 bursts in XTE J1814--338  \citep[for results on these bursts see][]{strohmayer03:apj596} 
for which we excluded 16 s before and 240 s after the onset of each burst, the amount of time excluded depending on 
the length of the bursts. No bursts were found in XTE J0929--314.

\subsection{Color-Color Diagrams}
\label{sec.anal_col}

We used the 16--s time-resolution Standard 2 mode to calculate the colors.
For each of the PCA's five PCU detectors separately we calculated a hard color, defined as 
the count rate in the energy band 9.7$-$16.0  keV divided by that in the 
6.0$-$9.7 keV band, and a soft color, defined similarly as  
3.5$-$6.0 keV/2.0$-$3.5 
keV. For each detector we also calculated the intensity, the count rate in the 
2.0$-$16.0 keV band. To obtain the count rates in these exact energy 
ranges we interpolated linearly between count rates in the PCU channels based on the channel to energy calibration. 
We subtracted the background contribution in each band using 'pcabackest' 
version 3.0, with as input the ``Mission-long Faint Model File'' if the count rate in the 
full energy band becomes less than 40 counts s$^{-1}$ during an observation (faint data), 
and otherwise the ``Mission-long Bright Model File'' (bright data). We 
averaged up the background-subtracted 16--s count rates to obtain count rates 
per observation, and we calculated the colors separately for each PCU. 
An observation spanned between 1 and 30 ks, and within an observation the colors generally
varied by 3 to 9 times the 1 $\sigma$ standard deviation of the 16--s colors.
For faint data we filtered out data taken within thirty minutes of the peak of the South Atlantic 
Anomaly and data with high electron contamination.

The voltage settings of the PCUs on board {\it RXTE} were changed on three occasions 
defining the first four {\it RXTE} gain epochs. The fifth {\it RXTE} gain epoch started 
when a micro-meteorite created a small hole in the front window of PCU0.
In order to correct for the changes in effective detector area between the different 
{\it RXTE} gain epochs and for the gain drifts within these epochs, as well as for the 
differences in effective area between the PCUs themselves we calculated the 
colors for each PCU of the Crab nebula, which can supposed to be constant 
in its colors, in the same manner. We then averaged the 
16--s Crab colors and intensity per PCU for each day. For each PCU we 
divided the obtained color and intensity points per observation by the 
corresponding Crab values that are closest in time but within the same gain 
epoch. We then averaged the colors and intensity over all PCUs. Note 
that starting May 12, 2000, the propane layer on PCU0, which functions as an 
anti-coincidence shield for charged particles, was lost, defining the 
beginning of gain epoch 5. Therefore we applied the additional filter
recommended by the RXTE science team to PCU0 data from epoch 5.

\subsection{Timing}
\label{sec.anal_timing}

For the Fourier timing analysis we used the 122$-\mu$s time-resolution Event modes and 
the 0.95$-\mu$s time-resolution Good Xenon modes. We used data from all energy channels.
We constructed power spectra per observation (see above) using data segments of 256 s and 1/8192 s time bins 
such that the lowest available frequency is 1/256 Hz and the Nyquist frequency 4096 Hz; the 
normalization of \citet{leahy83:apj266} was used. We subtracted a Poisson noise level using the method 
of \citet{klein04:prep}, which is built on the analytical function from \citet{zhang95:apj449}.
The resulting power spectra were then converted to squared fractional rms. 
Within the observations the power spectra did not change within the errors.
When possible several observations, adjacent in time, and for which the power spectra 
remain the same, were added up to improve the statistics. These are called groups.
Occasionally, several observations, not adjacent in time, for which the power spectra are
the same, were added up into groups. The groups for which this was done will be noted in the 
text (see \S \ref{sec.results}). Before fitting, all frequency bins (generally about 20) containing 
the pulse spike were removed from the power spectra.

As a fit function we used the multi-Lorentzian function; a sum of 
Lorentzian components \citep[see e.g.][]{belloni02:apj572,vstr02:apj568}. 
Note, that we no longer constrained the Lorentzian \citep{vstr02:apj568,vstr03:apj596} that fits the band-limited noise
component to be zero-centered, but instead let all Lorentzian parameters float. 
This was motivated by the work of \citet{pottschmidt03:aa407}, who found that for the band-limited noise
component in the black hole candidate Cyg X--1, a non zero-centered Lorentzian provided better fits
(usually we find a $Q$ significantly larger than zero but $\lesssim  0.3$ for this component, see below). 
We generally only included those Lorentzians in the fit whose power can be measured to an accuracy of better than 33\%.
The few exceptions where we deviated from this will be noted in the text (\S \ref{sec.results}).
We plot the power spectra in the power times frequency representation ($\nu{\rm P}_{\nu}$), where the 
power spectral density ${\rm P}_{\nu}$ is multiplied by its Fourier frequency $\nu$. For a multi-Lorentzian fit function 
this representation helps to visualize a characteristic frequency, $\nu_{\rm max}$, the frequency 
where each Lorentzian component contributes most of its variance per logarithmic frequency interval, as in 
$\nu{\rm P}_{\nu}$ the Lorentzian's maximum occurs at $\nu_{\rm max}$
($\nu_{\rm max} = \sqrt{\nu_0^2 + \Delta^2}$, where $\nu_0$ is the centroid and $\Delta$ the HWHM 
of the Lorentzian; \cite{belloni02:apj572}).  We represent the Lorentzian relative width by $Q$ defined as $\nu_0/2\Delta$.

\section{RESULTS}
\label{sec.results}

\begin{figure}
\figurenum{1}
\epsscale{0.6}
\plotone{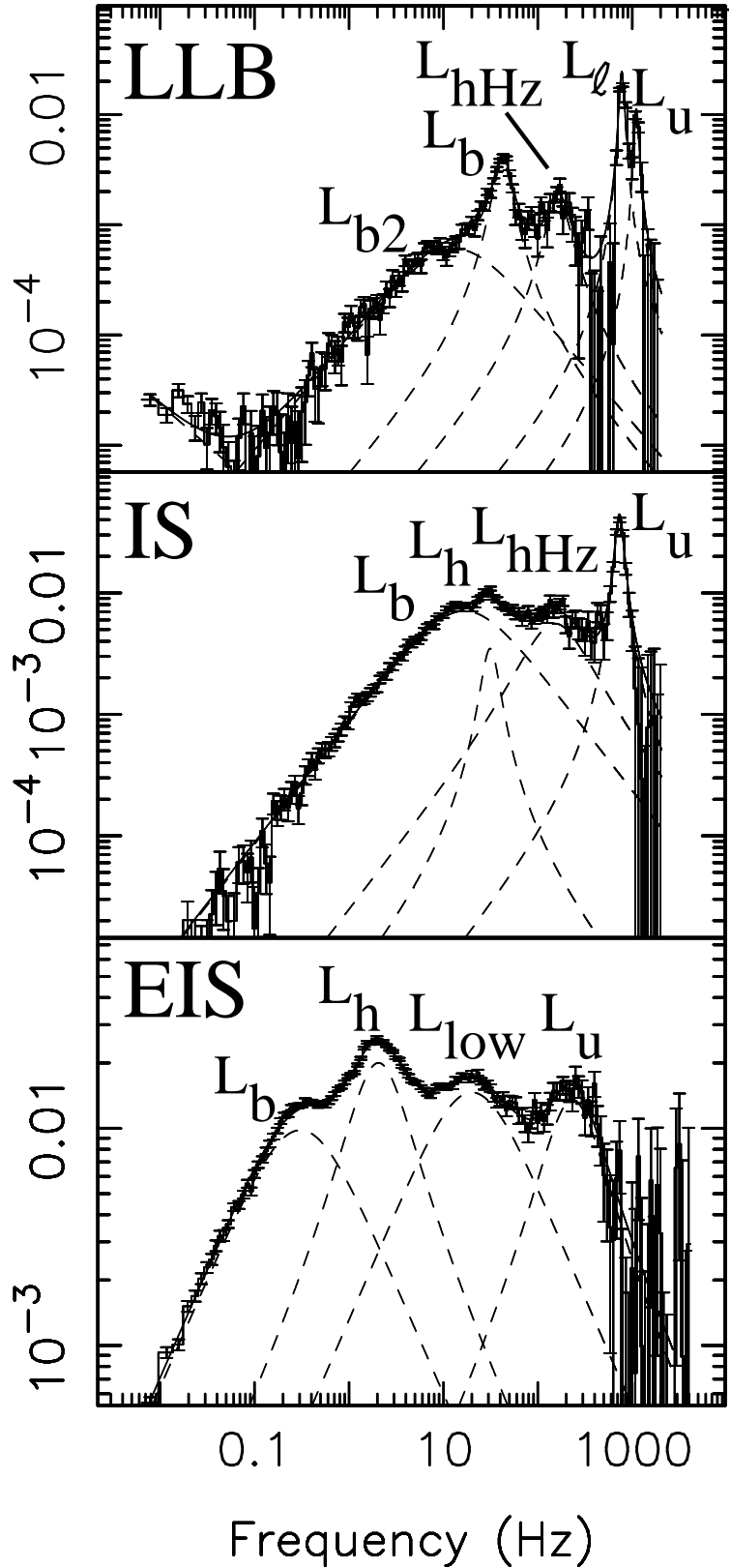}
\caption{The multi-Lorentzian components of the low 
mass X-ray binary scheme of \citet{belloni02:apj572} 
and \citet{vstr03:apj596}. The states in which the 
components occur are indicated; lower left banana (LLB), 
island state (IS), and extreme island state (EIS). 
Note that L$_{LF}$ and L$_{LF/2}$
are not shown. They generally occur in the extreme island state as 
a narrow QPO with a frequency between that of L$_{b}$ and L$_{h}$.}
\label{fig.naming}
\end{figure}

We find that three to seven Lorentzian components are needed to fit the power spectra
of the millisecond pulsars. All Lorentzian components can be identified in the low 
mass X-ray binary scheme of \citet{belloni02:apj572} and \citet{vstr03:apj596}. 
In Figure \ref{fig.naming} we illustrate the main multi-Lorentzian components of this scheme.
Note, that some updates to the
scheme based on results of this paper are made (see \S \ref{sec.identify}).
The fit parameters are listed in Tables \ref{tbl.numax}
($\nu_{\rm max}$), \ref{tbl.qvalues} ($Q$), and \ref{tbl.rms} (fractional rms amplitude).
Of the fits for the accreting millisecond pulsars listed in Tables \ref{tbl.numax}, \ref{tbl.qvalues}, and \ref{tbl.rms} 
47\% has a $\chi^2$/dof below 1.1, 27\% has a $\chi^2$/dof between 1.1 and 1.3, 23\% between 1.3 and 1.7, and 3\% (group 4) 
has $\chi^2$/dof = 2.2. The degrees of 
freedom range between 91 and 142. Note, that the $\chi^2$/dof values of the multi-Lorentzian fit function are quite high. However, they are
comparable to previous results of this fit function \citep{vstr02:apj568,vstr03:apj596}, and other
fit functions tried in previous studies \citep[see][for a comparison between the multi-Lorentzian fit 
function and the broken powerlaw fit function ]{belloni02:apj572,vstr02:apj568} do not give better results. 
Note that this is not
a problem as here we are only measuring the relevant characteristic frequencies and not trying to find the 
true model for the power spectra.
The fit groups contain data from several {\it RXTE} observations. The
observation IDs of each fit group are listed in Table \ref{tbl.obsid}.  

\subsection{SAX J1808.4--3658: the 2002 Outburst}
\label{sec.1808_2002}

\begin{figure}
\figurenum{2}
\epsscale{0.38}
\begin{tabular}{ccc}
\plotone{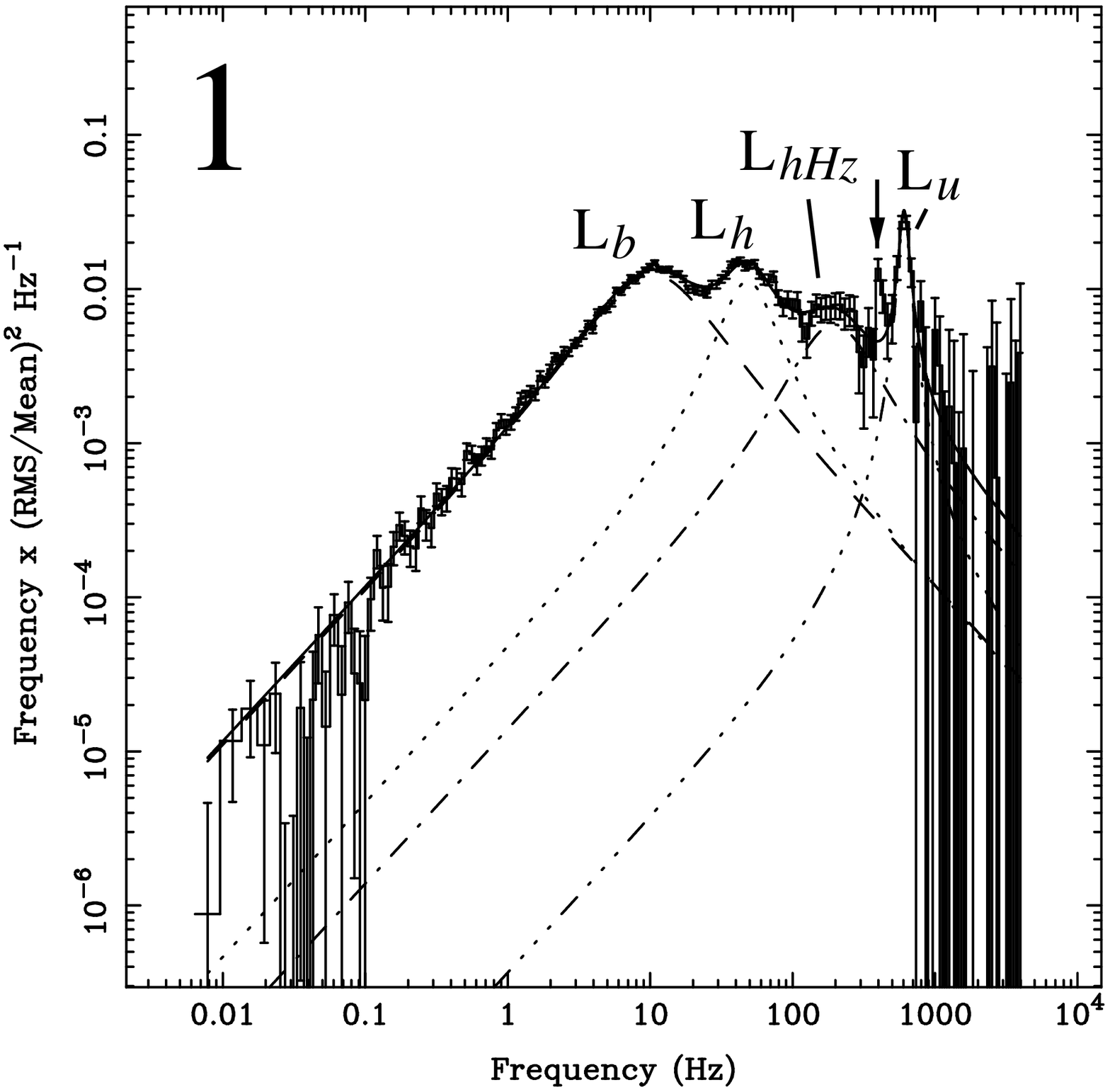} & 
\plotone{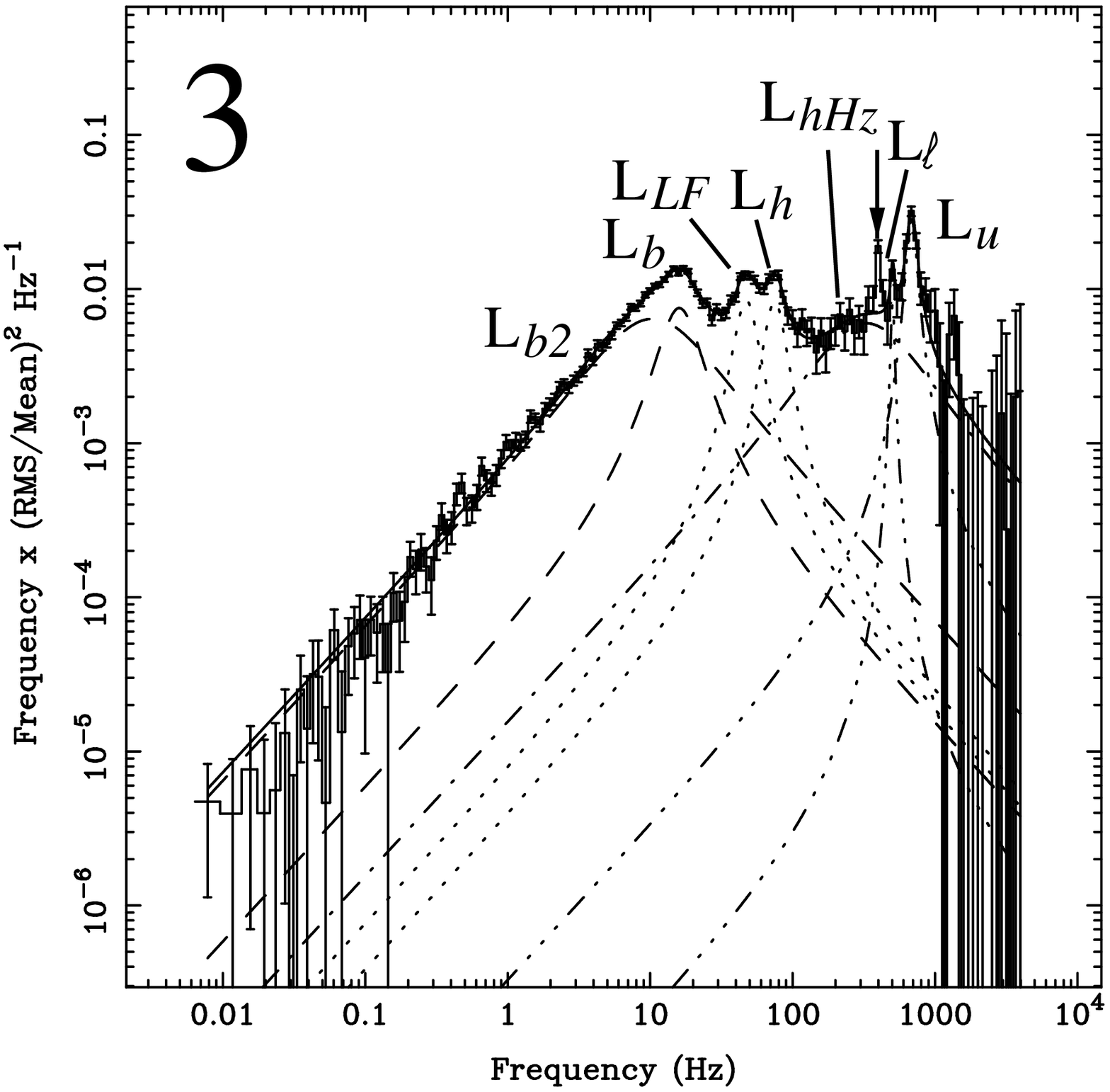} & \\
\plotone{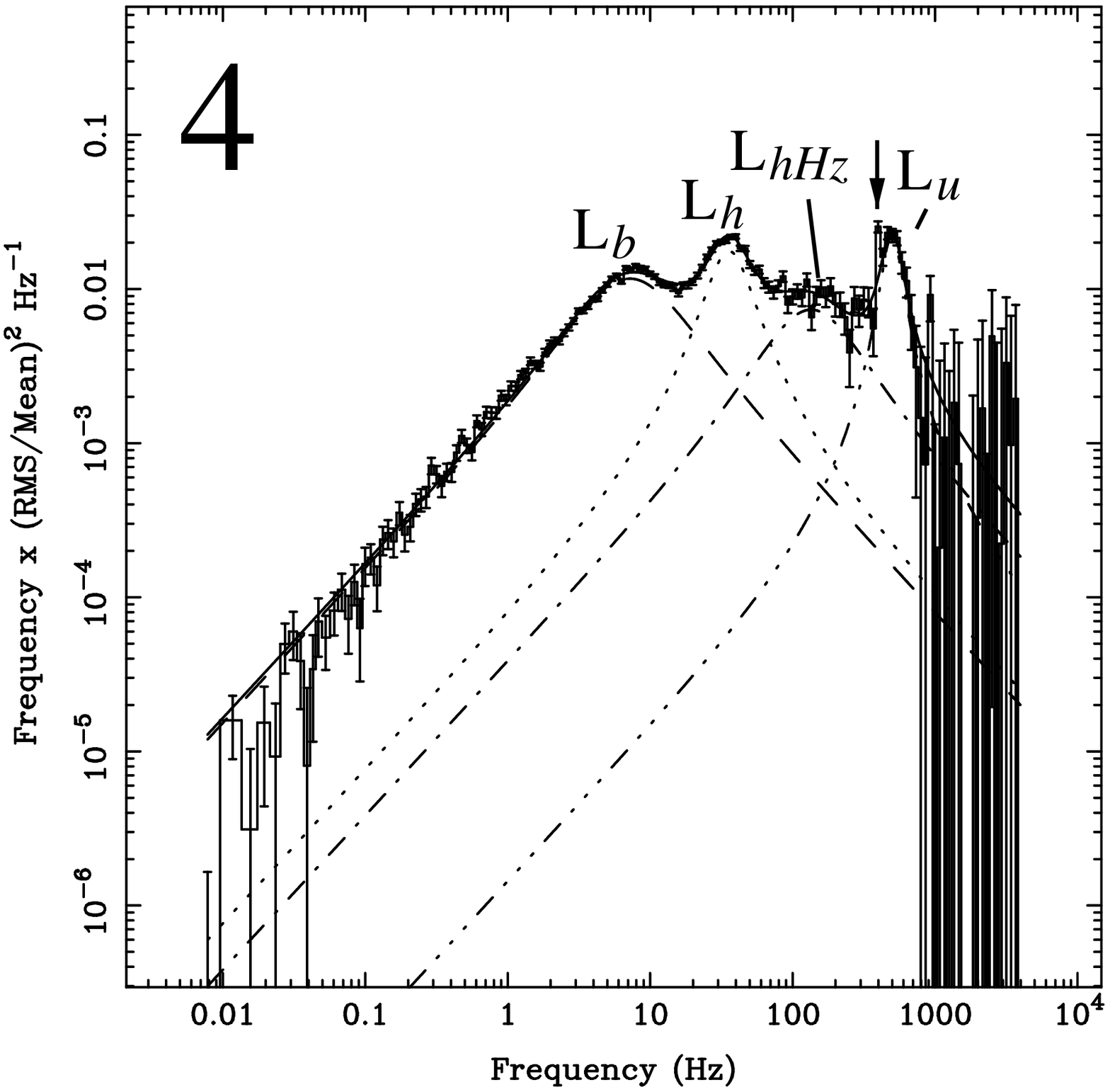} & 
\plotone{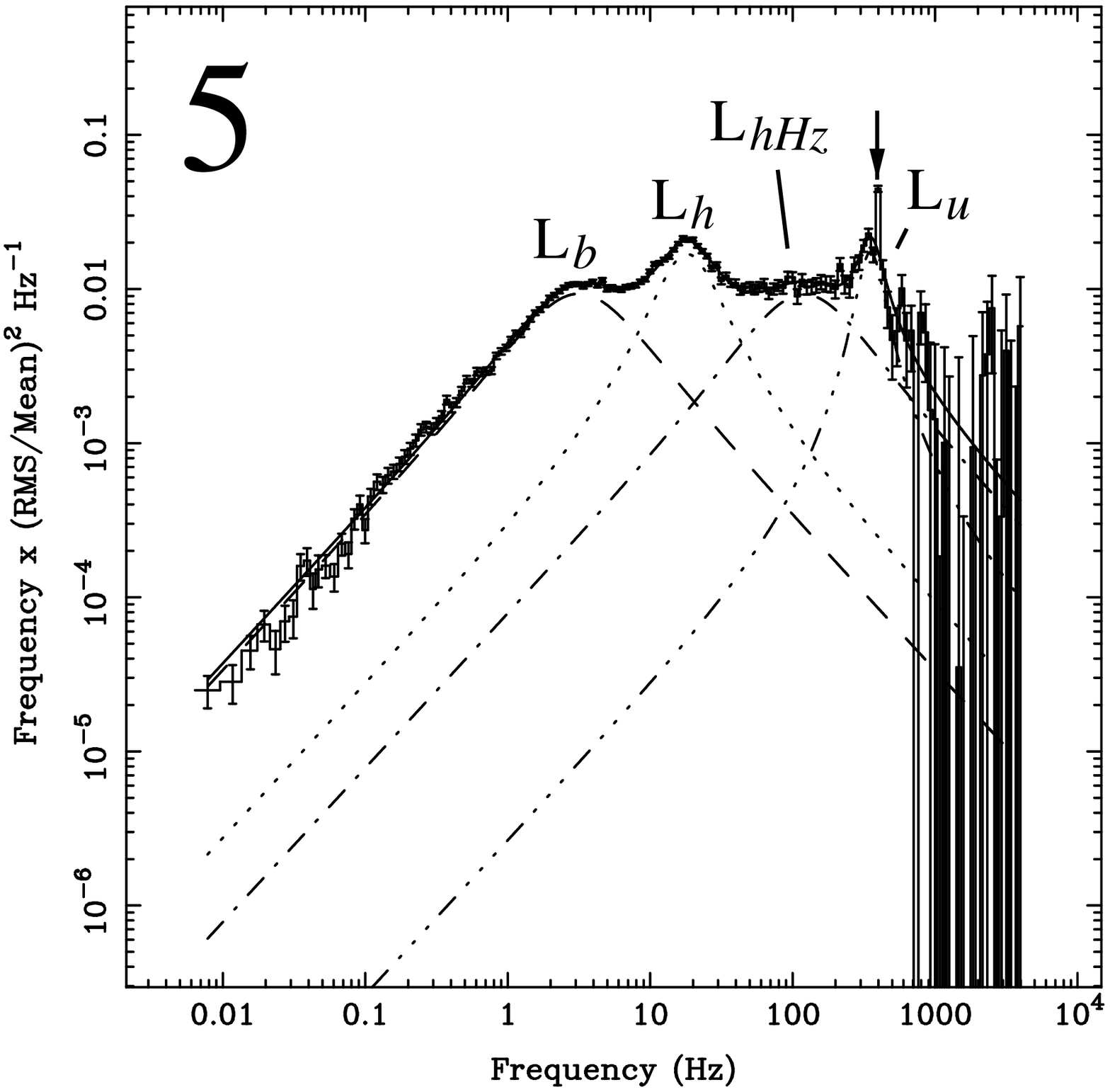} & \\
\plotone{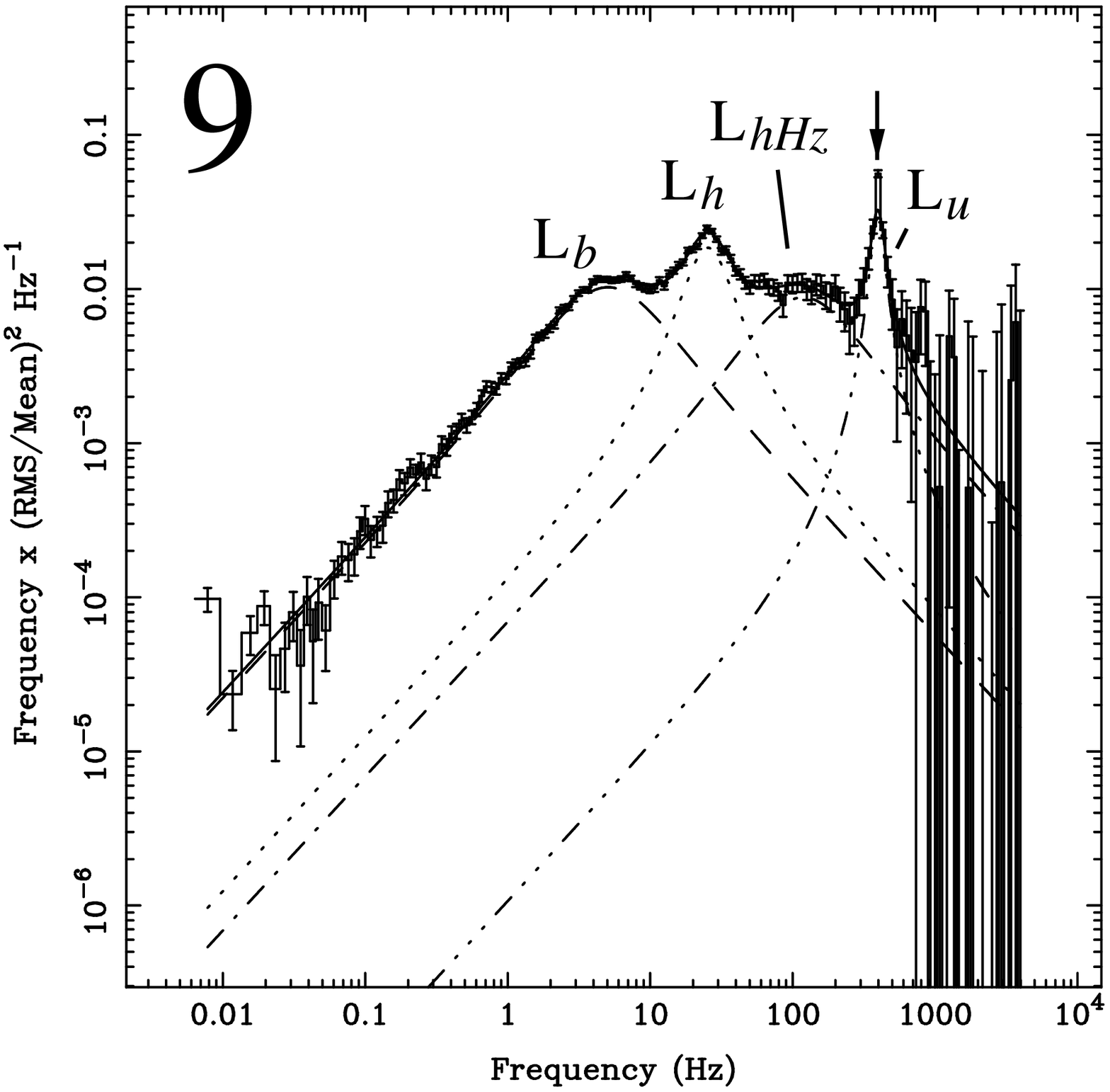} & 
\plotone{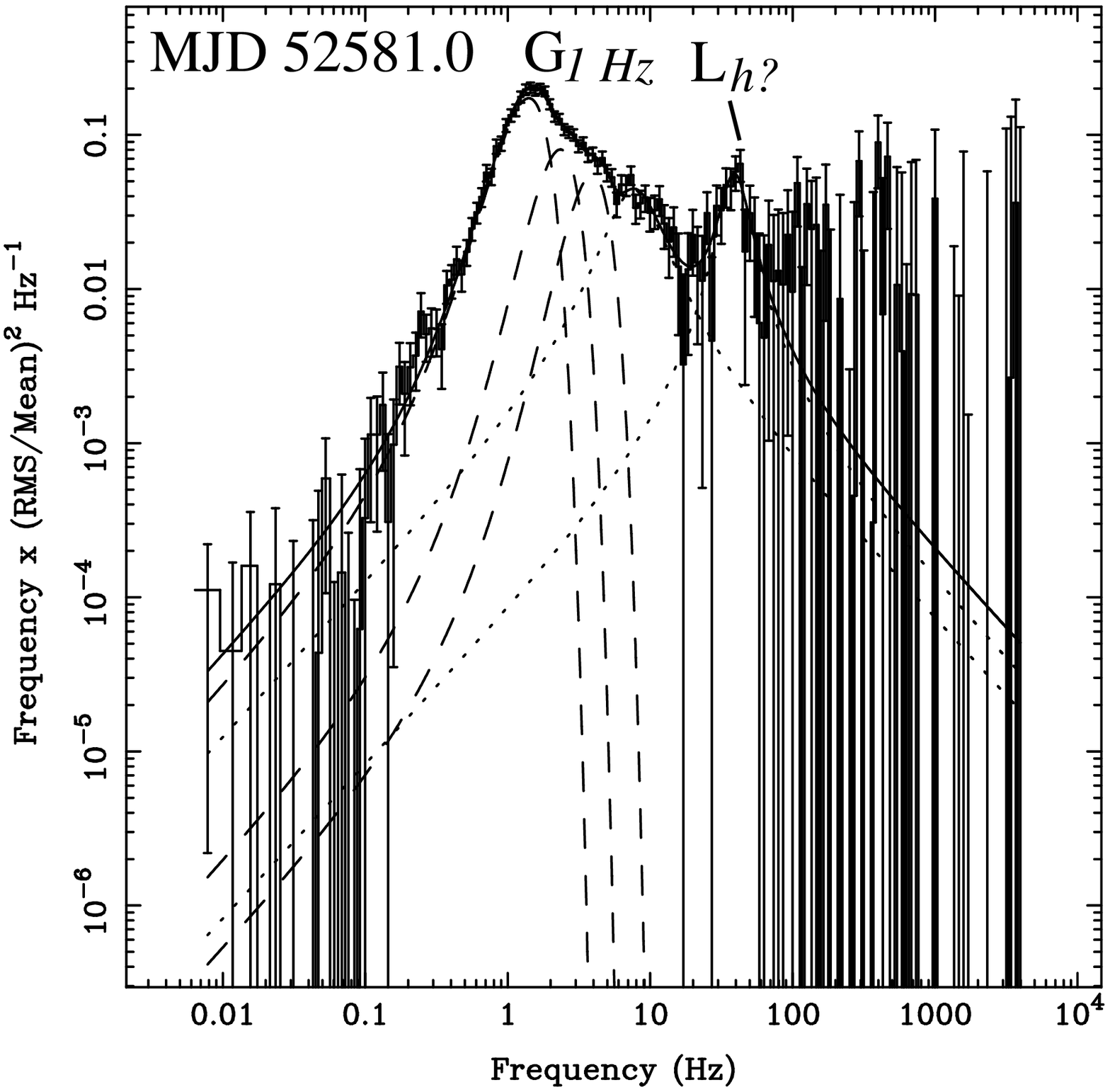} & \\
\end{tabular}
\caption{Six representative power spectra and fit functions in the power spectral density 
times frequency representation for the 2002 outburst of SAX J1808.4--3658. 
The different lines in the first 5 panels mark the individual Lorentzian components of the fit 
that are also indicated in the plots. 
The dashed lines in the sixth panel mark the Gaussian components of the fit, the dotted lines 
mark the Lorentzian components.
The arrows indicate the 401 Hz pulsar spike that was excluded during the fit.
The group numbers are also indicated in the plots.}
\label{fig.pds_1808_2002}
\end{figure}

\begin{figure}
\figurenum{3}
\epsscale{0.7}
\plotone{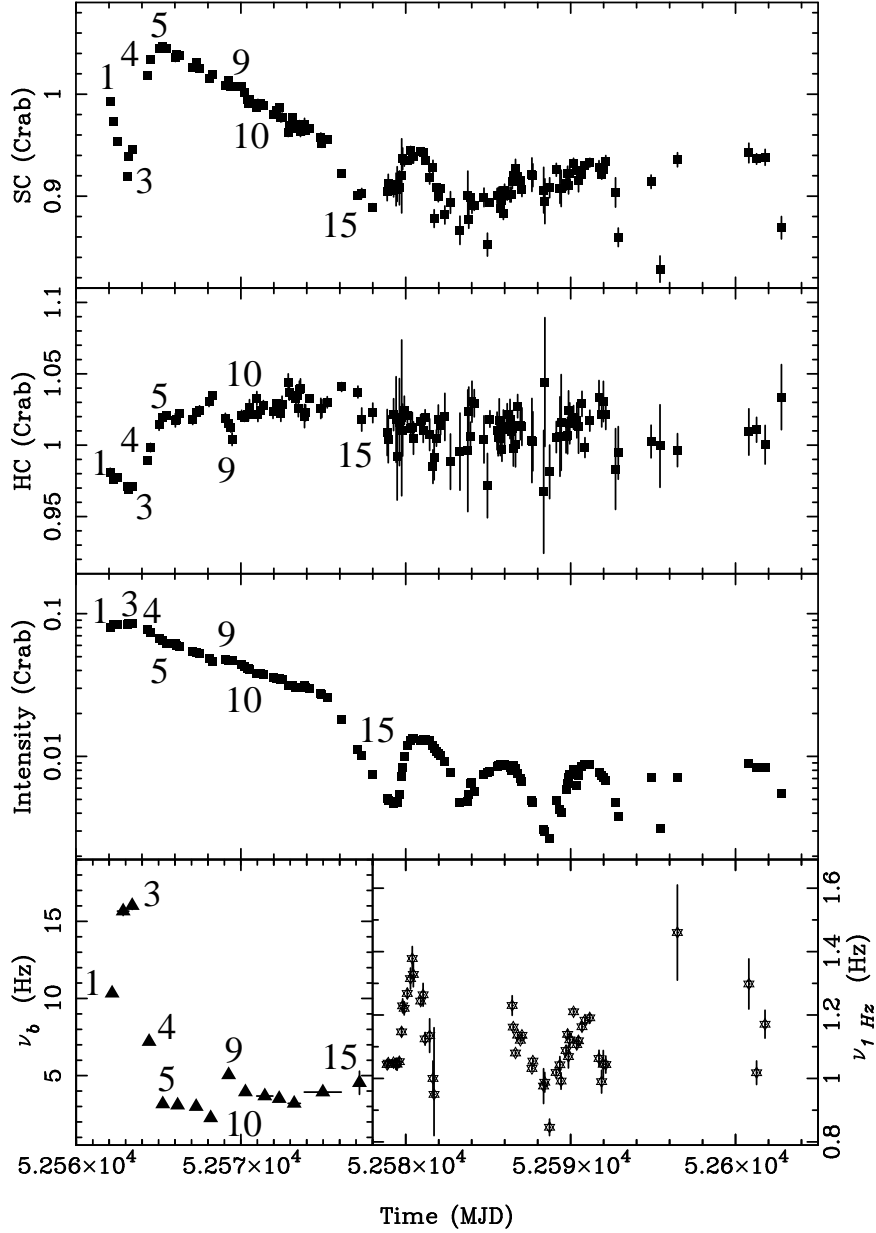}
\caption{Hard color (top panel), soft color (second panel from the top), 
intensity (third panel from the top), and the characteristic frequencies of 
two of the power spectral components (bottom panel) plotted versus time for 
the 2002 outburst of SAX J1808.4--3658. In the bottom panel the triangles mark
$\nu_{b}$ and the stars the centroid frequency of the main Gaussian peak used 
to fit the 1 Hz complex (see \S \ref{sec.1808_2002}). For clarity some of the 
group numbers are indicated. Errors on time in the bottom plot 
indicate the addition of several observations to improve the statistics 
(see \S \ref{sec.anal_timing}). }
\label{fig.nub_andcolors_1808_2002}
\end{figure}

In this paragraph we 
describe the color and timing behavior of SAX J1808.4--3658 during the 2002 outburst as a function of time 
with the help of Figures \ref{fig.pds_1808_2002} and \ref{fig.nub_andcolors_1808_2002}. 
In Figure \ref{fig.pds_1808_2002} we show six representative power spectra for the 2002 outburst. 
In Figure \ref{fig.nub_andcolors_1808_2002} we plot hard color,
soft color, intensity \citep[see also][]{wijnands03:nat424} and the frequency of one of two timing components (depending on epoch) versus time. 
We use the terminology of \citet{belloni02:apj572} and \citet{vstr03:apj596} in identifying the timing features (each Lorentzian 
component is indicated with an L, for example the upper kilohertz QPO is called L$_u$).
In the first group {\bf (group 1)} a broad L$_{b}$ at $\sim10$ Hz, L$_{h}$ 
at $\sim50$ Hz, L$_{hHz}$ at $\sim200$ Hz and L$_{u}$ at $\sim600$ Hz are observed (see Fig. \ref{fig.pds_1808_2002}). 
As the intensity increases to a maximum, both the hard color and the soft color drop to a 
minimum (Fig. \ref{fig.nub_andcolors_1808_2002}). The power 
spectra {\bf (groups 2 and 3)} now show L$_{u}$ at a higher characteristic frequency than in group 1. 
L$_{hHz}$ is also present but at a rather high frequency ($\sim300$ Hz). 
The low frequency part of the power spectrum is complex; L$_{h}$ is at a higher frequency than in group 1, and 
below L$_{h}$ another low frequency QPO appears that can be identified with L$_{LF}$ (see \S \ref{sec.identify}). L$_{b}$ 
changes from a broad band-limited noise component into a QPO and L$_{b2}$ appears at a lower characteristic 
frequency (see \S \ref{sec.identify}). Group 3 is the only group where the lower kilohertz QPO, L$_\ell$, can be 
detected \citep[see also][]{wijnands03:nat424}. In the fit we fixed the $Q$ value to the value that was found by 
\citet{wijnands03:nat424}. This was done because with the rebinning used in our fit we only had a few points 
in this peak.
After the maximum, the intensity starts to decrease while the hard color and the soft color increase. The power 
spectrum of {\bf group 4} is similar to that of group 1, showing L$_{b}$, L$_{h}$, L$_{hHz}$, and L$_{u}$, 
at somewhat lower characteristic frequencies.  While the intensity continues to decrease, the power spectrum 
{\bf (group 5)} still shows these components with $\nu_b$, $\nu_h$, 
and $\nu_u$ even lower. The characteristic frequency 
of L$_{u}$ is now below the spin frequency of 401 Hz. While the intensity keeps 
decreasing and the soft color starts decreasing, the power spectrum remains 
about the same for {\bf groups 5--8}. After that there is a slight increase in intensity accompanied by a decrease in hard color
and an increase of $\nu_b$, $\nu_h$, and $\nu_u$ {(\bf group 9)}. Then, while intensity
and soft color decrease again, the power spectra of {\bf groups 10--15} are very similar to that of group 5. It is not 
influenced by the break in the intensity and soft color versus time curves that 
occurs at MJD 52575. In group 15, L$_{hHz}$ is no longer observed with an upper limit consistent with this component 
still being present at similar strength (see Table \ref{tbl.rms}).
After that no power spectral components are detected until MJD 52578 due to bad statistics \citep[see][]{wijnands03:nat424}. After 
MJD 52578 the behavior of the source has changed completely; the intensity fluctuates on time scales of days, and the power spectra show 
a strong QPO around 1 Hz (see Fig. \ref{fig.pds_1808_2002}). This behavior is similar to that observed during what appeared to be 
the end stage of an outburst in 2000 \citep{wijnands01:apj560,wijnands04:bepposax}. The 1 Hz QPOs could not be fitted well with 
Lorentzians. Instead, several 
Gaussians were required. In addition to the 1 Hz QPO sometimes another QPO is present at around 30 Hz. As an illustration 
the centroid frequency of the main Gaussian used to fit the 1 Hz QPO is plotted in the bottom panel of 
Figure \ref{fig.nub_andcolors_1808_2002}. The full behavior of the 1 Hz QPO is beyond the scope of this paper and a detailed analysis 
of this phenomenon is in progress.

\begin{figure}
\figurenum{4}
\epsscale{1.0}
\plotone{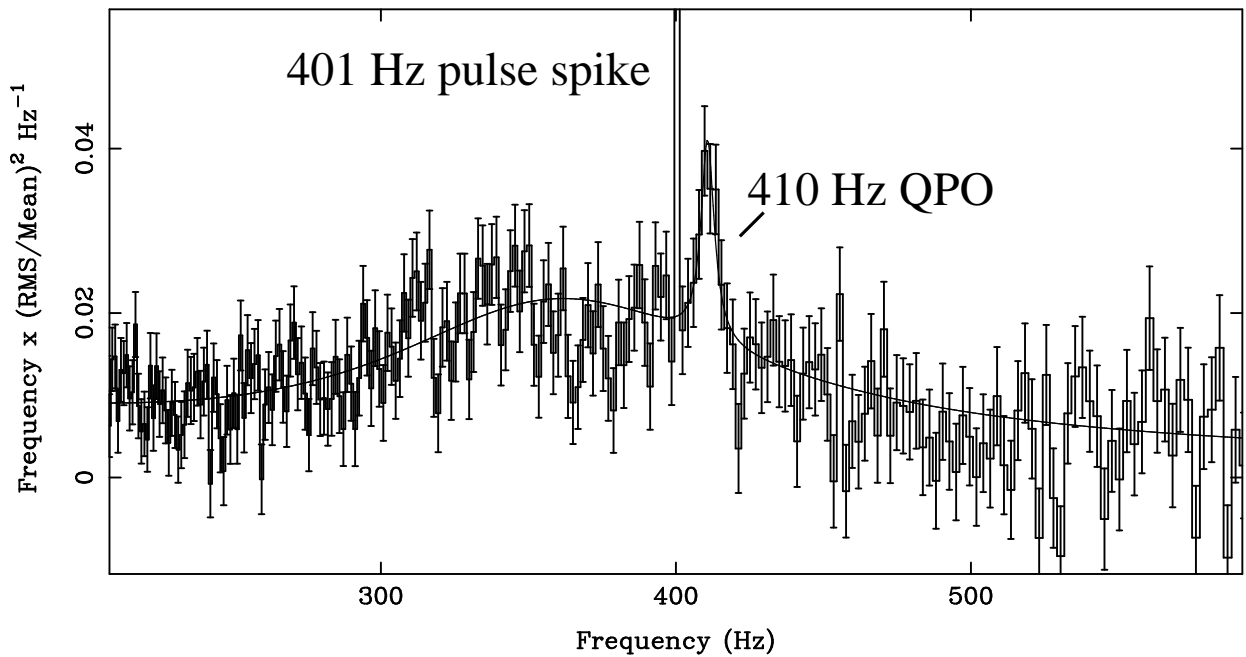}
\caption{The combined power spectrum and fit function of groups 5--6 and 9--11 in 
the power spectral density 
times frequency representation. 
The 410 Hz QPO and the $\sim401$ Hz pulse spike, that was excluded 
during the fit, are indicated.}
\label{fig.410hz}
\end{figure}

During the 2002 outburst \citet{wijnands03:nat424} detected a narrow QPO near 410 Hz. 
To search for these 
410 Hz QPOs we created power spectra similar to those of \S \ref{sec.anal_timing} but now in the 3--60 keV energy range.
This is the same energy range for which \citet{wijnands03:nat424} found these QPOs. We removed the Poisson level and the pulse spike 
and fitted the power spectra with the multi-Lorentzian fit function (see also \S \ref{sec.anal_timing}). 
To be able to detect the narrow 410 Hz QPO we had to use a finer rebinning than in \S \ref{sec.anal_timing}. We sometimes had 
to add several groups, adjacent in time, and for 
which the power spectra remained the same, to detect the 410 Hz QPO. We find the 410 Hz QPO in group 9 and in the combined 
groups 5--6, and 10--11. In Figure \ref{fig.410hz} we plot the combined power spectrum of groups 5--6, 9 and 10--11. 
The characteristic frequency of the 410 Hz QPO, the frequency difference between the $\sim401$ Hz pulse spike and the 
410 Hz QPO, and L$_{h}$ are listed in Table \ref{tbl.410hz}. We do not detect 410 Hz QPOs during the 1998 outburst.

\subsection{SAX J1808.4--3658: the 1998 Outburst}
\label{sec.1808_1998}

\begin{figure}
\figurenum{5}
\epsscale{0.38}
\plotone{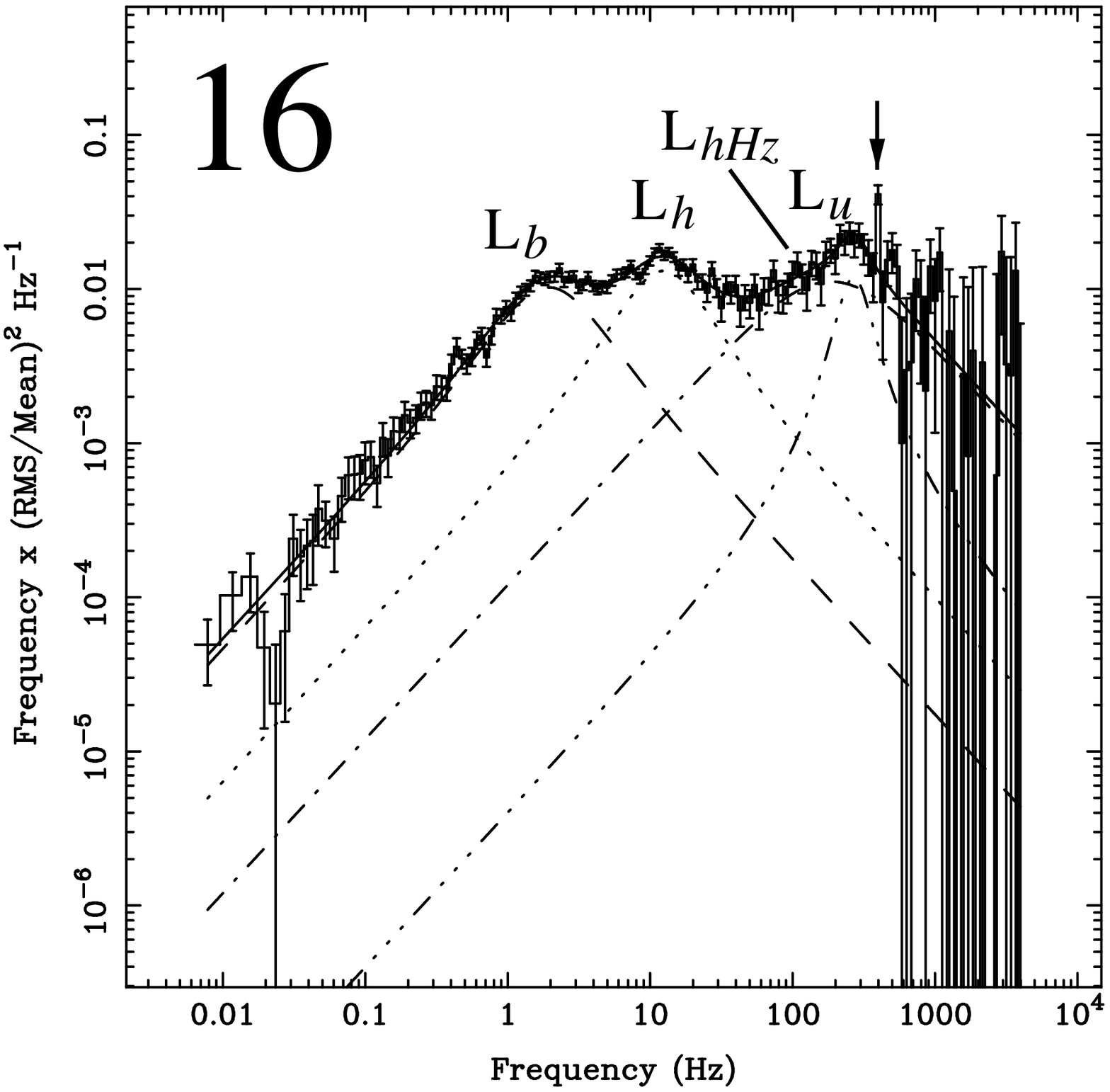} \\
\plotone{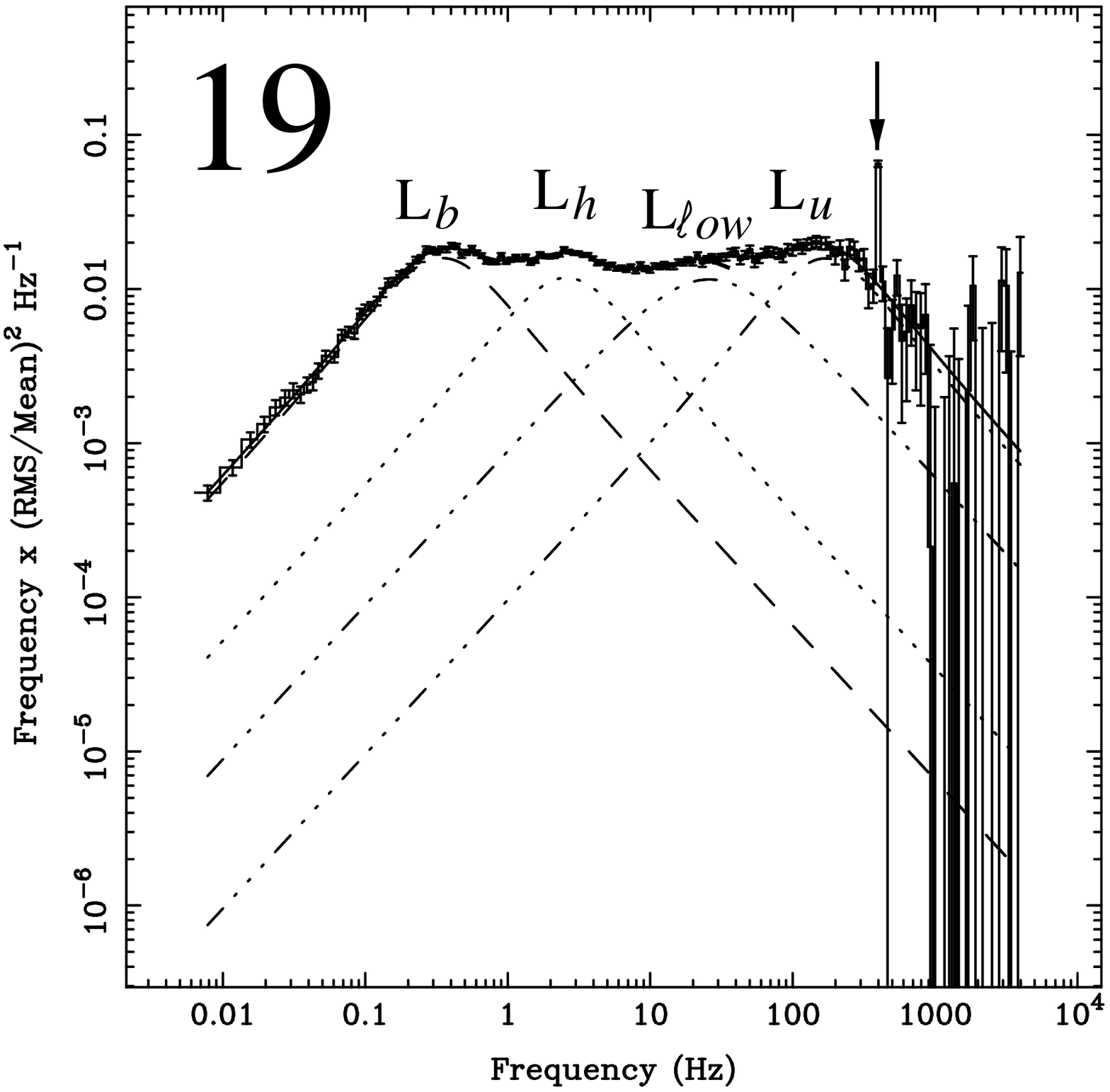} \\
\plotone{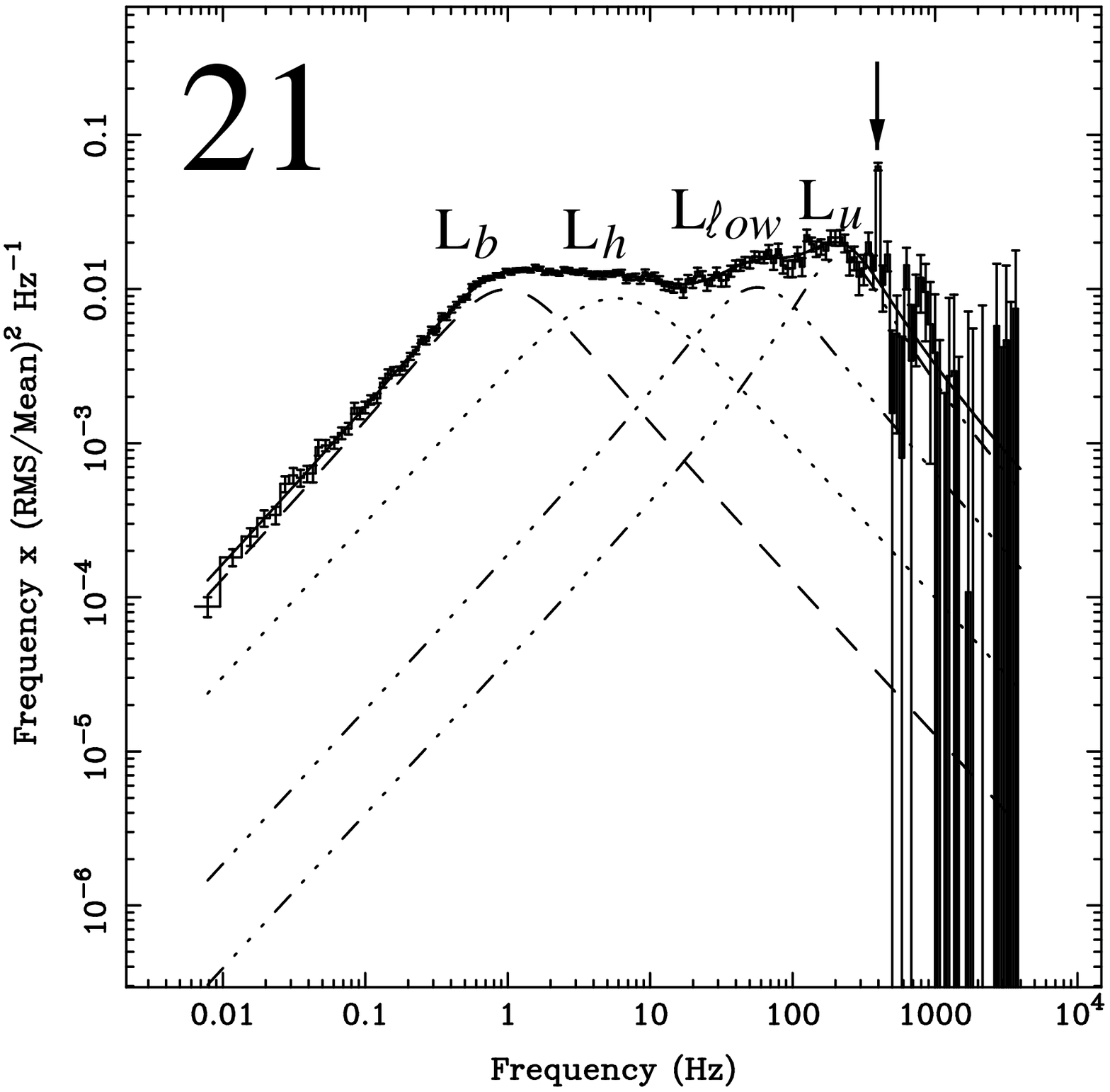} 
\caption{Three representative power spectra and fit functions in the power spectral density 
times frequency representation for the 1998 outburst of SAX J1808.4--3658. 
The different lines mark the individual Lorentzian components of the fit 
that are also indicated in the plots. The arrows indicate the 401 Hz pulsar spike that was excluded 
during the fit. The group numbers are also indicated in the plots.}
\label{fig.pds_1808_1998}
\end{figure}

\begin{figure}
\figurenum{6}
\epsscale{0.8}
\plotone{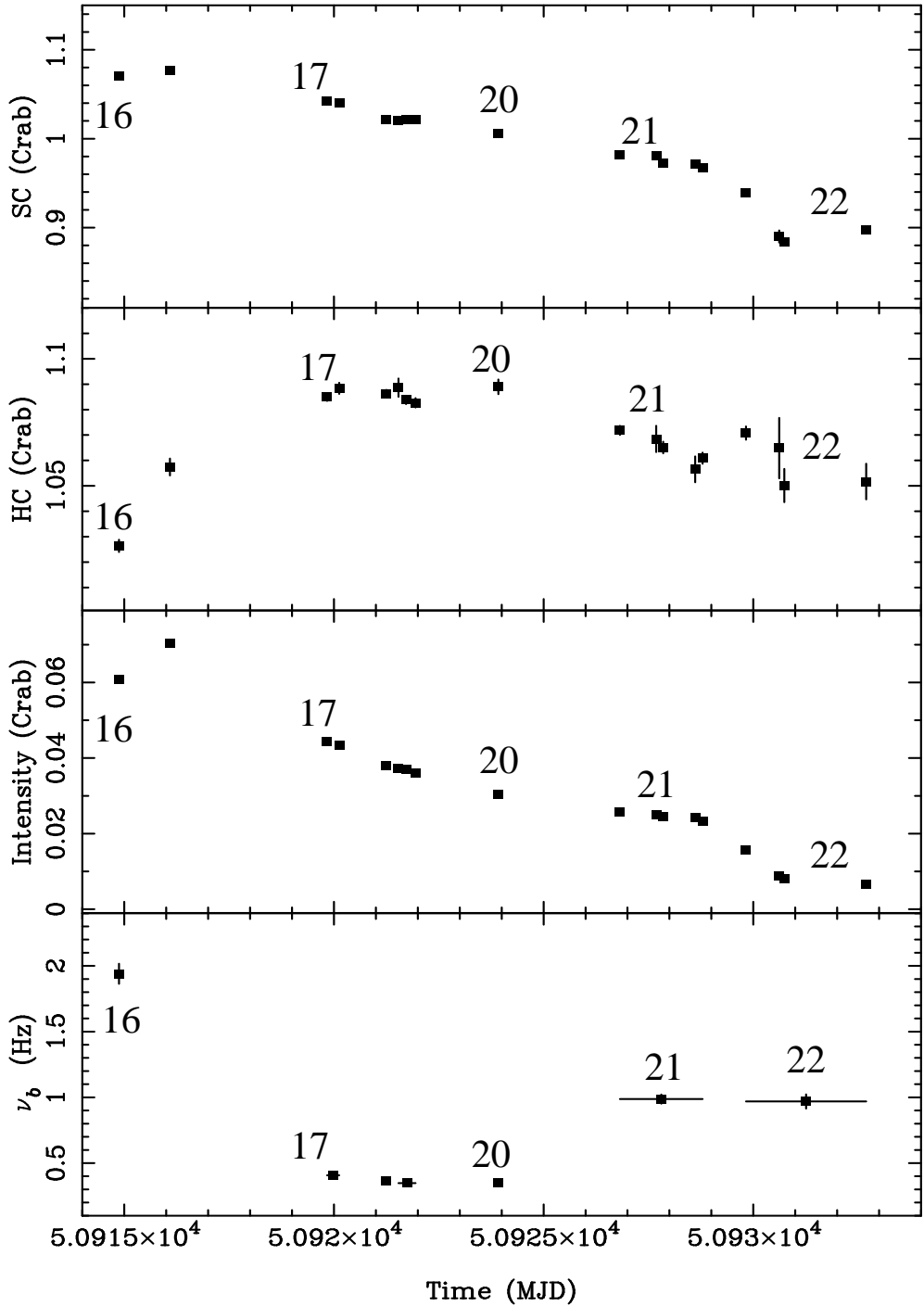}
\caption{Hard color (top panel), soft color (second panel from the top), 
intensity (third panel from the top), and $\nu_b$ (bottom panel) plotted versus time for 
the 1998 outburst of SAX J1808.4--3658. For clarity some of the 
group numbers are indicated. Errors on time in the bottom plot 
indicate the addition of several observations to improve the statistics 
(see \S \ref{sec.anal_timing}). }
\label{fig.nub_andcolors_1808_1998}
\end{figure}

In this paragraph we describe the color and timing behavior of SAX J1808.4--3658 during the 1998 outburst as a function of time 
with the help of Figures \ref{fig.pds_1808_1998} and \ref{fig.nub_andcolors_1808_1998}. 
The 1998 outburst of SAX J1808.4--3658 has been studied extensively previously; the broad-band timing properties where studied 
by \citet{wijnands98:apj507}, and the X-ray energy spectrum was studied by \citet{gilfanov98:aa338}, \citet{heindl98:apj506}, and \citet{gierdonebarret02:mnras331}. 
In {\bf group 16} the power spectrum 
shows L$_{b}$ at $\sim2$ Hz, L$_{h}$ at $\sim10$ Hz, L$_{hHz}$ at $\sim190$ 
Hz and L$_{u}$ at $\sim270$ Hz (see Fig. \ref{fig.pds_1808_1998}). The characteristic frequencies are all lower than during 
the 2002 outburst. Note, that the power of L$_{hHz}$ and L$_{u}$ could only be measured to an accuracy of better 
than 50\% (and not 33\%, see \S \ref{sec.anal_timing}). However, omitting either one of these components leads to one broad 
Lorentzian fitting both L$_{hHz}$ and L$_{u}$. For the second observation (at MJD 50916) there is not enough 
data left to create a power spectrum after the screening criteria have been applied (see \S \ref{sec.observ}). 
After a gap of four days, the intensity and soft color are lower and the hard color is higher. Then, as 
the intensity and soft color decrease, the power spectrum and hard color remain the same during {\bf groups 17--20}. The power 
spectrum (see group 19 in Fig. \ref{fig.pds_1808_1998}) shows L$_{b}$, L$_{h}$, and L$_{u}$ at the lowest characteristic 
frequencies observed for SAX J1808.4--3658. It also shows L$_{\ell ow}$ at about 30 Hz. In {\bf group 21} the intensity, soft 
and hard color are lower and $\nu_{b}$, $\nu_{h}$, $\nu_{\ell ow}$, and $\nu_{u}$ are higher (see Fig. \ref{fig.pds_1808_1998}). At 
MJD 50929, a change of slope in the intensity and soft color curves occurs.
This change of slope was also observed during the 2002 outburst (see 
Fig. \ref{fig.nub_andcolors_1808_1998}).  {\bf Group 22} was fitted with only two Lorentzians, L$_{b}$ and a broad Lorentzian 
fitting the higher frequencies. Due to lack of statistics L$_{h}$, L$_{\ell ow}$, and L$_{u}$ could not be identified. 

\subsection{XTE J1751--305}
\label{sec.1751}

\begin{figure}
\figurenum{7}
\epsscale{0.38}
\plotone{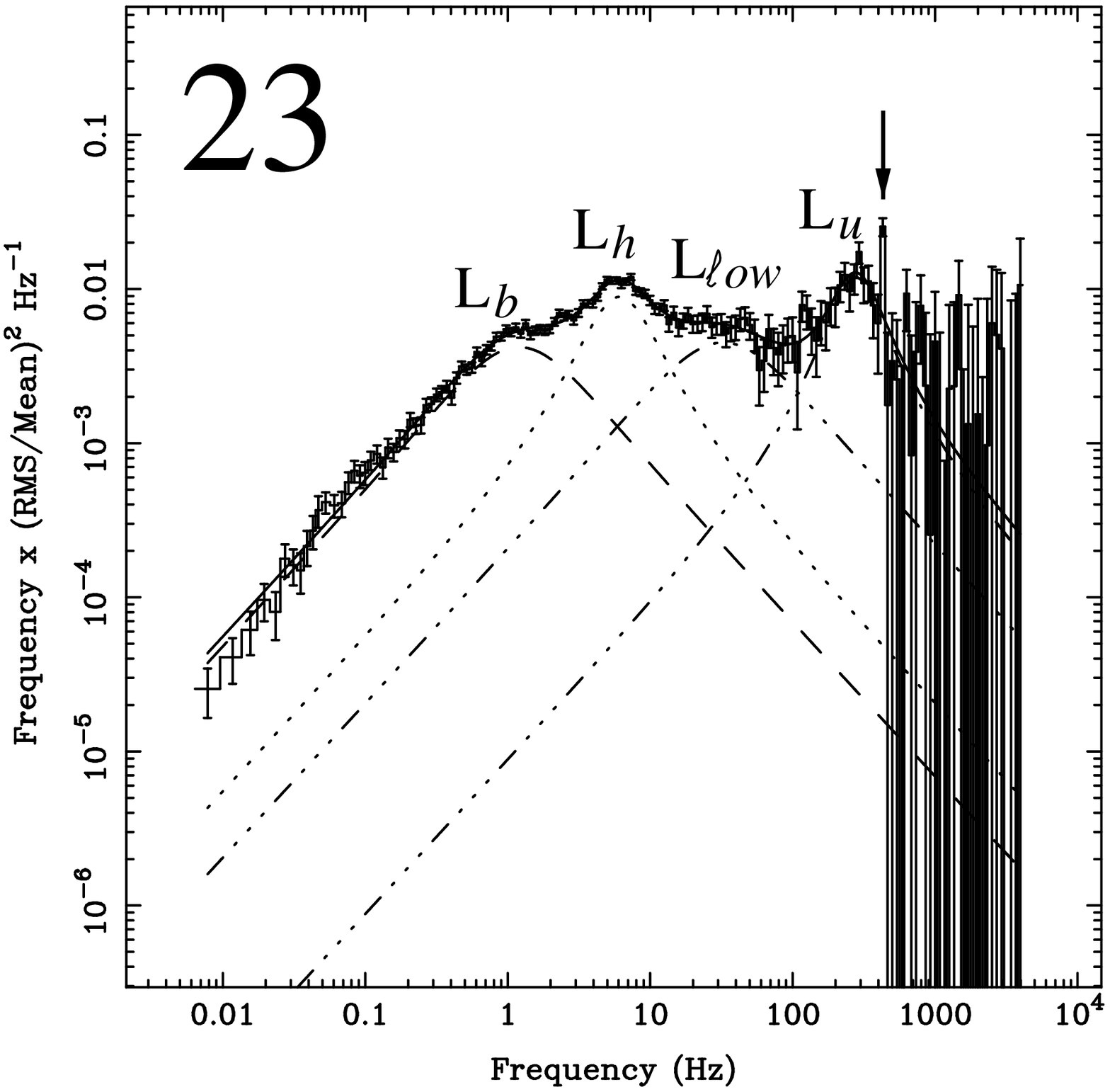} \\
\plotone{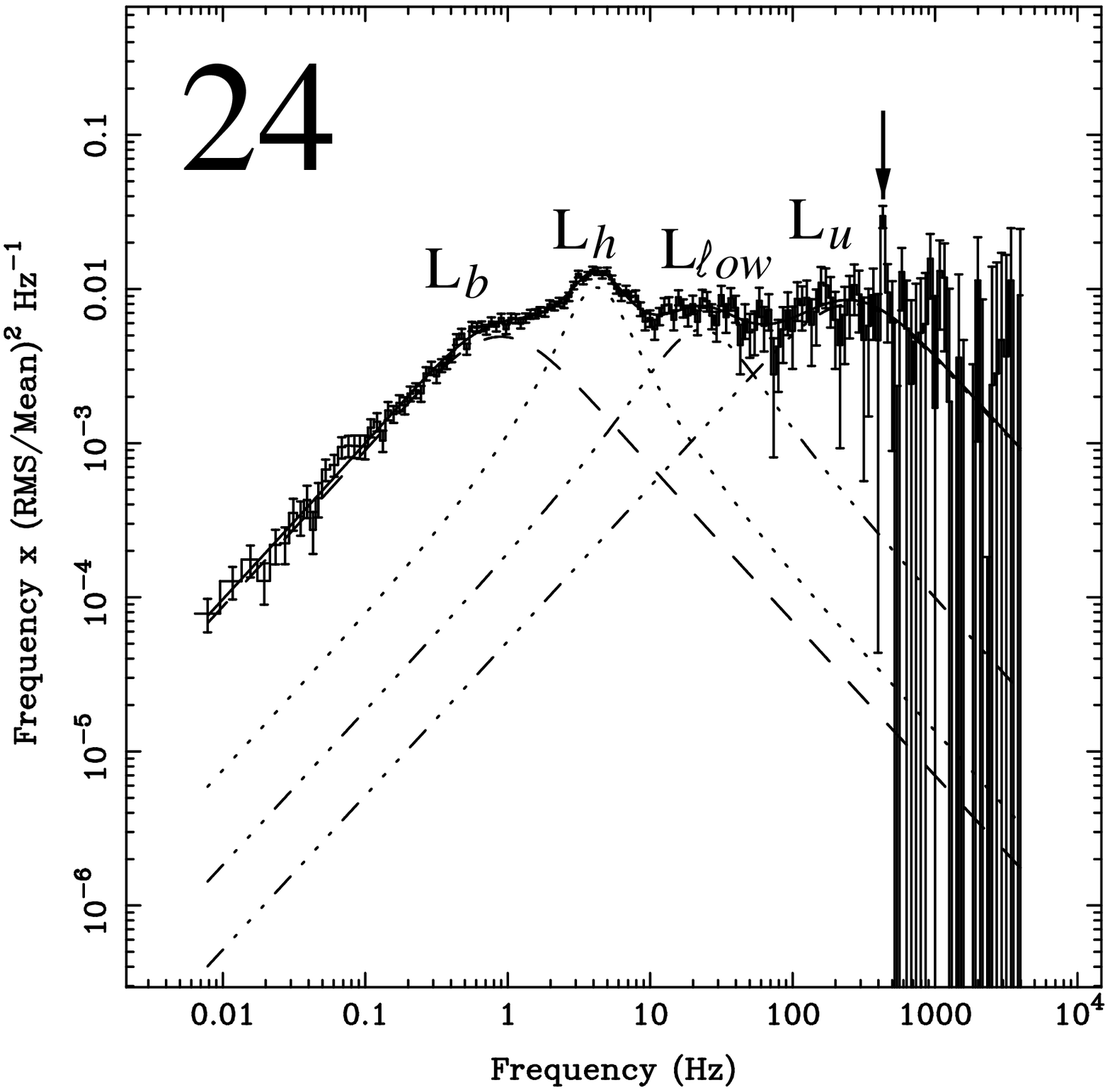} \\
\plotone{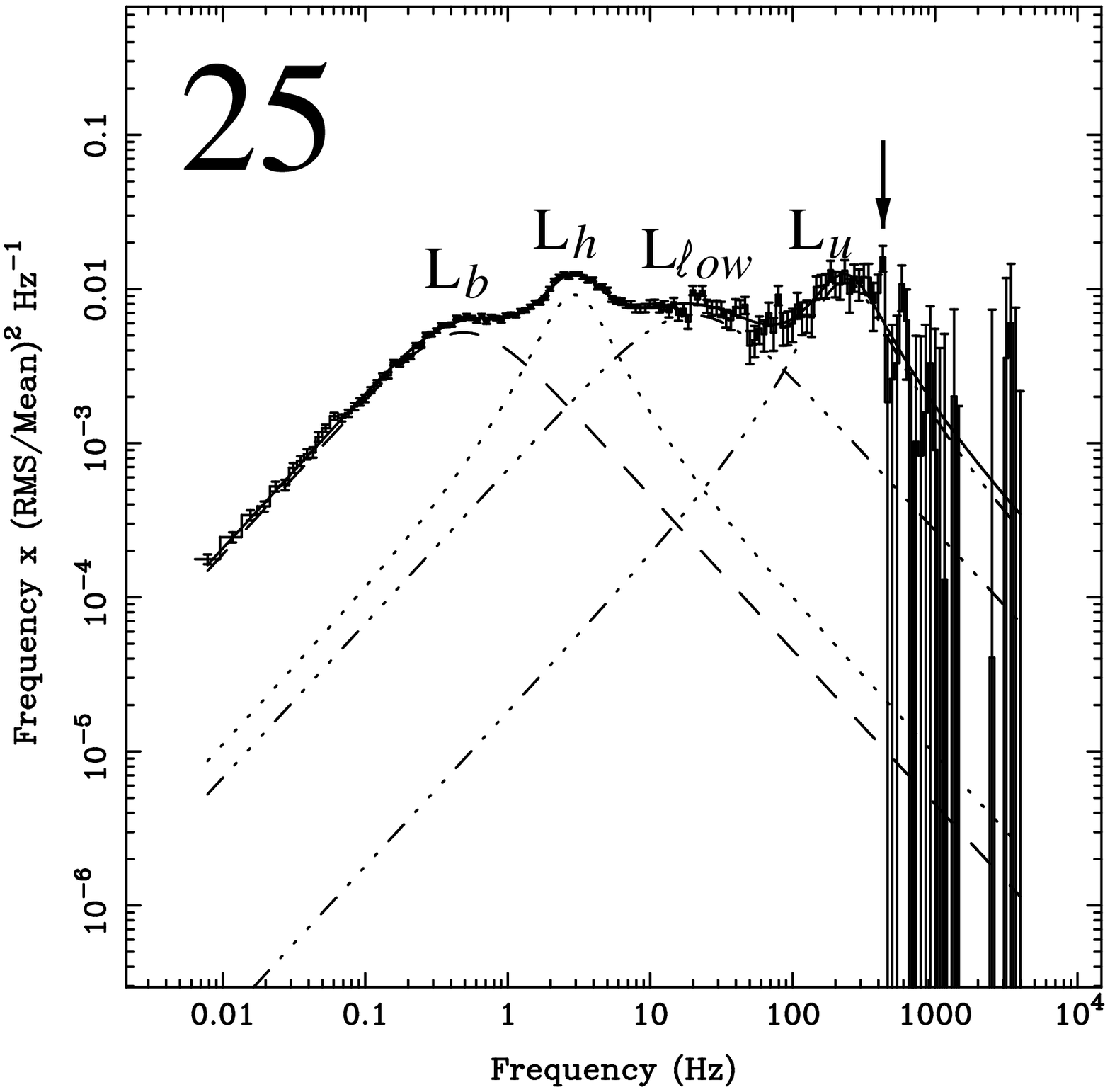} 
\caption{Three representative power spectra and fit functions in the power spectral density 
times frequency representation of XTE J1751--305. 
The different lines mark the individual Lorentzian components of the fit 
that are also indicated in the plots. The arrows indicate the 435 Hz pulsar spike that was excluded 
during the fit. The group numbers are also indicated in the plots.}
\label{fig.pds_1751}
\end{figure}

\begin{figure}
\figurenum{8}
\epsscale{0.8}
\plotone{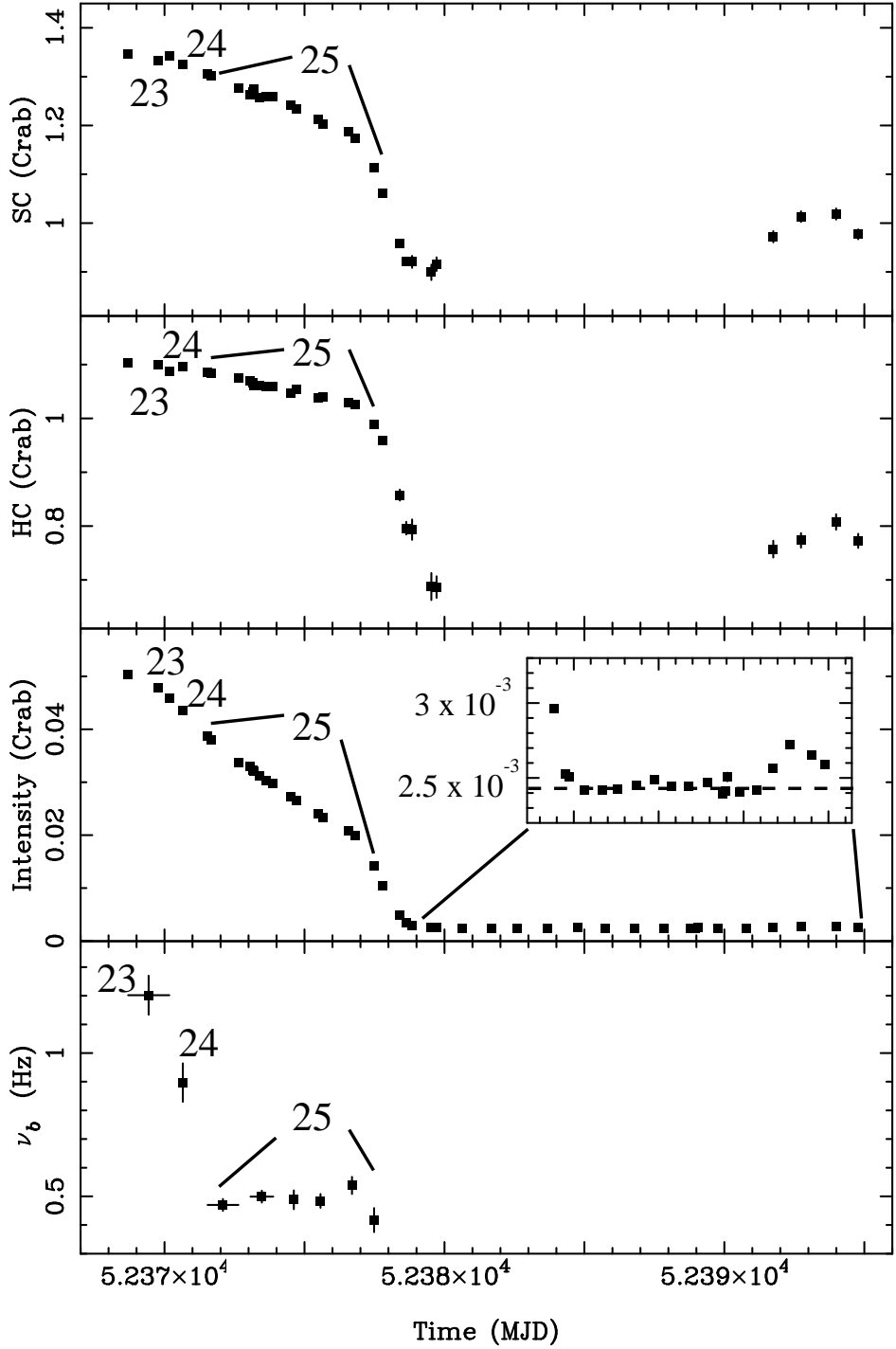}
\caption{Hard color (top panel), soft color (second panel from the top), 
intensity (third panel from the top), and $\nu_b$ (bottom panel) plotted versus time for 
the 2002 outburst of XTE J1751--305. Group numbers are indicated. Errors on time 
in the bottom plot 
indicate the addition of several observations to improve the statistics 
(see \S \ref{sec.anal_timing}). }
\label{fig.nub_andcolors_1751}
\end{figure}

In this paragraph we describe the color and timing behavior of XTE J1751--305 during its 2002 outburst as a function of time 
with the help of Figures \ref{fig.pds_1751} and \ref{fig.nub_andcolors_1751}. In Figure \ref{fig.pds_1751} we show three 
representative power spectra. In Figure \ref{fig.nub_andcolors_1751} we plot hard color, soft color, intensity and $\nu_b$ versus 
time. Note that in our analysis we subtract the background with {\it RXTE} background models (see \S \ref{sec.anal_col}). 
In addition to this there is also a significant background component for XTE J1751--305 due to nearby sources and Galactic diffuse 
emission \citep{markwardt02:apj575}. We draw the baseline (the dashed line in the insert of the intensity curve in 
Fig. \ref{fig.nub_andcolors_1751}) through the same observations as was done by \citet{markwardt02:apj575} and omit the soft and hard 
color points that are obtained when the source is actually undetected.

In the first observation {\bf (group 23)} XTE J1751--305 shows L$_{b}$ at $\sim1$ Hz, L$_{h}$ at $\sim6$ Hz, L$_{\ell ow}$ at $\sim30$ 
Hz, and L$_{u}$ at $\sim280$ Hz (see Fig. \ref{fig.pds_1751}). Then with intensity, hard color and soft color decrease (see Fig. 
\ref{fig.nub_andcolors_1751}), {\bf group 24} shows the same power spectral components as group 23, but at lower characteristic 
frequencies. One day later the characteristic frequencies are even lower. Then the power spectrum remains the same for 6
days. To further improve the power spectral statistics we combined these 6 days into {\bf group 25}. At MJD 52575, the intensity, 
hard and soft color curves show a change in slope similar to that in the intensity and soft color curves of the 1998 and 2002 outbursts
of SAX J1808.4--365 (see \S \ref{sec.1808_2002} and \ref{sec.1808_1998}). After the observations that span up group 25, the power 
spectrum still shows power but 
the statistics are not sufficient to identify the power spectral components. Between MJD 52380 and MJD 52391 the source is not detected, 
then a small reflare is observed (see the insert of the intensity curve in Fig. \ref{fig.nub_andcolors_1751}), for which it is uncertain 
whether it is from XTE J1751--305 \citep{markwardt02:apj575}. The statistics are too low to detect any power spectral components during the 
reflare.

\subsection{XTE J0929--314}
\label{sec.0929}

\begin{figure}
\figurenum{9}
\epsscale{0.38}
\plotone{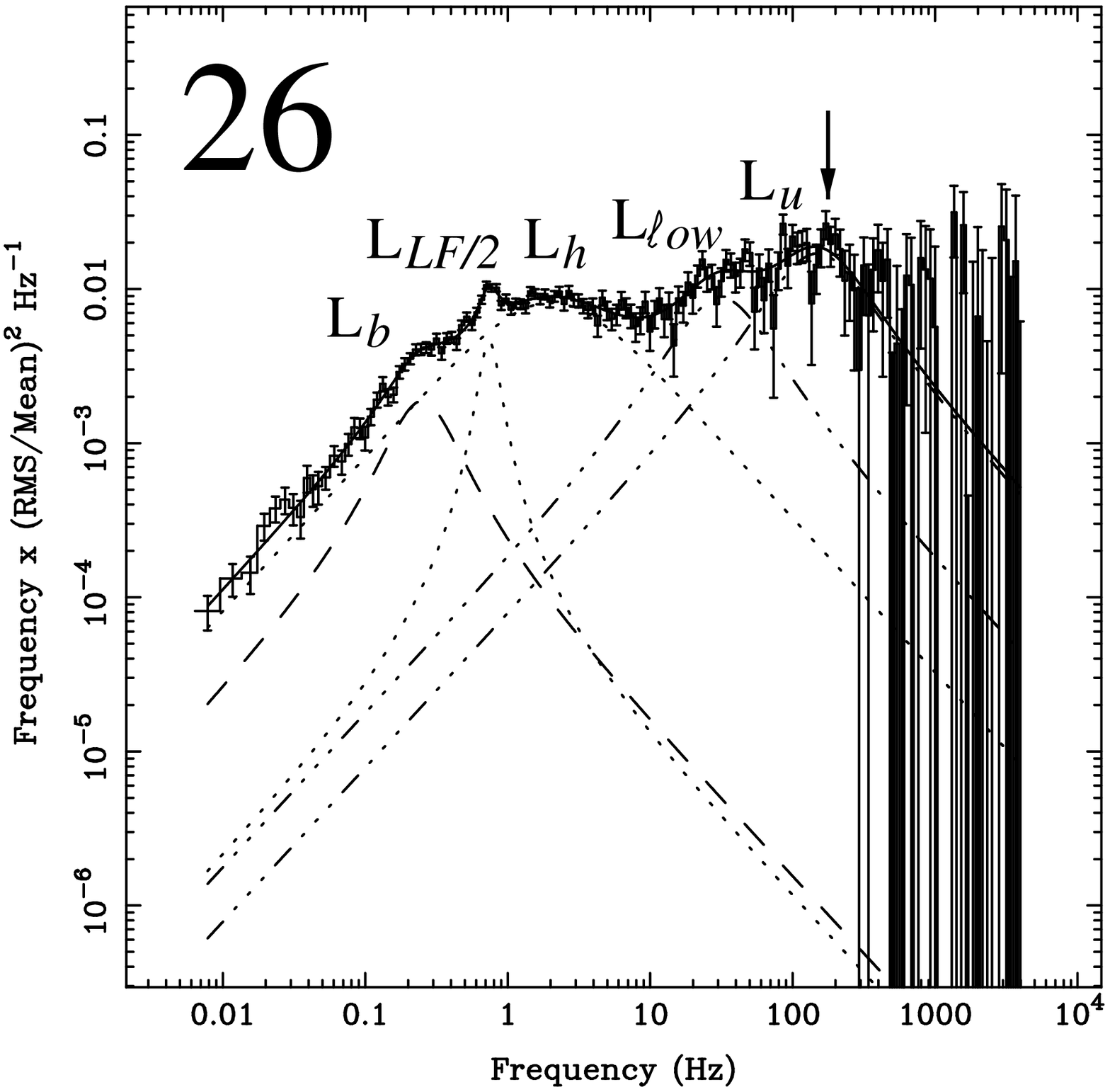} \\
\plotone{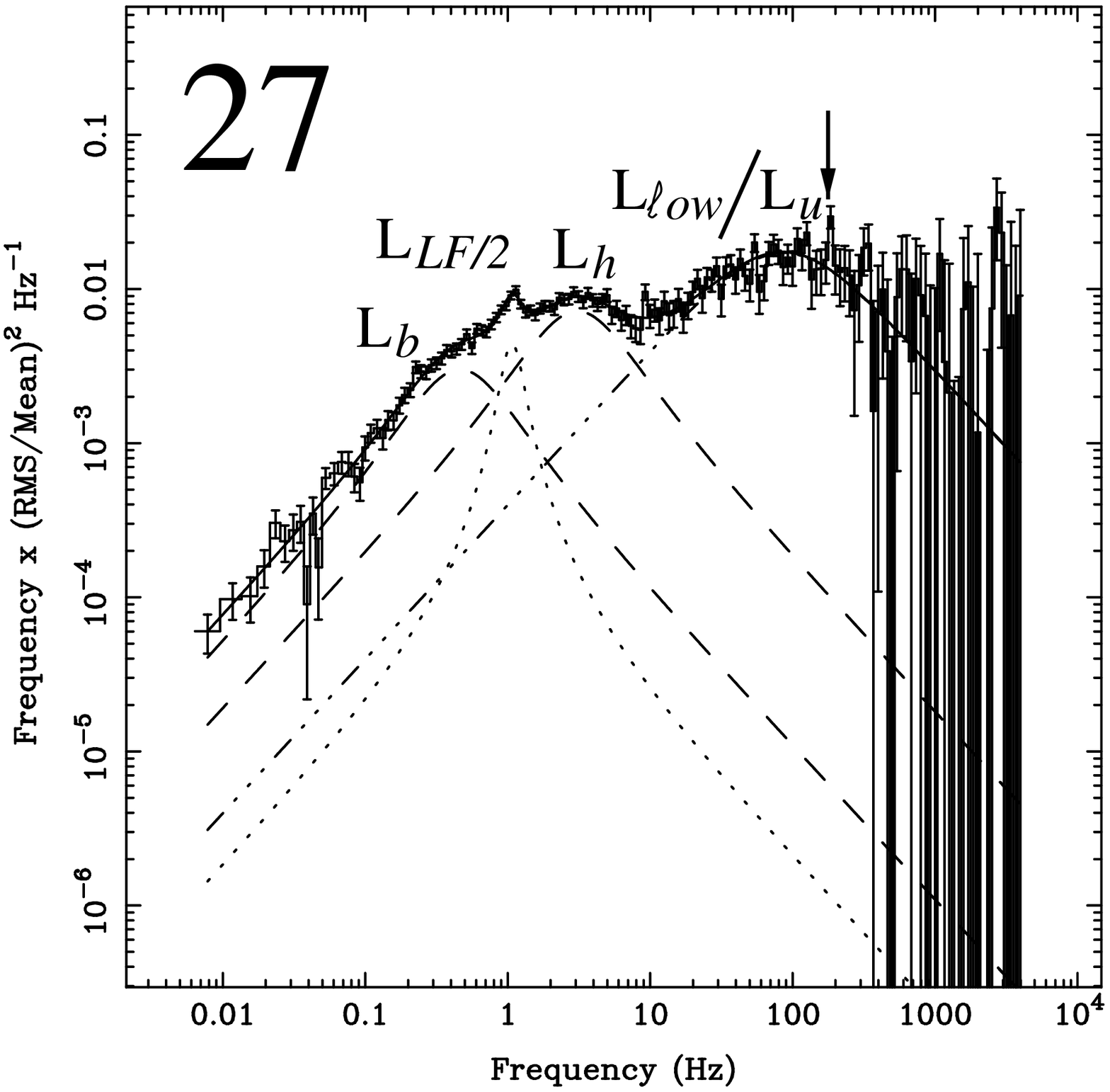} \\
\plotone{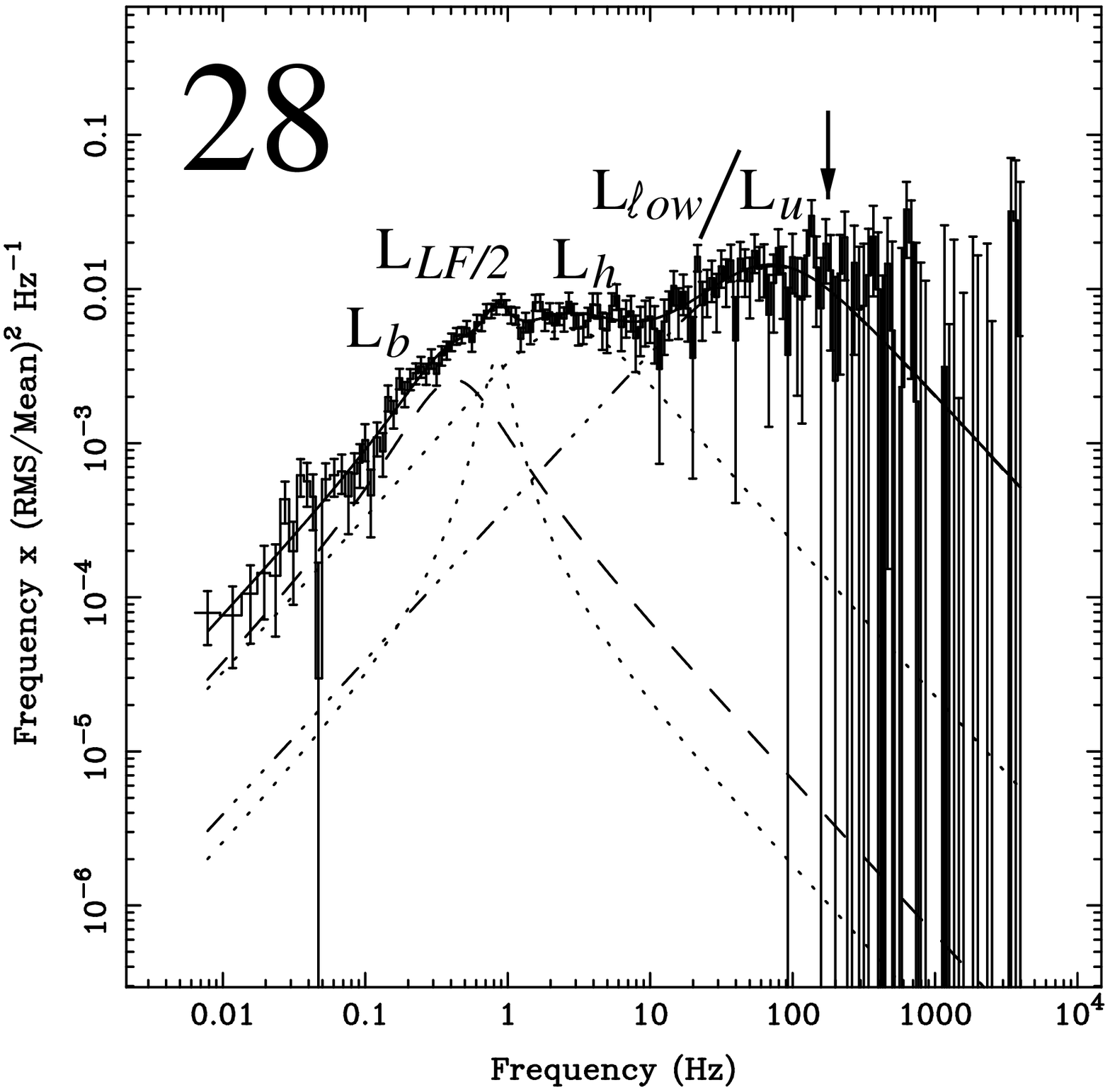} 
\caption{Three representative power spectra and fit functions in the power spectral density 
times frequency representation of XTE J0929--314. 
The different lines mark the individual Lorentzian components of the fit 
that are also indicated in the plots. The arrows indicate the 185 Hz pulsar spike that was excluded 
during the fit. The group numbers are also indicated in the plots.}
\label{fig.pds_0929}
\end{figure}

\begin{figure}
\figurenum{10}
\epsscale{0.7}
\plotone{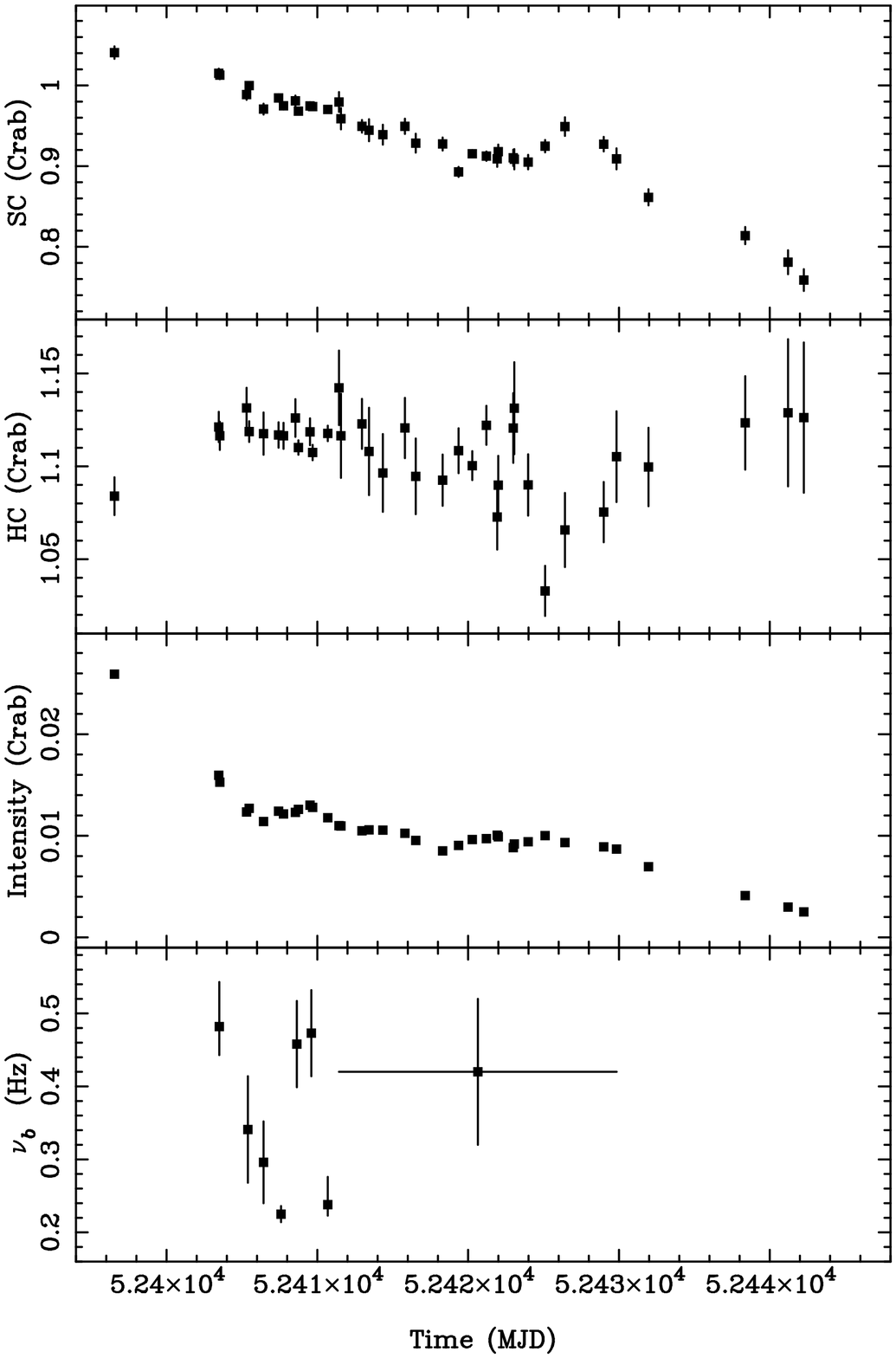}
\caption{Hard color (top panel), soft color (second panel from the top), 
intensity (third panel from the top), and $\nu_b$ (bottom panel) plotted versus time for 
the 2002 outburst of XTE J0929--314. Errors on time in the bottom plot 
indicate the addition of several observations to improve the statistics 
(see \S \ref{sec.anal_timing}). }
\label{fig.nub_andcolors_0929}
\end{figure}

In this paragraph we describe the color and timing behavior of XTE J0929--314 during its 2002 outburst with the help of 
Figures \ref{fig.pds_0929} and \ref{fig.nub_andcolors_0929}. In Figure \ref{fig.pds_0929} we show three representative power 
spectra. In Figure \ref{fig.nub_andcolors_0929} we plot hard color, soft color, intensity and $\nu_b$ versus time. 
For the first observation (at MJD 52396) there are not enough data left to create a power spectrum after the screening 
criteria have been applied (see \S \ref{sec.observ}). Then, after a gap of seven days, 
the intensity and soft color are lower and the hard color is higher (see Fig. \ref{fig.nub_andcolors_0929}). As the intensity 
and soft color keep declining the hard color becomes approximately constant. In the next seven days (MJD 52403--52410) the power 
spectrum shows at low frequencies L$_{b}$, L$_h$, and L$_{LF/2}$. L$_{LF/2}$ is a narrow QPO ($Q \approx 3$) with a 
characteristic frequency that varies between 0.6 and 1.3 Hz. At high frequencies the power spectrum shows one broad Lorentzian that probably 
(see below) fits both L$_{\ell ow}$ and L$_u$. To improve the statistics at high frequencies we add up the observations with 
$0.64 < \nu_{LF/2} < 0.86$ into 
group 26 and the observations with $0.99 < \nu_{LF/2} < 1.33$ into group 27. The power spectrum of {\bf group 26} 
shows L$_{b}$ at $\sim0.25$ Hz, L$_h$ at $\sim2$ Hz, and L$_{LF/2}$ at $\sim0.7$ Hz (Fig. \ref{fig.pds_0929}). At high frequencies 
L$_{\ell ow}$ and L$_u$ could now be identified. {\bf Group 27} is similar to group 26, except that, as mentioned above 
at high frequencies the fit could not distinguish between L$_{\ell ow}$ and L$_u$ (Fig. \ref{fig.pds_0929}). After that, from MJD 52411--52428, 
the power spectra still show power but the statistics are not sufficient to identify the power spectral components. If we
add up all observations from MJD 52411--52428 we obtain {\bf group 28}. This group is similar to group 27, 
and just as in group 27 the fit does not distinguish between L$_{\ell ow}$ and L$_u$ (Fig. \ref{fig.pds_0929}). 
After MJD 52428 the statistics become insufficient and the power spectra no longer show significant power.

\subsection{XTE J1814--338}
\label{sec.1814}

\begin{figure}
\figurenum{11}
\epsscale{0.5}
\plotone{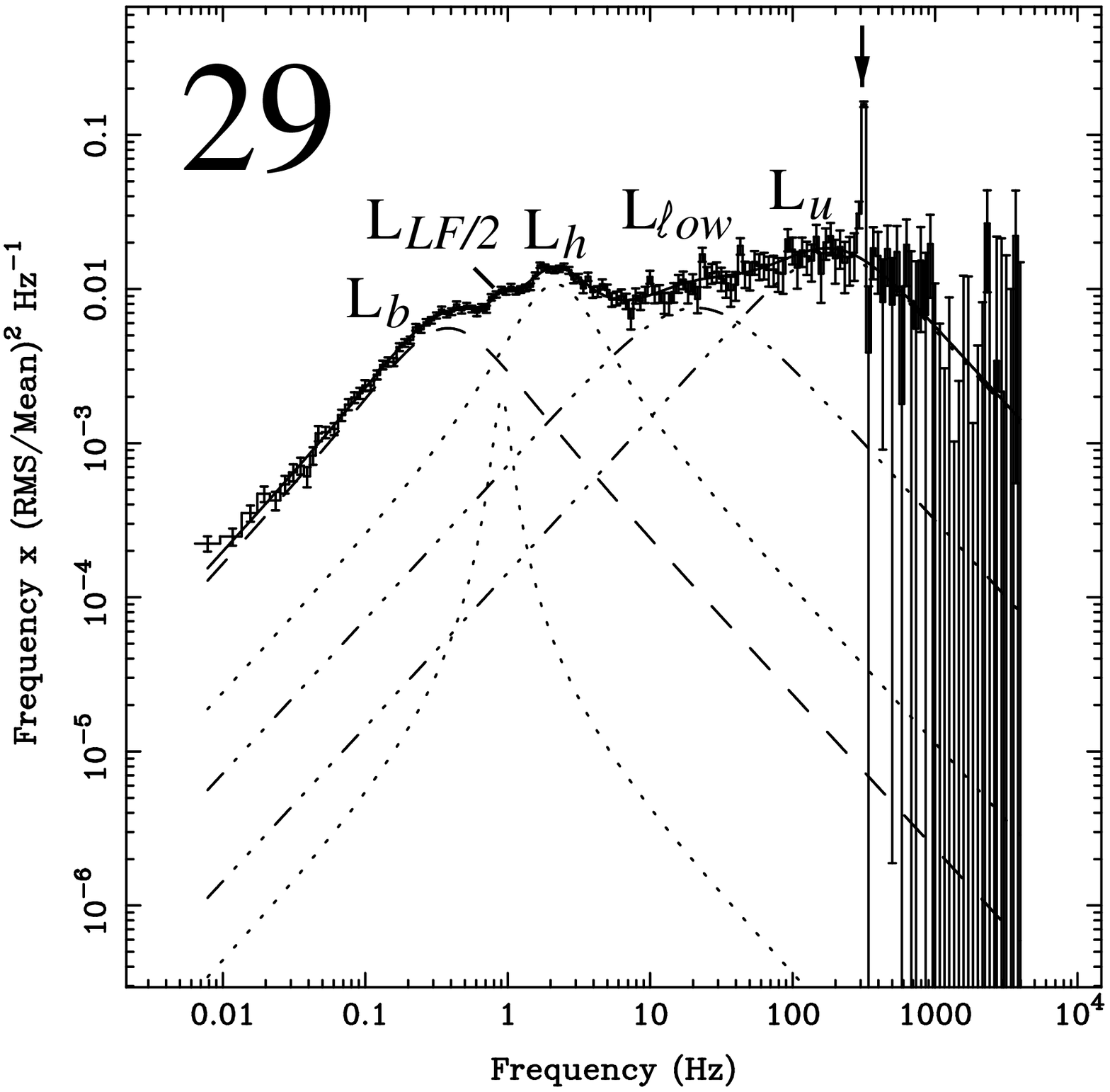} \\
\plotone{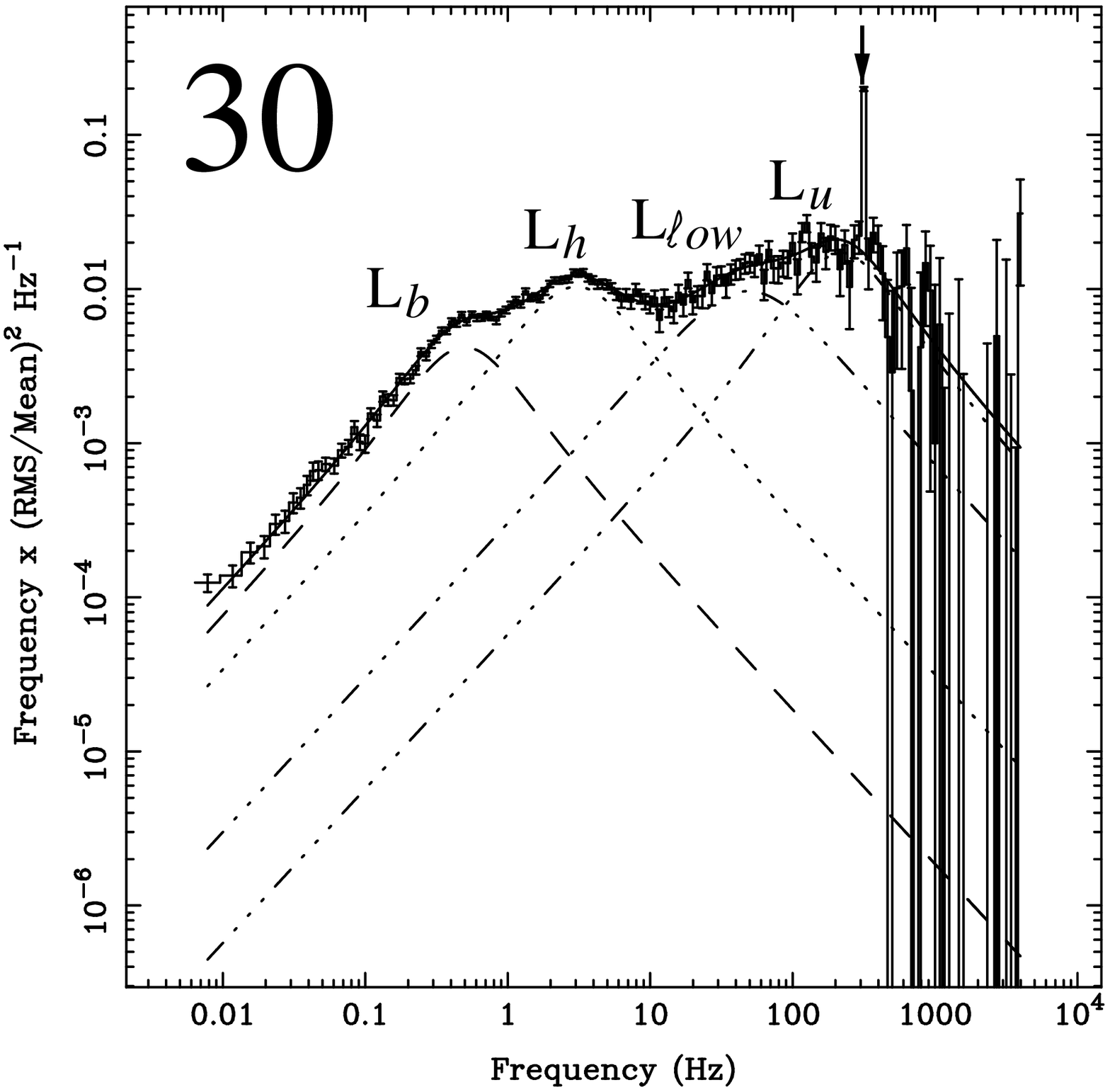}
\caption{Two representative power spectra and fit functions in the power spectral density 
times frequency representation of XTE J1814--338. 
The different lines mark the individual Lorentzian components of the fit 
that are also indicated in the plots. The arrows indicate the 314 Hz pulsar spike that was excluded 
during the fit. The group numbers are also indicated in the plots.}
\label{fig.pds_1814}
\end{figure}

\begin{figure}
\figurenum{12}
\epsscale{0.7}
\plotone{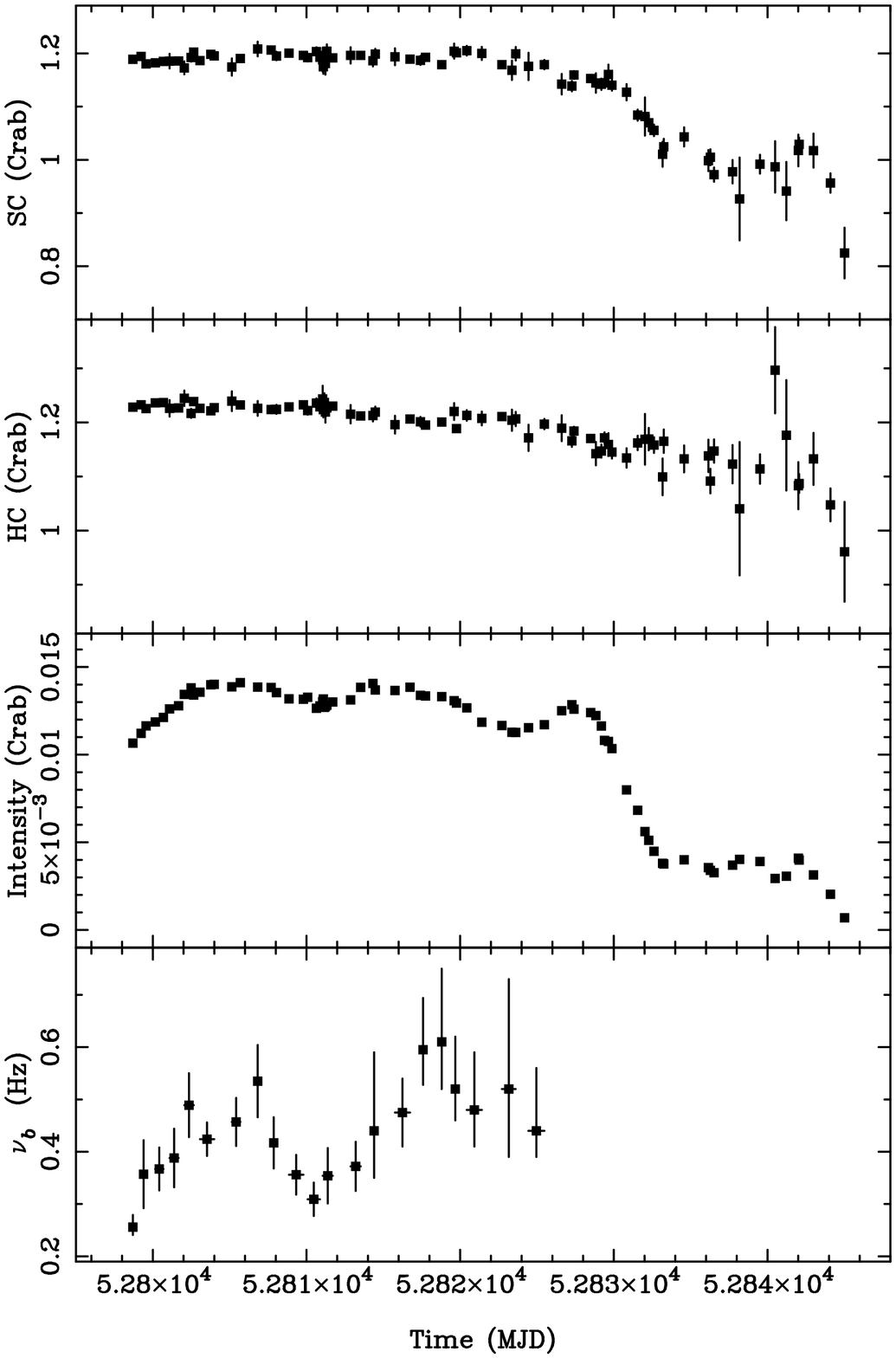}
\caption{Hard color (top panel), soft color (second panel from the top), 
intensity (third panel from the top), and $\nu_b$ (bottom panel) plotted versus time for 
the 2003 outburst of XTE J1814--338. Errors on time in the bottom plot 
indicate the addition of several observations to improve the statistics 
(see \S \ref{sec.anal_timing}). }
\label{fig.nub_andcolors_1814}
\end{figure}

In this paragraph we describe the color and timing behavior of XTE J1814--338 during its 2003 outburst as a function of time 
with the help of Figures \ref{fig.pds_1814} and \ref{fig.nub_andcolors_1814}. In Figure \ref{fig.pds_1814} we show three representative 
power spectra. In Figure \ref{fig.nub_andcolors_1814} we plot hard color, soft color, intensity and $\nu_b$ versus time. From MJD
52798 to MJD 52825 the power spectra at low frequencies show L$_{b}$ at $\sim0.5$ Hz and L$_{h}$ at $\sim3$ Hz. The characteristic
frequencies vary systematically (see Fig. \ref{fig.nub_andcolors_1814}), $\nu_b$ between 0.25 and 0.61 Hz, and $\nu_h$, which correlates 
with $\nu_b$, between 1.4 and 3.7 Hz. At high frequencies the power spectra are fitted with one broad Lorentzian that probably 
(see below) fits both L$_{\ell ow}$ and L$_u$. To improve the statistics at high frequencies we add up the observations with 
$0.31 < \nu_{b} < 0.39$ into group 29 and the observations with $0.41 < \nu_{b} < 0.61$ into group 30. {\bf Group 29}
shows L$_{b}$ at $\sim0.38$ Hz, L$_h$ at $\sim2.2$ Hz, L$_{\ell ow}$ Hz at $\sim21$ Hz and L$_u$ at $\sim190$ Hz (see 
Fig. \ref{fig.pds_1814}). It also shows the narrow L$_{LF/2}$ at $\sim0.9$ Hz. {\bf Group 30} shows, except for L$_{LF/2}$, the same 
components as group 29 (see Fig. \ref{fig.pds_1814}). In group 30 all characteristic frequencies are higher; 
$\nu_b \approx 0.5$ Hz, $\nu_h \approx 3$ Hz, $\nu_\ell \approx 50$ Hz, and $\nu_u \approx 220$ Hz.
After MJD 52825 the power spectra still show power between MJD 52826 and MJD 52829, but the statistics are not sufficient to 
identify the power spectral components. After  MJD 52830 the statistics become insufficient and the power spectra no longer show 
significant power.

\section{Timing and Colors}

Now that we have described and identified all power spectral components in the accreting pulsars SAX J1808.4--365, XTE J1751--305,
XTE J0929--314, and XTE J1814--338, we can link the timing properties to the position in the color-color diagram. We can also compare 
the timing properties and X-ray spectral properties to the classic island and banana states described by \citet{hasinger89:aa225} and compare 
with other sources.
An addition to the classic states described by \citet{hasinger89:aa225} is the extreme island state. 
At the lowest count rates, an extension of the island state is sometimes seen 
with a harder spectrum and stronger band-limited noise than the "canonical" 
island state. The term "extreme island state" has been used to designate such 
a state \citep[][]{prins97:aa319,reig00:apj530}. In \citet{reig00:apj530} this 
extreme island state was detached from the "normal" island state. For 4U 1608--52 
\citep[][]{vstr03:apj596} and now here for SAX J1808.4--3658 the change from 
extreme to "normal" island state is more gradual, so the question arises where 
to put the boundary. Here we have used the presence of L$_{\ell ow}$. In the extreme 
island state the power spectrum shows L$_{\ell ow}$ and in the island state L$_{\ell ow}$ is absent.

\subsection{SAX J1808.4--3658}
\label{sec.1808_colors}

\begin{figure}
\figurenum{13}
\epsscale{0.48}
\plotone{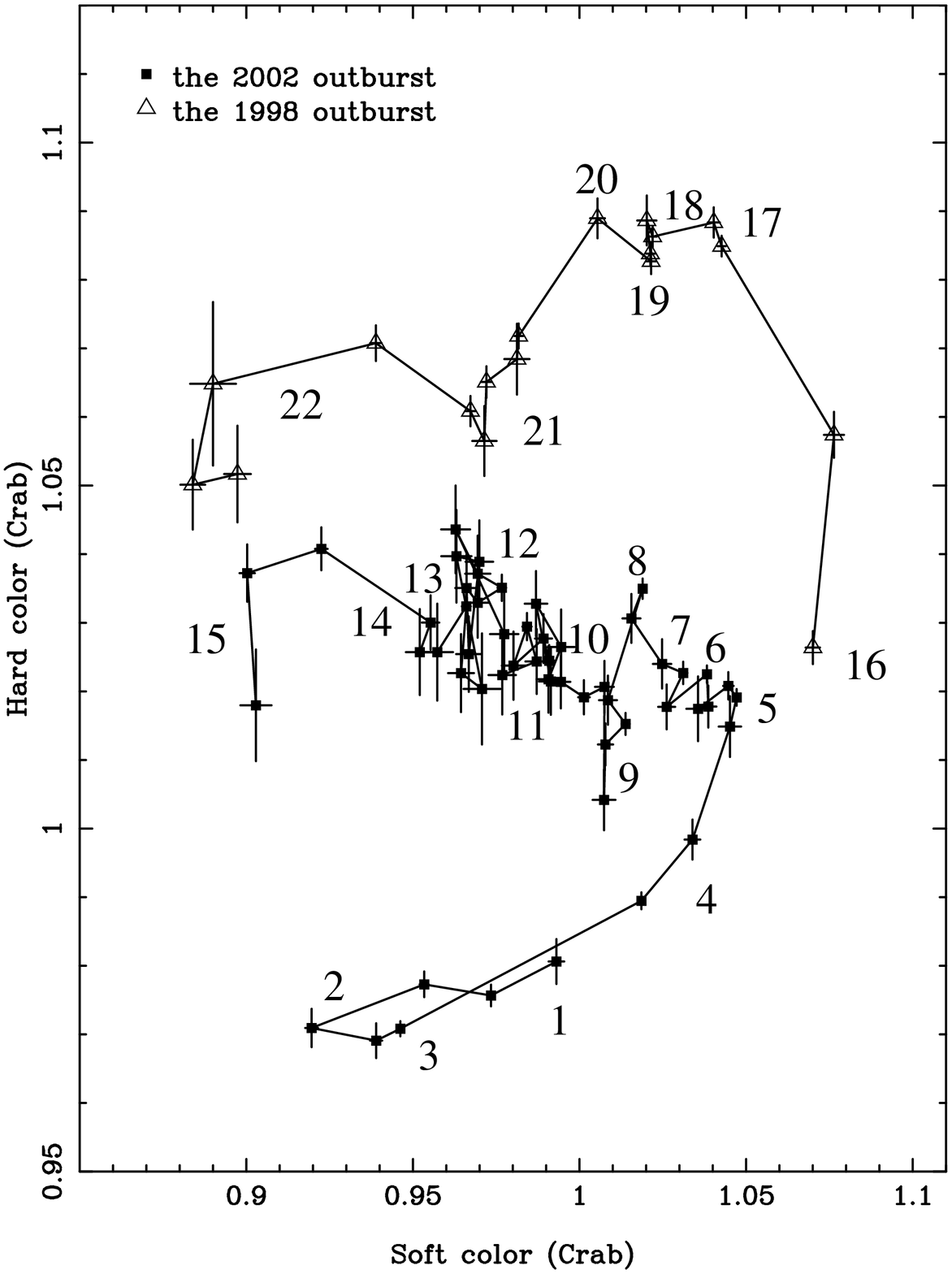}
\plotone{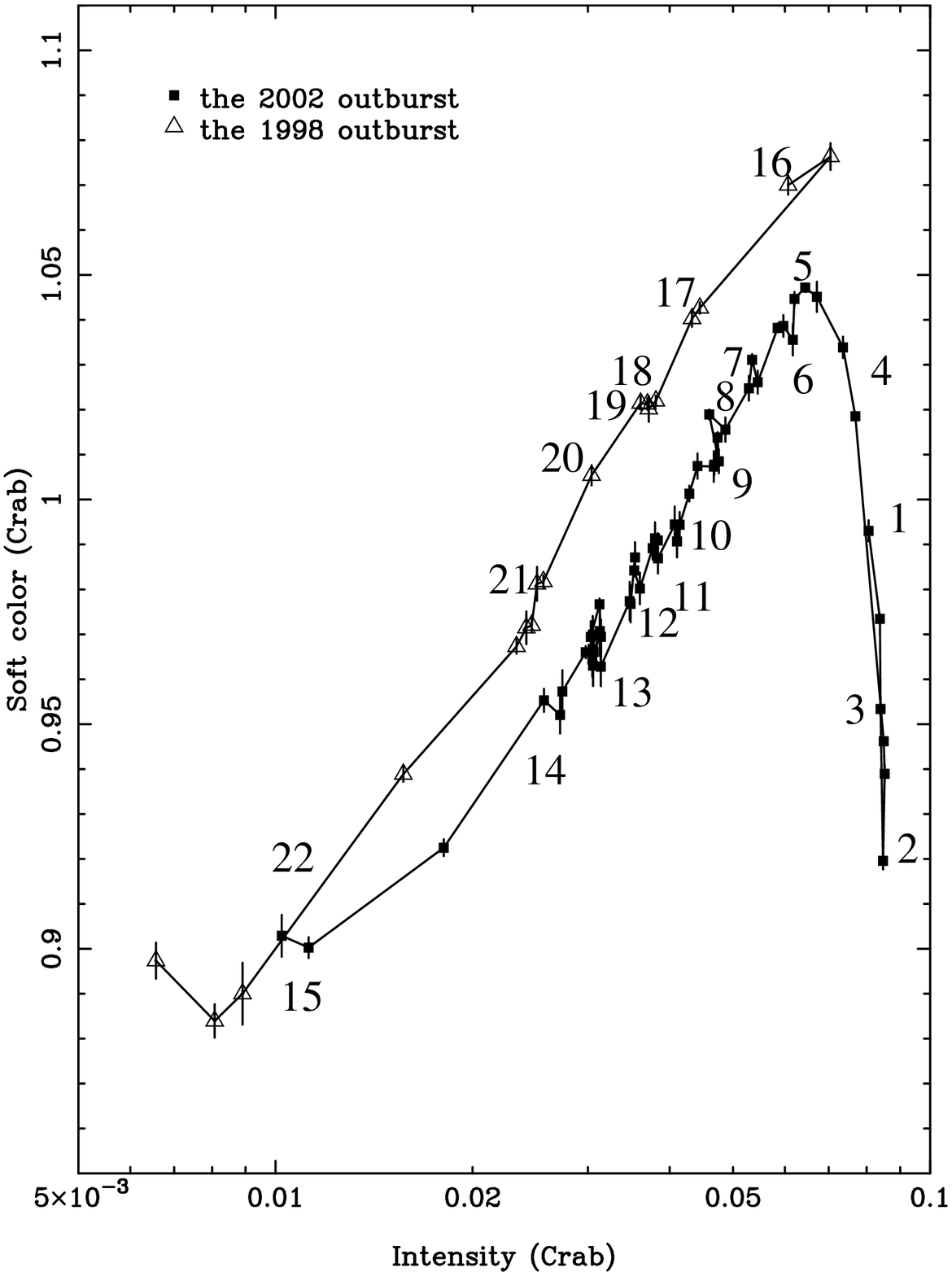}
\caption{Color-color diagram (left panel) and soft color vs. intensity diagram (right panel) of SAX J1808.4--3658. 
The 1998 outburst is marked 
by open triangles, the 2002 outburst by filled squares. The line connects observations that are adjacent in time.
The group numbers of the power spectra are indicated. Colors are in units of 
Crab (see \S \ref{sec.observ}).}
\label{fig.cc_1808}
\end{figure}

In Figure \ref{fig.cc_1808} we plot the color-color and soft color vs. intensity diagram of both the 2002 and 1998 outburst of 
SAX J1808.4--365. For the 
2002 outburst we only plot observations from before MJD 52578, at which time the behavior of the source changed completely 
(see \S \ref{sec.1808_2002}). At the start of the 2002 outburst {\bf (group 1)} the power spectrum is that 
of an island state; it shows strong broadband noise with low $\nu_{\rm max}$ and no very low frequency noise (VLFN). 
{\bf Group 2 and 3} show several power spectral characteristics of the lower left banana, the characteristic frequencies are higher, 
L$_\ell$ just appears, and the low frequency part of the power spectrum becomes complex. Only the VLFN that is typical for the banana 
state has not yet appeared. In {\bf group 4} the source is back in the island state. After that the source remains in an 
island state. The characteristic frequencies decrease until {\bf group 5}, and after that only small changes in the characteristic 
frequencies occur. During the whole 2002 outburst SAX J1808.4--365 did not go into an extreme island state. 
In an extreme island state
the power spectra show L$_{b}$, L$_h$, L$_{\ell ow}$ and L$_u$ all at very low characteristic frequencies \citep[][]{vstr03:apj596}. 
In 4U 1608--52 \citep[][]{vstr03:apj596} the boundary between island and extreme island state lies at a $\nu_b$ of about 2--3 Hz. In the 2002
outburst of SAX J1808.4--3658 L$_{\ell ow}$ was not observed and $\nu_b$ was always above 2 Hz.
During the 1998 outburst the characteristic frequencies of all components were lower than during the 2002 outburst.
In the first group {\bf (group 16)} of the 1998 outburst SAX J1808.4--365 was in the island state. After 
that the source was in an extreme island state showing L$_{\ell ow}$ and $\nu_b \leq 1$. 

\begin{figure}
\figurenum{14}
\epsscale{0.8}
\plotone{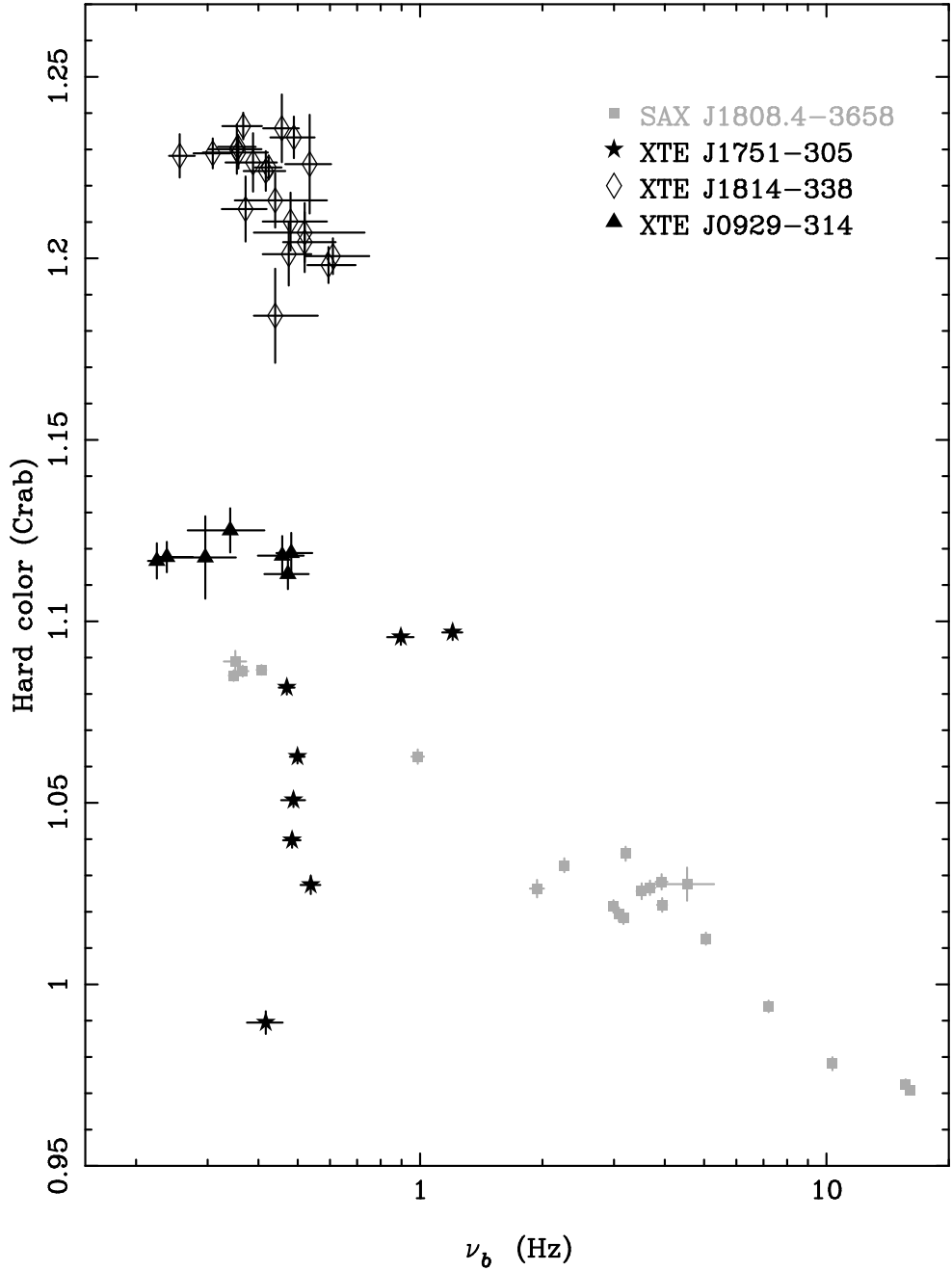}
\caption{Hard color plotted versus $\nu_{b}$ for SAX J1808.4--3658, 
XTE J1751--305, XTE J0929--314, and XTE J1814--338. The sources
are indicated in the plot.}
\label{fig.hc_vs_nub}
\end{figure}

The determination of the atoll source states based on the power spectral results is in agreement with the positions 
in the color-color diagram. SAX J1808.4--365 behaves similarly to the atoll sources 4U 1705--44, Aql X--1, and 
4U 1608--52 \citep{muno02:apj568,gierlinski02:mnras331,barret02:apj576,olive03:apj583,vstr03:apj596,reig04:apj} which in the extreme island state show 
horizontally extended branches in the color-color diagram.
At the beginning of the 2002 outburst SAX J1808.4--365 showed a state transition 
from the island state {\bf (group 1)} to {\bf groups 2 and 3} and back to the island state {\bf (group 4)}. This 
transition is seen in the color-color diagram as a lowering of the hard and soft color and a slight increase in intensity. 
At the lowest hard color (groups 2 and 3) the source has almost reached the lower left banana (see above). 
It seems unlikely that the actual banana branch is reached as the gap between group 2 and 3 is only 3000 s.
Then the hard color increases again and the source goes back to an island state. 
As noted previously \citep{mendez99:apj511,vstr03:apj596}, hard color appears to be the parameter best 
correlated to the power spectral characteristics in 
the island states, and this is confirmed here.
During the whole transition, and also in the other groups, the hard color is anti-correlated with the characteristic 
frequencies of the power spectral components (see Figure \ref{fig.hc_vs_nub}).
This behavior is typical for atoll sources (4U 0614+09, \citealt{vstr00:apj540}; 4U 1728--34, \citealt{disalvo01:546}; 
4U 1705--44, \citealt{olive03:apj583}; 4U 1608--52, \citealt{vstr03:apj596}).
Note that compared to other atoll sources the relative change in hard color during the transition is only subtle. 
For SAX J1808.4--3658 the relative change in hard color is about 5\% were it is $\geq 30\%$ for 
4U 1728--34 \citep[][]{disalvo01:546}, 4U 1705--44 \citep[][]{olive03:apj583}, and 4U 1608--52 \citep[][]{vstr03:apj596}. 
The relative change in soft color is about 10\%, comparable to the relative soft color changes for 4U 1728--34, 4U 1705--44, 
and 4U 1608--52.  
The behavior after the state transition can be described in the scheme proposed for the atoll sources by \citet{vstr03:apj596} (see their 
figure 12). 
The source remained in an island state in which the characteristic frequencies are low and hardly change after 
{\bf group 5}. Within the island state the soft color changes in correlation with intensity.
As the intensity decays the soft color decreases and a horizontally extended island branch appears. Note, that 
the constant frequencies make this branch very different from the Z source horizontal branch.
In other sources horizontally extended branches where only observed in the {\it extreme} island state \citep[][]{vstr03:apj596}, here we 
observe the same phenomenon but now in an island state. However, note that the characteristic frequencies in this island state are low 
and only the absence of L$_{\ell ow}$ distinguish it from an extreme island state (see above)
The small changes 
in characteristic frequency that do occur are accompanied by anti-correlated changes in hard color (see Fig. \ref{fig.hc_vs_nub}) and show 
no relation to soft color or intensity. In the 1998 
outburst the same phenomenon occurs; in {\bf groups 17--20} the power spectra remain in one extreme island state and as the intensity decreases so does the soft color and again a horizontally extended extreme island branch appears. In {\bf groups 21--22} the 
characteristic 
frequencies are higher and another horizontally extended extreme island branch is formed at a lower hard color. Thus in total 
three horizontally extended island branches (5--15, 21--22, and 17--20) appear above each other in the left panel of 
Figure \ref{fig.cc_1808}, and the frequencies on each of them anti-correlate with hard color. 
This anti-correlation with hard color also holds between the 1998 and the 2002 outbursts. For a large 
part (from group 5 of the 2002 outburst) both outbursts cover the same intensity range (see Fig \ref{fig.cc_1808}). 
At similar intensities the characteristic frequencies in the 1998 outburst are much lower, while the hard color is higher.

\subsection{XTE J1751--305}

\begin{figure}
\figurenum{15}
\epsscale{0.48}
\plotone{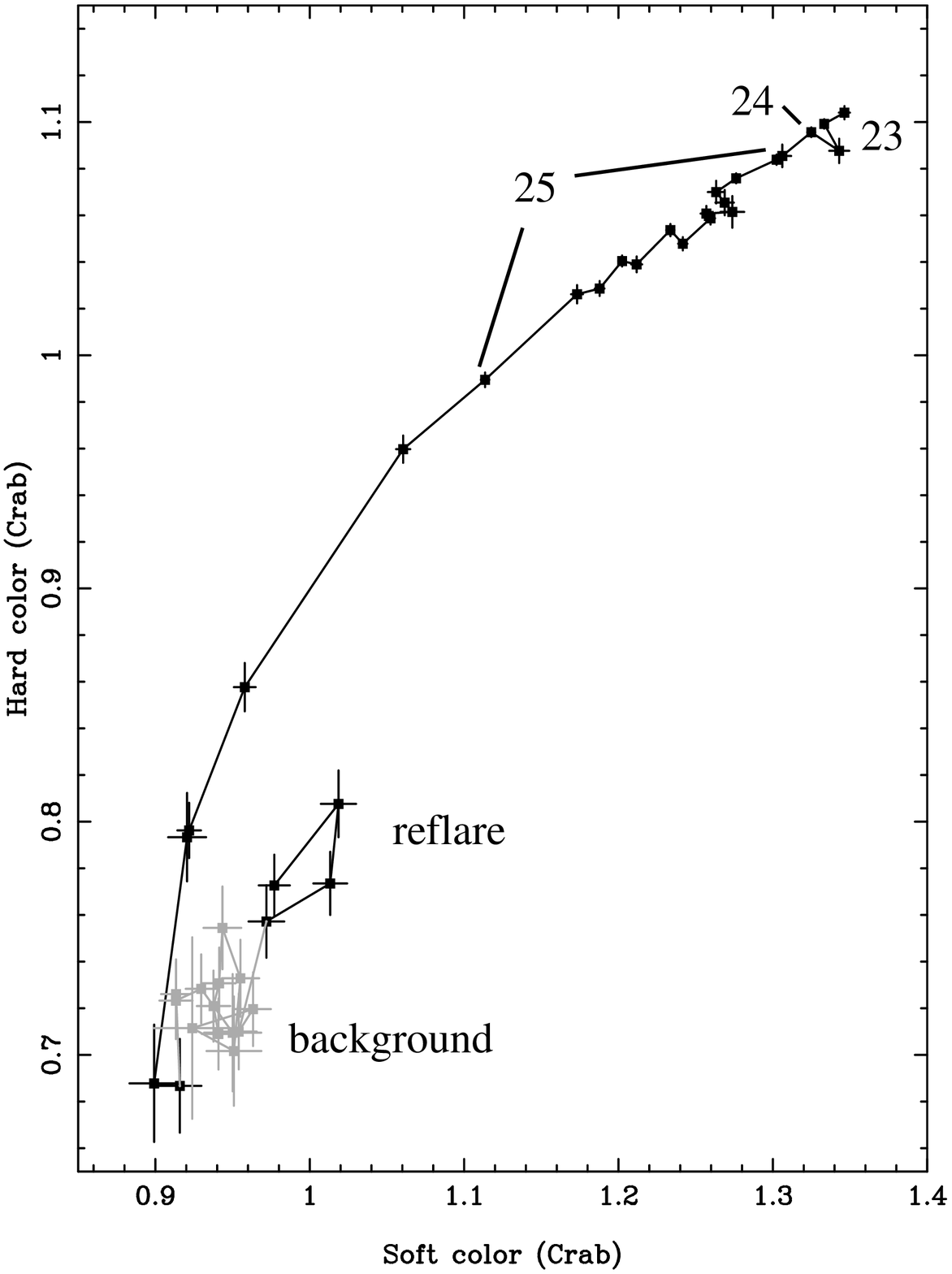}
\plotone{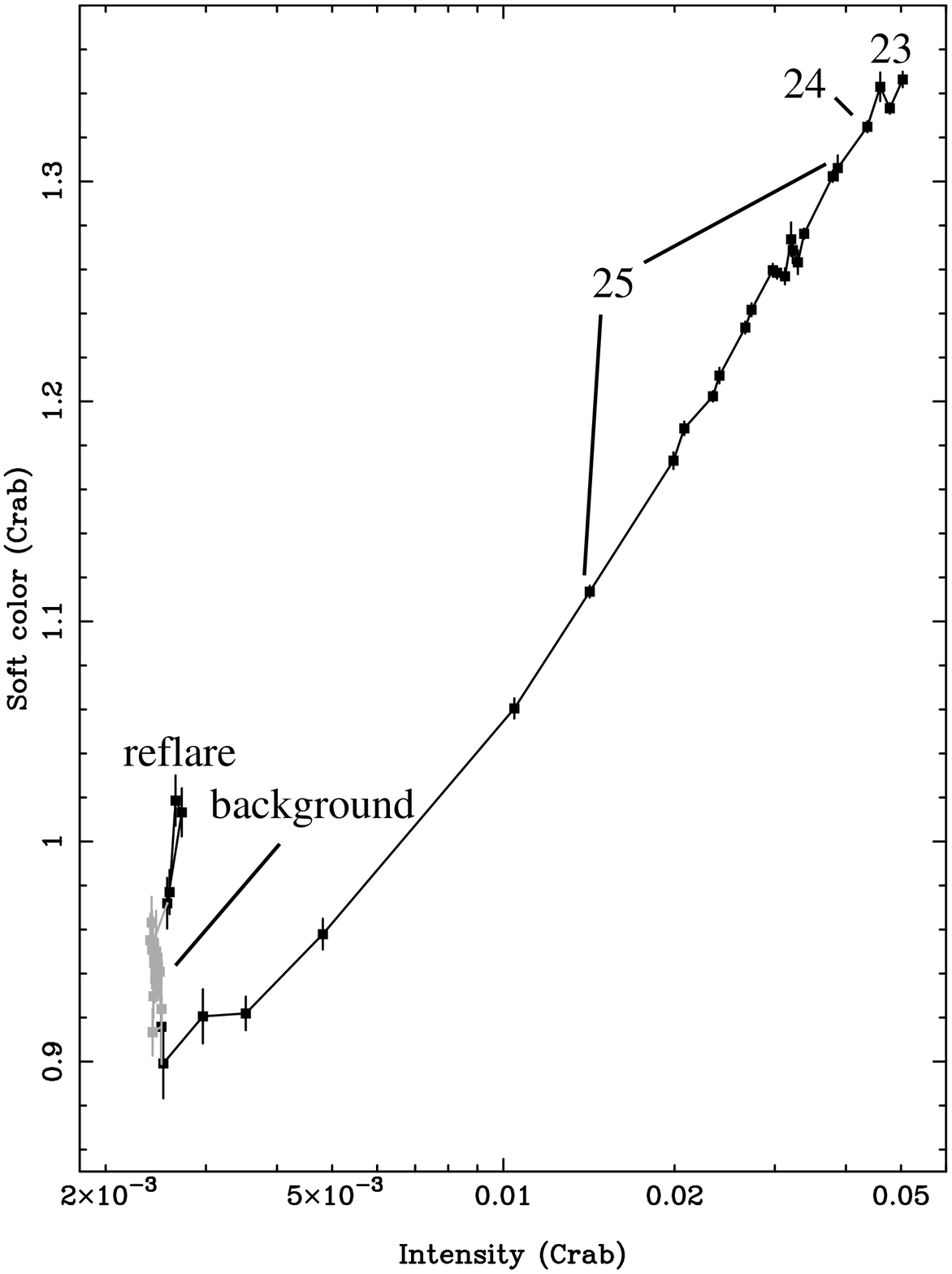}
\caption{Color-color diagram (left panel) and soft color vs. intensity diagram (right panel) of XTE J1751--305. 
The black point mark the 2002 outburst and the reflare, the grey points mark the background points that are due to 
nearby sources and Galactic diffuse emission (see \S \ref{sec.1751})
The line connects observations that are adjacent in time.
The group numbers of the power spectra are indicated. Colors are in units of 
Crab (see \S \ref{sec.observ}).}
\label{fig.cc_1751}
\end{figure}

In Figure \ref{fig.cc_1751} we plot the color-color and soft color vs. intensity diagram of XTE J1751--305.
The power spectra of XTE J1751--305 show L$_{b}$, L$_h$, L$_{\ell ow}$ and L$_u$ all at low characteristic frequencies.
This is typical for an extreme island state. The source stays in an extreme island state during the whole outburst while
the characteristic frequencies change between $\nu_b = 1.2$ and 0.4 Hz. 
Although the power spectra are typical for those of the extreme island of the atoll sources, the behavior in the color-color 
diagram is not. During {\bf group 25} the characteristic frequencies remain the same. 
In the atoll sources (and in SAX J1808.4--365, see above) the hard color would then remain the same while the intensity and soft color
change in a correlated fashion. This would lead to an extended horizontal branch in the color-color diagram. 
Here the hard color is not constant but is correlated with intensity in a similar way as the soft color. This correlation between
hard color, soft color and intensity exists during the whole outburst. Also, unlike in the atoll sources and SAX J1808.4--3658,
no anti-correlation occurs between hard color and the characteristic frequencies of the power spectral components 
(see Fig. \ref{fig.hc_vs_nub}). In Figure \ref{fig.cc_1751} we also plot the colors of the 
reflare and those of the background due to nearby sources and Galactic diffuse 
emission (see \S \ref{sec.1751}). The points of the reflare do not lie on the the curve of the main outburst, but on an extension of 
the background. This suggests that the reflare is caused by one of the background sources and not by XTE J1751--305.

\subsection{XTE J0929--314}
\label{sec.0929_colors}

\begin{figure}
\figurenum{16}
\epsscale{0.48}
\plotone{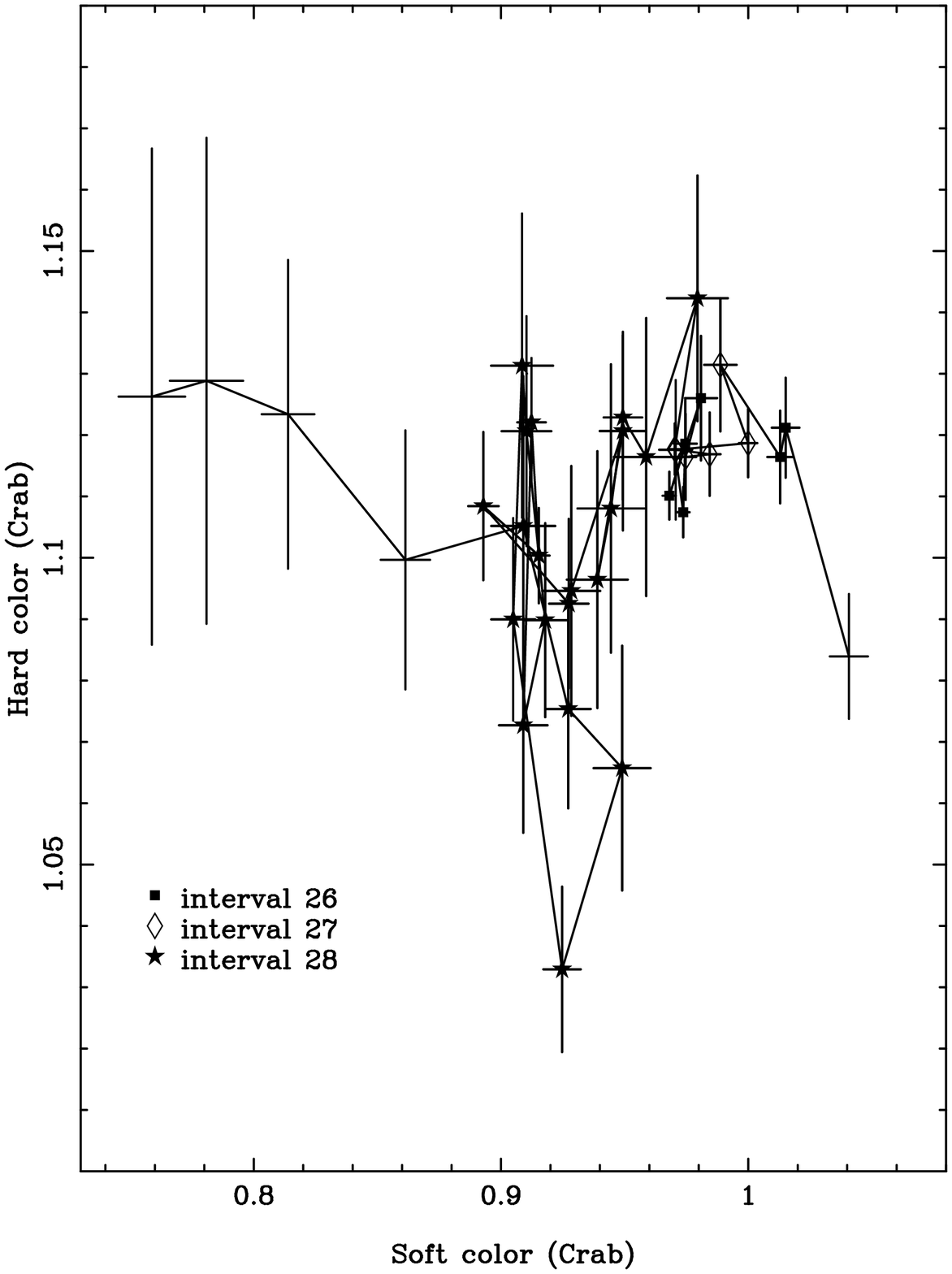}
\plotone{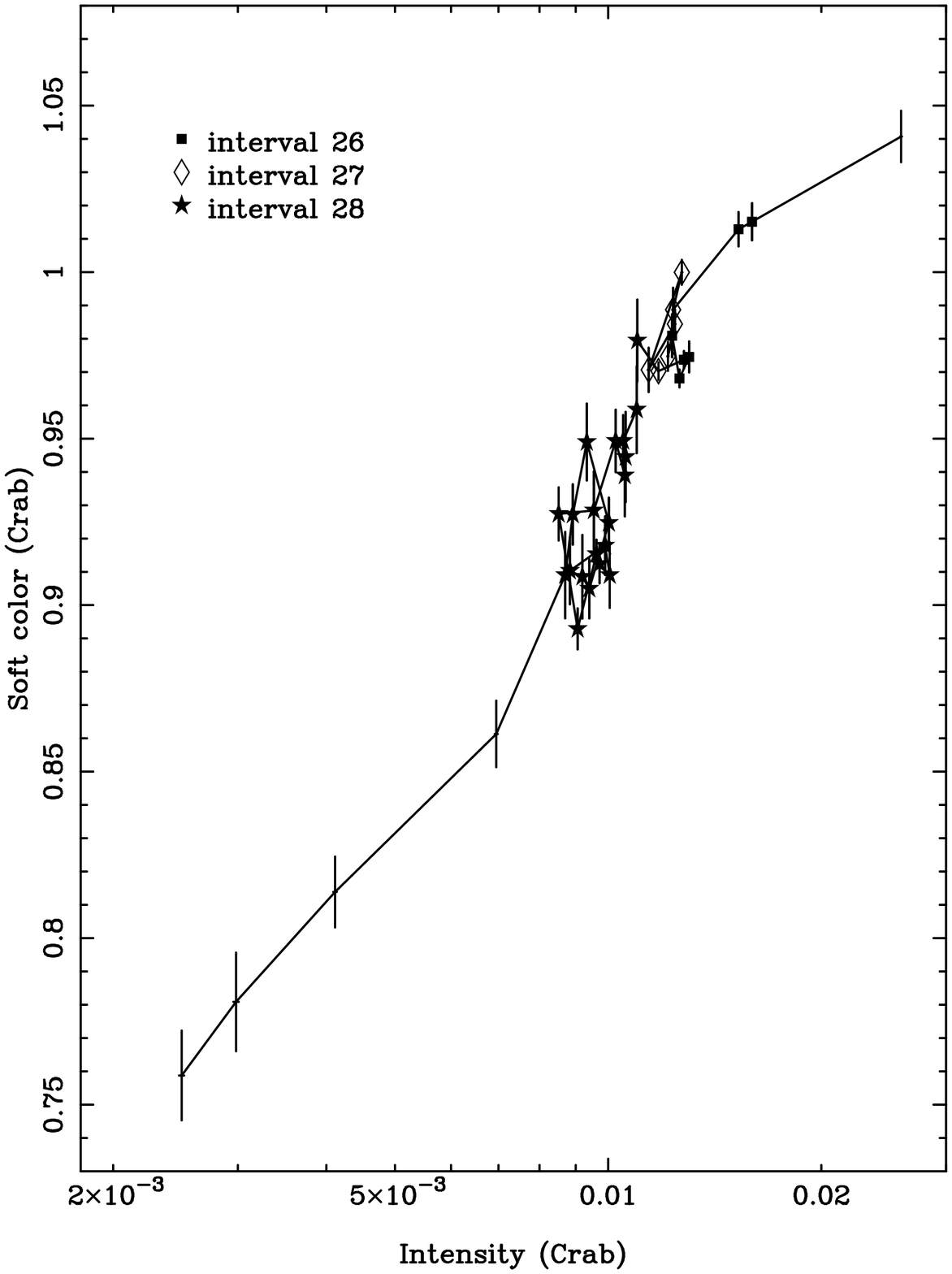}
\caption{Color-color diagram (left panel) and soft color vs. intensity diagram (right panel) of XTE J0929--314. 
The filled squares mark group 26, the open diamonds group 27, and the filled stars group 28.
The line connects observations that are adjacent in time.
Colors are in units of Crab (see \S \ref{sec.observ}).}
\label{fig.cc_0929}
\end{figure}

In Figure \ref{fig.cc_0929} we plot the color-color and soft color vs. intensity diagram of XTE J0929--314. 
The filled squares mark group 26, the open diamonds group 27, and the filled stars group 28. The source remains
in an extreme island state during the whole outburst, showing L$_{b}$, L$_h$, L$_{\ell ow}$ and L$_u$ all at low characteristic 
frequencies. The power spectra also show the narrow L$_{LF/2}$. This component is also found in 4U 1608--52 \citep{vstr03:apj596} and in
XTE J1814--338 (see \S \ref{sec.1814_colors}), also in the
extreme island state. So, the timing behavior is that of the extreme island state of the atoll sources. The behavior 
in the color-color diagrams is also similar to that of an extreme island state. Within the extreme island state the soft color 
changes in correlation with intensity. As the intensity decreases the soft color decreases and a horizontally extended island 
branch appears (see Fig. \ref{fig.cc_0929}). The hard color and the characteristic frequencies of the power spectral 
components are not correlated with intensity (and thus soft color). In the atoll sources the hard color is anti-correlated 
with the characteristic frequencies of the power spectral components. Here that anti-correlation is not evident, but the changes 
in hard color and characteristic frequencies are only small (see Fig. \ref{fig.hc_vs_nub}).

\subsection{XTE J1814--338}
\label{sec.1814_colors}

\begin{figure}
\figurenum{17}
\epsscale{0.48}
\plotone{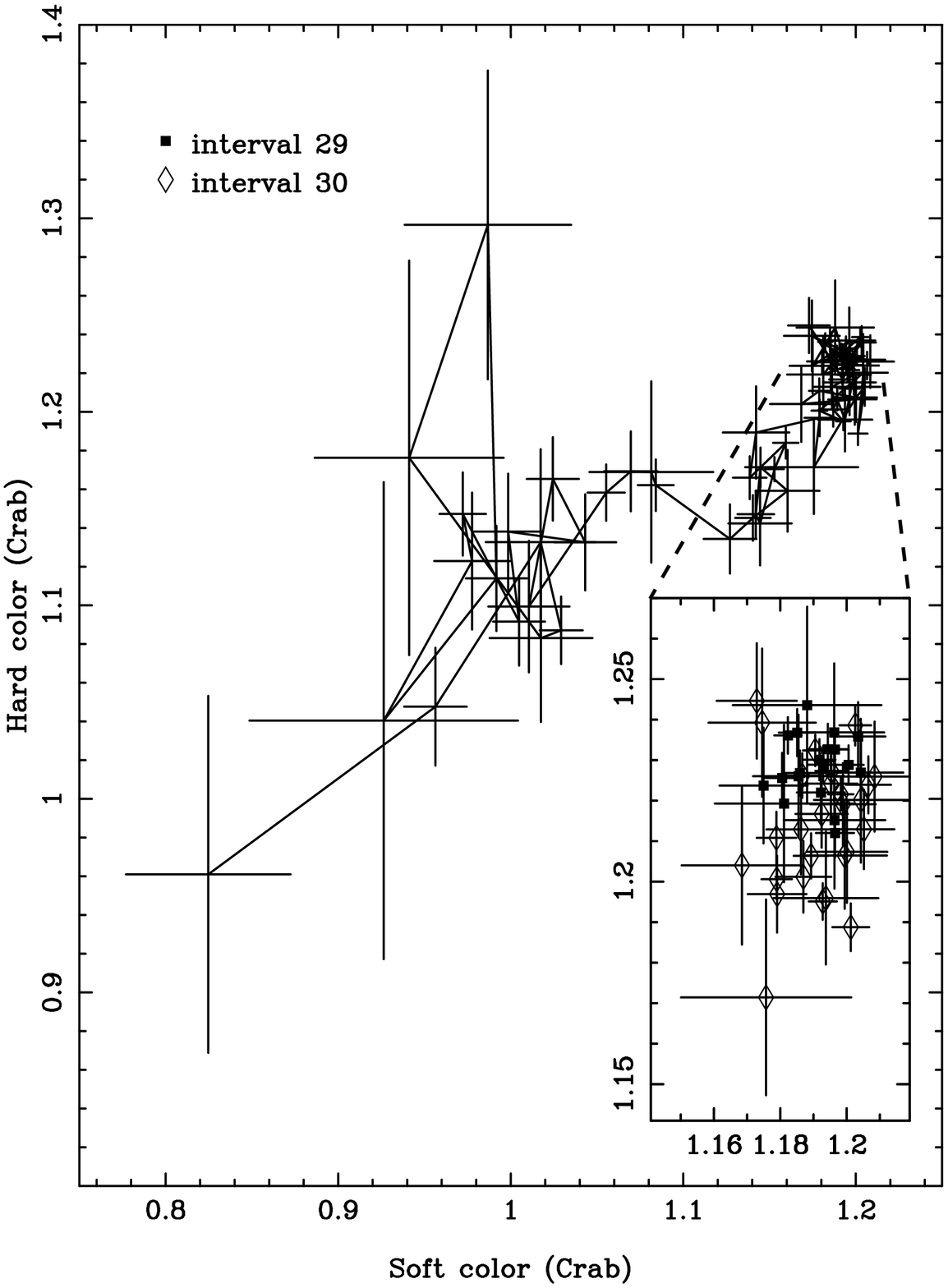}
\plotone{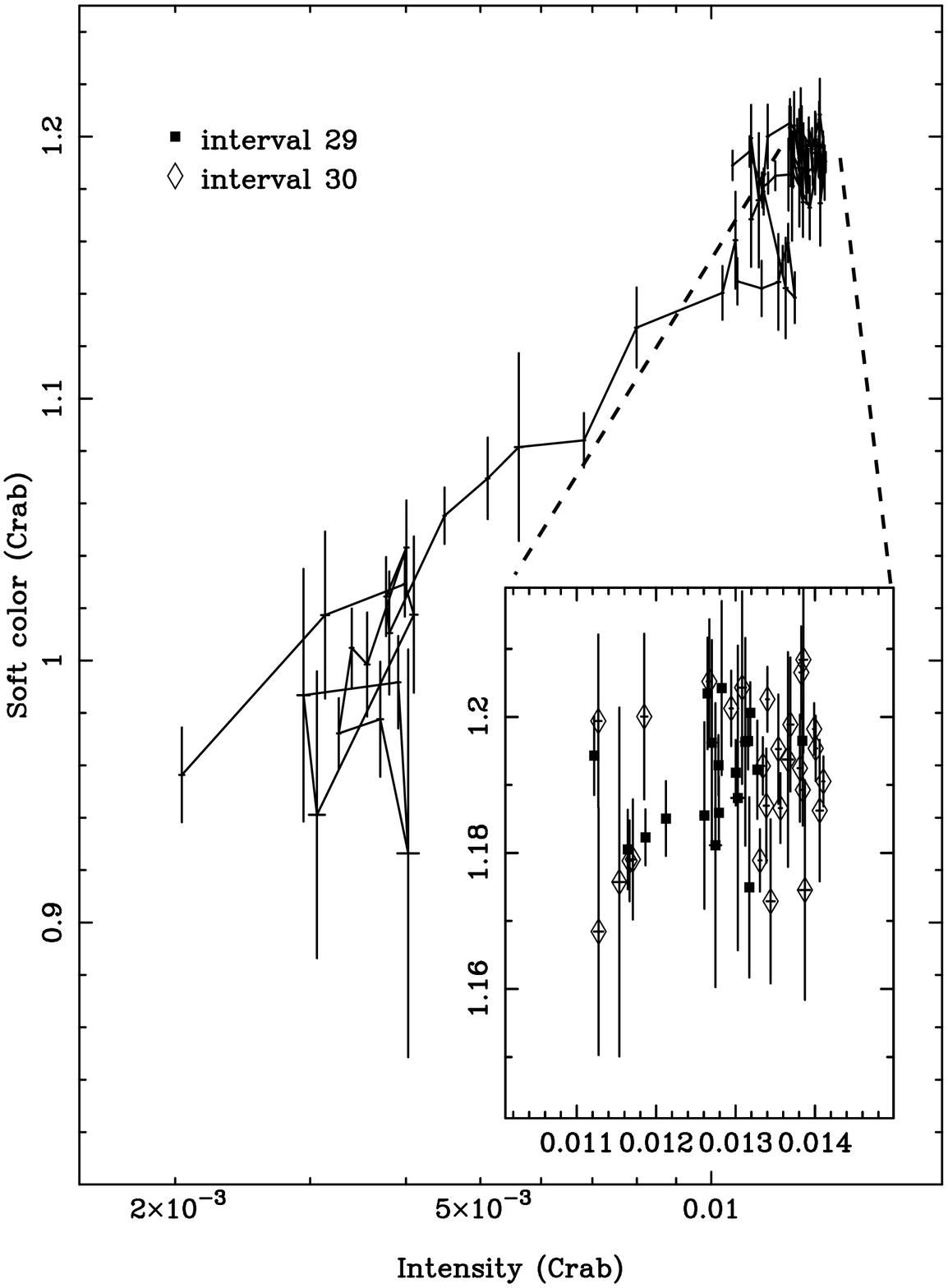}
\caption{Color-color diagram (left panel) and soft color vs. intensity diagram (right panel) of XTE J1814--338. 
The filled squares mark group 29, the open diamonds group 30.
The line connects observations that are adjacent in time. The insert in both panels is a blow up of that part of the diagrams 
for which we also obtained timing results.
Colors are in units of 
Crab (see \S \ref{sec.observ}).}
\label{fig.cc_1814}
\end{figure}

In Figure \ref{fig.cc_1814} we plot the color-color and soft color vs. intensity diagram of XTE J1814--338. 
The filled squares mark group 29, and the open diamonds group 30. The source remains
in an extreme island state during that part of the outburst for which power spectral fits could be made.
It shows L$_{b}$, L$_h$, L$_{\ell ow}$ and L$_u$ all at low characteristic frequencies. {\bf Group 29} also shows the narrow 
L$_{LF/2}$. This component was also found in 4U 1608--52 \citep{vstr03:apj596} and XTE J0929--314 (see \S \ref{sec.0929_colors}), 
also in the extreme island state. The timing properties could only be measured in a small range of the color-color diagrams 
(see Fig. \ref{fig.cc_1814}). Within this range, the characteristic frequencies, colors and intensity only vary by a
small amount, and therefore it is hard to compare this extreme island state of XTE J1814--338 with that of the atoll sources.
Within the extreme island state of XTE J1814--338 the soft color (and intensity) does not change enough for the source to draw up 
a horizontal branch as is often seen in atoll sources in the extreme island state. 
The range in characteristic frequencies within the extreme island state is too small to observe the anti-correlation
between hard color and the characteristic frequencies of the power spectral components that is seen in the atoll sources 
(see Fig. \ref{fig.hc_vs_nub}).
 
\section{DISCUSSION}

\subsection{Correlations of Timing Features}
\label{sec.compare_timing}

The frequencies of the variability components of the atoll sources, 
4U 0614+09, 4U 1608-52, and 4U 1728-34, follow a universal scheme 
of correlations when plotted versus $\nu_{u}$. In Figure \ref{fig.freq_freq} 
we present such a plot for SAX J1808.4--3658, XTE J1751--305, XTE J0929--314, 
and XTE J1814--338 (colored points). We also include (black points)
results for the atoll sources 4U 0614+09, 4U 1728-34 \citep{vstr02:apj568}, 
4U 1608-52 \citep{vstr03:apj596}, and Aql X--1 \citep{reig04:apj}. For Aql X--1 we 
only include the results for the extreme island state for which $\nu_{u}$ 
could unambiguously be identified. For 4U 1728-34 we also include a recent 
result by \citet*{migliari03:mnras345}, where for the first time twin kilohertz 
QPOs were observed in the island state. As a third group of sources we have 
included the low luminosity bursters 1E 1724--3045, GS 1826--24, and 
SLX 1735--269 (grey points) described in \citet{belloni02:apj572}. \citet{belloni02:apj572} 
constructed power spectra only up to 512 Hz, which led to an ill-constrained L$_u$.
Here we re-analyze the timing properties of groups C--R of \citet{belloni02:apj572} using 
the same method as described in \S \ref{sec.anal_timing} 
(Tables \ref{tbl.numax}, \ref{tbl.qvalues}, and \ref{tbl.rms}). With 
0.0039--4096 Hz power spectra this leads to a much better constrained L$_u$. 

SAX J1808.4--3658 shows similar relations between its characteristic frequencies 
as the atoll sources do. However, with 
respect to these sources, the relations of SAX J1808.4--3658 are shifted.
XTE J0929--314 shows a similar shift, although this is only 
based on one power spectrum. XTE J1751--305 and XTE J1814--338 as well as the 
low luminosity bursters show relations consistent with those of the atoll 
sources. The shift occurs only between the 
characteristic frequencies of the low frequency components on the one hand 
and $\nu_{u}$ (and probably $\nu_\ell$) on the other (see below):
the mutual relations between the characteristic frequencies of the 
{\it low frequency components} are identical for all sources (see Figure \ref{fig.nu_vs_nub}). 

To determine the actual shift factors between the frequency relations of SAX J1808.4--3658, 
XTE J0929--314, and those of the atoll sources we consider the $\nu_i$ vs. $\nu_u$ relations, where  
$\nu_i$ is either $\nu_b$, $\nu_h$, or $\nu_{\ell ow}$. We only use that part of the relations 
for which $\nu_u < 600$ Hz, as the behavior of the low-frequency components above 600 Hz is 
complex (see also \S \ref{sec:physical}). We take a $\nu_i$ vs. $\nu_u$ relation of one source 
(for example SAX J1808.4--3658) together with that of the atoll sources and perform a power law fit.
As our points have errors in both coordinates, we use the FITEXY routine of Press \citet{press92:comp274}
that performs a straight-line fit to data with errors in both coordinates. We take the logarithm of both 
$\nu_i$ and $\nu_u$ so that fitting a straight-line becomes equivalent to fitting a power law. 
Before fitting we multiply $\nu_u$ (or $\nu_i$) with a shift factor. We let this  shift factor
run between 0.1 and 3 with steps of 0.001. The fit with the minimal $\chi^2$ then gives the 
shift factor. The errors in the shift factor use $\Delta\chi^{2}$ = 1.0 
(corresponding to a 68\% confidence level). In Table \ref{tbl.shftfactors} 
we list the shift factors both in $\nu_u$ and $\nu_i$ for the different relations and sources. Next to the 
sources that show a clear shift in Figure \ref{fig.freq_freq} (SAX J1808.4--3658 and XTE J0929--314) we also 
include the other millisecond pulsars as well as the low luminosity bursters. 
We also determine the shift factor for the single $\nu_\ell$ observation of SAX J1808.4--3658. We determine
the shift factor if only $\nu_\ell$ is shifted, and the shift factor if both $\nu_\ell$ and $\nu_u$ are shifted.
For the latter we first multiply $\nu_u$ with 1.454, the weighted average of the $\nu_u$ shift factors for the $\nu_b$,
 $\nu_h$, and $\nu_{\ell ow}$ vs. $\nu_u$ relations of SAX J1808.4--3658 (see below).

The shift factors in $\nu_u$ for SAX J1808.4--3658 are between 1.4 and 1.5 when determined with the $\nu_b$ and 
$\nu_h$ vs. $\nu_u$ relation. For the $\nu_{\ell ow}$ vs. $\nu_u$ relation the factor is higher, but this 
is based on fewer points which results in a larger error on the shift factor. From this point on we take the shift 
in $\nu_u$ to be the weighted average of the three which results in 1.454$\pm$0.009. After shifting $\nu_u$ with this factor,
the additional shift needed for the single $\nu_\ell$ point is 1.359, similar to the shift factors found for $\nu_u$. 
For XTE J0929--314 the $\nu_u$ shift factors determined with the $\nu_b$ and $\nu_h$ vs. $\nu_u$ relation are similar 
to those in SAX J1808.4--3658, but note that these results are based on only one point from XTE J0929--314. The 
$\nu_{\ell ow}$ vs. $\nu_u$ relation, just like for SAX J1808.4--3658, gives a higher value but with large error.
For some of the other sources small shifts might exist, but unlike for SAX J1808.4--3658 the frequency vs. frequency relations 
for these sources have up to know only been observed over a small frequency range. The frequency relations should be 
measured over larger frequency ranges to confirm or reject the idea that small shifts occur.

The simplest explanation for this shift is that some physical difference between these sources 
affects $\nu_{u}$ and $\nu_\ell$. To make the relations of SAX J1808.4--3658 and XTE J0929--314 
coincide with those of the atoll sources, $\nu_{u}$ and $\nu_\ell$ must be multiplied with {$\sim1.5$}. 
In Figure \ref{fig.freq_freq_shifted} we replot Figure \ref{fig.freq_freq}, but 
now with the $\nu_{u}$ and $\nu_\ell$ of SAX J1808.4--3658 and XTE J0929--314 multiplied by 1.454.
The alternative, that $\nu_\ell$ and the characteristic frequencies of all the 
low frequency components, L$_{b}$, L$_h$, L$_{\ell ow}$, and the narrow QPOs 
(see \S \ref{sec.identify} for a discussion about the narrow QPOs) are affected is more complex.
In this case to make the relations coincide the characteristic frequencies of all the low frequency components 
have to be multiplied with about 0.35, $\nu_\ell$ would have to be multiplied with about 0.8, 
and $\nu_{hHz}$ would have to remain the same. It is simpler if $\nu_{u}$ 
and $\nu_\ell$ are the shifted frequencies. 
Note, that the resulting $\chi^2$/dof of the power law fits (see Table \ref{tbl.shftfactors}) are independent of 
whether $\nu_{u}$ or the frequencies of the low frequency components are shifted.

We tried to further test our idea that the shift occurs in $\nu_{u}$ and not in the  
frequencies of all the low frequency components, by plotting $Q$ versus $\nu_{\rm max}$ for both 
L$_u$ and L$_h$ (see Figure \ref{fig.nu_vs_q}). The $Q_u$ vs. $\nu_{u}$ plot showed a shift in $\nu_{u}$ similar to that in 
Figure \ref{fig.freq_freq}, while the $Q_h$ vs. $\nu_{h}$ plot did not show a clear shift,
consistent with the simpler interpretation above. 
However, the scatter in the $Q$ versus $\nu_{\rm max}$ correlations is much larger 
than of the frequency-frequency correlations in Figure \ref{fig.freq_freq}. Neither plots of fractional 
rms versus $\nu_{\rm max}$ nor linking Figure \ref{fig.freq_freq} to the atoll source states led to extra clues. 

Might we have wrongly identified the highest frequency Lorentzian component as L$_u$ while it really is  
L$_\ell$? There are a number of arguments against this, all based on 
similarities with the atoll source class of which SAX J1808.4--3658 seems part (see \S \ref{sec.1808_colors}).
First, L$_\ell$ was detected together with the component identified as L$_u$ in SAX J1808.4--3658 by \citet{wijnands03:nat424} 
(our group 3, see \S \ref{sec.1808_2002}) when the source was almost in the lower left banana (see \S \ref{sec.1808_colors}). 
In group 3 L$_\ell$ is weak and L$_u$ is strong 
(see Table \ref{tbl.rms}); this is common behavior for atoll sources when they are almost in the lower left 
banana \citep{mendez01:apj561,vstr02:apj568,vstr03:apj596}.
Second, in the island and extreme island states of SAX J1808.4--3658 the component identified as L$_u$ is a strong 
broad component that is present over a broad 
frequency range while L$_\ell$ is not. This is again common behavior for the atoll sources, where L$_\ell$ was 
observed in an island state only once, and then as a narrow and weak QPO \citep{migliari03:mnras345}.

\subsection{Identifying Timing Features}
\label{sec.identify}

The frequency relations of the accreting millisecond pulsars and the low luminosity bursters of Figure \ref{fig.freq_freq}, and 
especially the shift between the relations of SAX J1808.4--3658 and the atoll sources, provide new clues about the identification of 
some power spectral components. We assume here that the shift in Figure \ref{fig.freq_freq} is a shift in $\nu_{u}$; the conclusions
drawn below do not change if we use the alternative, a shift in $\nu_{\ell ow}$, $\nu_{h}$, and $\nu_{b}$ (see \S \ref{sec.compare_timing}).

The strong, broad highest-frequency Lorentzian in the extreme island state of the atoll sources could previously be identified as 
either L$_u$ or L$_{hHz}$ based on its frequency of about 200 Hz. The rms fractional amplitude 
vs. frequency relations of 4U 0614+09, 
4U 1608-52, and 4U 1728-34 gave a hint that this Lorentzian was L$_u$ \citep{vstr03:apj596}. Now we observe the same shift in frequency 
for this Lorentzian as for the component which in other states is unambiguously L$_u$ (see Fig. \ref{fig.freq_freq}), 
supporting the conclusion that it is indeed L$_u$. This 
conclusion is further strengthened by the results for the low luminosity bursters 1E 1724--3045, GS 1826--24, and SLX 1735--269.
These sources, which always have a power spectrum similar to that in the extreme island state of the atoll sources, 
show a clear correlation between the frequency identified as $\nu_{u}$, and $\nu_{\ell ow}$, $\nu_{h}$, and $\nu_{b}$ 
(Fig. \ref{fig.freq_freq}). This would not be expected if this frequency is really $\nu_{hHz}$, as in the atoll sources 
$\nu_{hHz}$ remains approximately 
constant while the other frequencies change. Note also that with the addition of the extreme island state results for Aql X--1, the 
atoll sources now also show these same correlations in the extreme island state (when $\nu_{u}$ is around 200 Hz).
So, we identify the broad low frequency features previously reported by \citet{wijnands98:apj507} from the 1998 outburst as 
manifestations of L$_u$.

The broad L$_{\ell ow}$ component found in the extreme island states of the atoll sources and in the low luminosity bursters
has now also been identified in the accreting millisecond pulsars. 
It was suggested that this component could be 
identified with the lower kilohertz QPO based on extrapolations of frequency-frequency relations 
\citep[see][]{psaltis99:apj520,belloni02:apj572,vstr02:apj568,vstr03:apj596}, however, it was also noted that it could be a different component \citep{vstr03:apj596}. Here we observe that 
if we make the relations of SAX J1808.4--3658 and XTE J0929--314 coincide with those of the atoll sources by multiplying $\nu_{u}$ 
with 1.454, the $\nu_{\ell ow}$ vs. $\nu_{u}$ relation of SAX J1808.4--3658 does indeed coincide with the atoll sources one, but the 
one $\nu_\ell$ point 
of SAX J1808.4--3658 is still offset from the atoll relation. We also have to multiply $\nu_\ell$ with 
a similar factor of about 1.5 to make 
the L$_\ell$ point of SAX J1808.4--3658 fall on the atoll relation.
If we apply the alternative shift (see \S \ref{sec.compare_timing}) we have to multiply $\nu_{\ell ow}$ with about 0.35, but $\nu_\ell$ with 
a different factor of about 0.8 to make the relations coincide.
Therefore it seems probable that L$_{\ell ow}$ and L$_\ell$ are different components.

For SAX J1808.4--3658 the characteristic frequency of L$_{hHz}$ seems to be slightly correlated with the characteristic frequencies 
of the other components (see Fig. \ref{fig.freq_freq}). The L$_{hHz}$ points also seem to lie on an extension of the L$_{\ell ow}$ relation
in Figure \ref{fig.freq_freq}. In the other atoll sources L$_{\ell ow}$ disappears and L$_{hHz}$ appears when the source goes from
the extreme island to the island state. This happens at $\nu_u \sim350$ Hz in Figure \ref{fig.freq_freq_shifted}. It might be that 
in SAX J1808.4--3658 L$_{\ell ow}$ is present in the island state, but is not detected. It is not detected as a separate component when its characteristic 
frequency is in the range of $\nu_{hHz}$, causing the correlation of L$_{hHz}$ with $\nu_u$ in Figure \ref{fig.freq_freq}.
However note that the scatter in the L$_{hHz}$ relation of the atoll sources is large and that a similar scatter in SAX J1808.4--3658
might accidentally cause the correlation of L$_{hHz}$ with $\nu_u$.

In the atoll sources 4U 0614+09, 4U 1608--52, and 4U 1728--34 the behavior of the band-limited noise
in the lower left banana state is complex. First, the characteristic frequency of 
the band-limited noise increases from $\sim0.3$ to $\sim15$ Hz. At this stage the band-limited 
noise is broad. Then at 
$\sim15$ Hz this noise component appears to ``transform'' into a narrow QPO whose frequency increases 
up to $\sim50$ Hz, while what appears to be another band-limited noise component 
appears at lower characteristic frequencies. 
In our previous papers \citep{vstr02:apj568,vstr03:apj596} we call 
the band-limited noise component that becomes a QPO, as well as the QPO it becomes, L$_{b}$ and the ``new'' 
broad band-limited noise appearing at lower frequency L$_{b2}$. In SAX J1808.4--3658 the same phenomenon
seems to occur in groups 2 and 3. Based on this similar behavior the $\sim15$ Hz Lorentzian can be identified as 
L$_{b}$ and the broad component at $\sim10$ Hz as L$_{b2}$.
 However, in Figure \ref{fig.freq_freq_shifted} for $\nu_{u} \gtrsim 600$ Hz 
the L$_{b}$ relation of SAX J1808.4--3658 follows the relation of the L$_{b2}$ component of the atoll sources. The 
L$_{b2}$ points of SAX J1808.4--3658 fall below that relation. This difference between L$_{b}$ in SAX J1808.4--3658
and in the other atoll sources can also be seen for $\nu_b \gtrsim 10$ Hz in Figure \ref{fig.nu_vs_nub}.
It seems that in this regime the frequency behavior of the band-limited noise in SAX J1808.4--3658 deviates from that 
in the other atoll sources.

In addition to the broad components L$_{b}$, L$_h$, L$_{\ell ow}$, the low luminosity bursters \citep{belloni02:apj572} and the atoll source 
4U 1608--52 \citep{vstr03:apj596,yoshida93:pasj45} show a narrow QPO that has $\nu_{\rm max}$ between $\nu_{b}$ and $\nu_{h}$. The QPO 
in 4U 1608--52 is not the same as that in the low luminosity bursters (called L$_{LF}$) but might be a sub-harmonic 
\citep{vstr03:apj596}. Therefore we will call it L$_{LF/2}$. L$_{LF}$ and L$_{LF/2}$ are not only present in neutron star X-ray 
binaries but could also be identified in the black hole candidate GX 339--4 \citep{belloni02:apj572,vstr03:apj596}. 
Also, a narrow QPO found in Cyg X--1 by \citet{pottschmidt03:aa407} might be L$_{LF/2}$ \citep{vstr03:apj596}. Until now these QPOs 
were only found at relatively low frequencies ($< 3$ Hz). In this work we find a QPO with a $\nu_{\rm max}$ 
between $\nu_{b}$ and $\nu_{h}$ at high frequencies ($\approx 47$ Hz) in SAX J1808.4--3658 
(groups 2 and 3), and at low frequencies (0.7--1.0 Hz) in XTE J1814--338 (group 29) and XTE J0929--314 
\citep[groups 23--25, previously reported by][]{markwardt02:apj575}. In Figure \ref{fig.narrowqpo} we plot our results of the
SAX J1808.4--3658, XTE J1814--338, and XTE J0929--314 on Figure 13 of \citet{vstr03:apj596}. 
Based on the power law relations of Figure \ref{fig.narrowqpo}
we can identify the QPOs in XTE J1814--338 and XTE J0929--314 as L$_{LF/2}$ and the QPOs in SAX J1808.4--3658 as L$_{LF}$.

\subsection{Timing-Color Correlations and States}

The timing properties of the accreting millisecond pulsars SAX J1808.4--3658, XTE J1751--305, XTE J0929--314, and XTE J1814--338 
are very similar to those of the atoll sources. Also based on its correlated timing and color behavior SAX J1808.4--3658 can be classified 
as a classical atoll source (see \S \ref{sec.1808_colors}). In the time span for which timing results could be obtained, XTE J0929--314 
and XTE J1814--338 showed little change in power spectra and colors. Therefore it is hard to classify these sources as atoll sources 
although their correlated timing and color behavior is consistent with that of a typical atoll source 
(see \S \ref{sec.0929_colors} and \S \ref{sec.1814_colors}). XTE J1751--305 
is different from a classical atoll source with respect to the behavior of its hard color (see \S \ref{sec.1751}). In atoll sources in 
the island and extreme island 
states the characteristic frequencies and hard color are anti-correlated with each other, and decoupled from soft color and intensity, which are 
correlated. The hard color in XTE J1751--305 is instead strongly correlated with soft color and intensity, and is not anti-correlated 
with any of the characteristic frequencies. 

To explain the one-to-one relation between soft color and intensity in an extreme island state of the atoll source 4U 1608--52
\citet{vstr03:apj596} used the following scenario: In the extreme island state of 4U 1608--52 the disk is far away from the neutron star
and cannot be seen in the PCA energy spectrum; the soft color therefore only depends on the spectral properties of the neutron 
star \citep{gierlinski02:mnras337}. As the accretion rate (intensity) increases, the neutron star surface becomes hotter and the soft color 
increases \citep{vstr03:apj596}. 
This scenario can also be applied to the horizontally extended extreme island state of SAX J1808.4--3658. In an energy spectral
study of the 1998 outburst of SAX J1808.4--3658, \citet{gierdonebarret02:mnras331} found that the soft component likely originates from the 
neutron star surface. 

In the horizontally extended island state branches  that are observed in the atoll sources (including SAX J1808.4--3658),
contrary to the horizontal branch of Z sources,
the timing properties remain similar when the intensity changes. As noted by \citet{olive03:apj583}, this aspect of the behavior 
can be explained by the scenario of \citet{vdk01:apj561} for the parallel track phenomenon in the intensity versus 
lower kilohertz QPO frequency diagram that occurs in the lower left banana of the atoll sources \citep[see e.g.][]{mendez00:texas}.
If we look at these island state branches in SAX J1808.4--3658 during the 
2002 outburst (groups 5--15) we observe two parallel tracks (see Figure \ref{fig.parallel}), similar to those in the lower left 
banana of the atoll sources. This is a further indication that the same 
physical phenomenon is causing the source to drift between parallel tracks in the lower left banana of the atoll sources 
and along the island state branches as suggested by \citet{vstr03:apj596}.
We observe one parallel track from group 5 to 8 and the next from group 9 to 13. Both parallel 
tracks last 4 days; this is 
longer than the parallel tracks in the lower left bananas which usually last about a day.

\subsection{Ingredients for a Physical Model}
\label{sec:physical}

\begin{figure}
\figurenum{18}
\epsscale{0.8}
\plotone{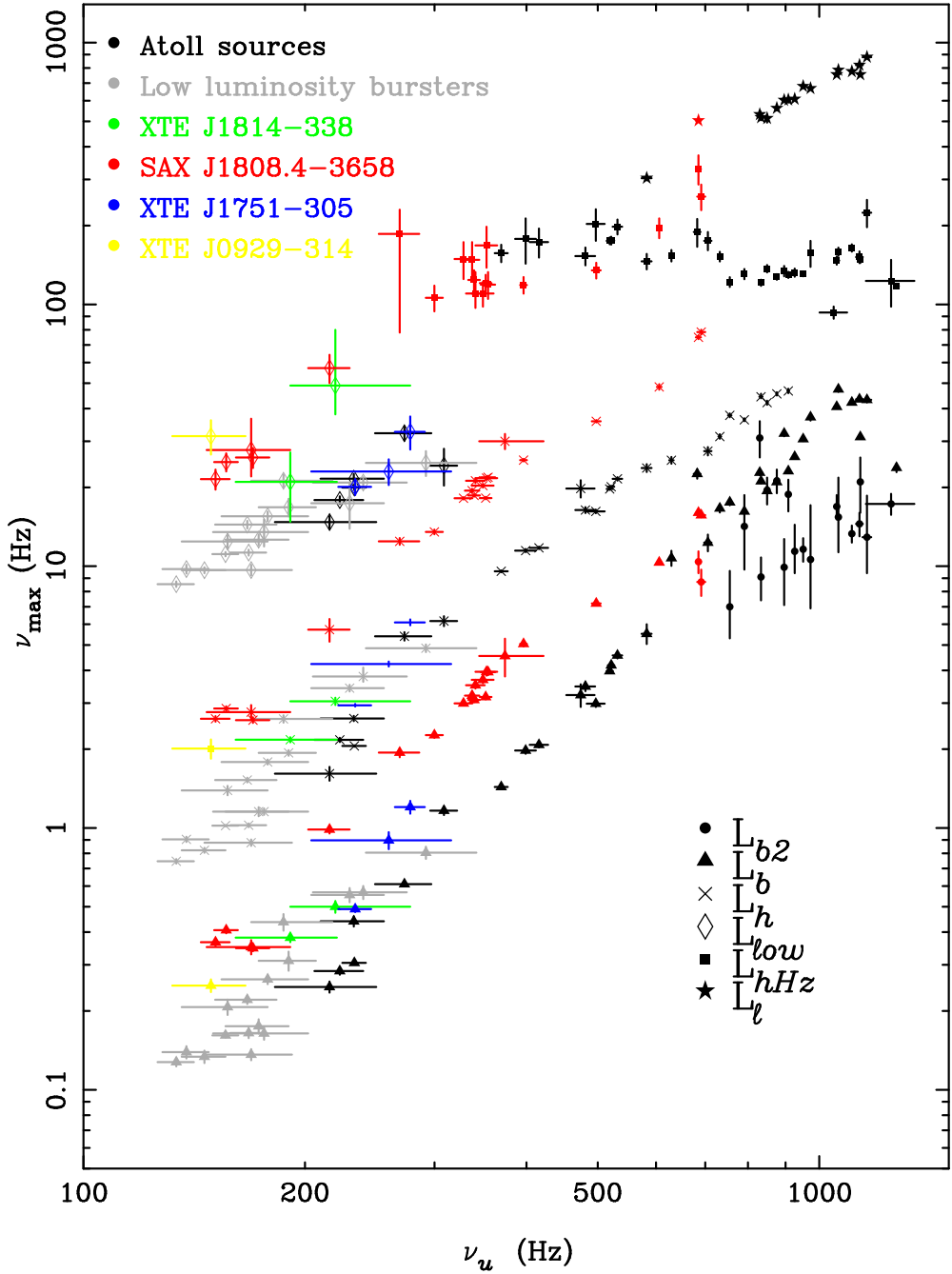}
\caption{
Correlations between the characteristic frequencies 
($\equiv \nu_{\rm max}$) of the power spectral components plotted versus $\nu_{u}$.
The colored points mark the accreting millisecond pulsars SAX J1808.4--3658 (red), 
XTE J1751--305 (blue), XTE J0929--314 (yellow), and XTE J1814--338 (green).
The black point mark the atoll sources 4U 0614+09, 4U 1608-52, 4U 1728-34, and Aql X--1. 
The grey point mark the low luminosity bursters 1E 1724--3045, GS 1826--24, and SLX 1735--269.
The solid dots mark L$_{b2}$, the triangles L$_{b}$, the crosses L$_h$, the squares L$_{hHz}$, 
the stars L$_\ell$ and the diamonds 
L$_{\ell ow}$.}
\label{fig.freq_freq}
\end{figure}

\begin{figure}
\figurenum{19}
\epsscale{0.8}
\plotone{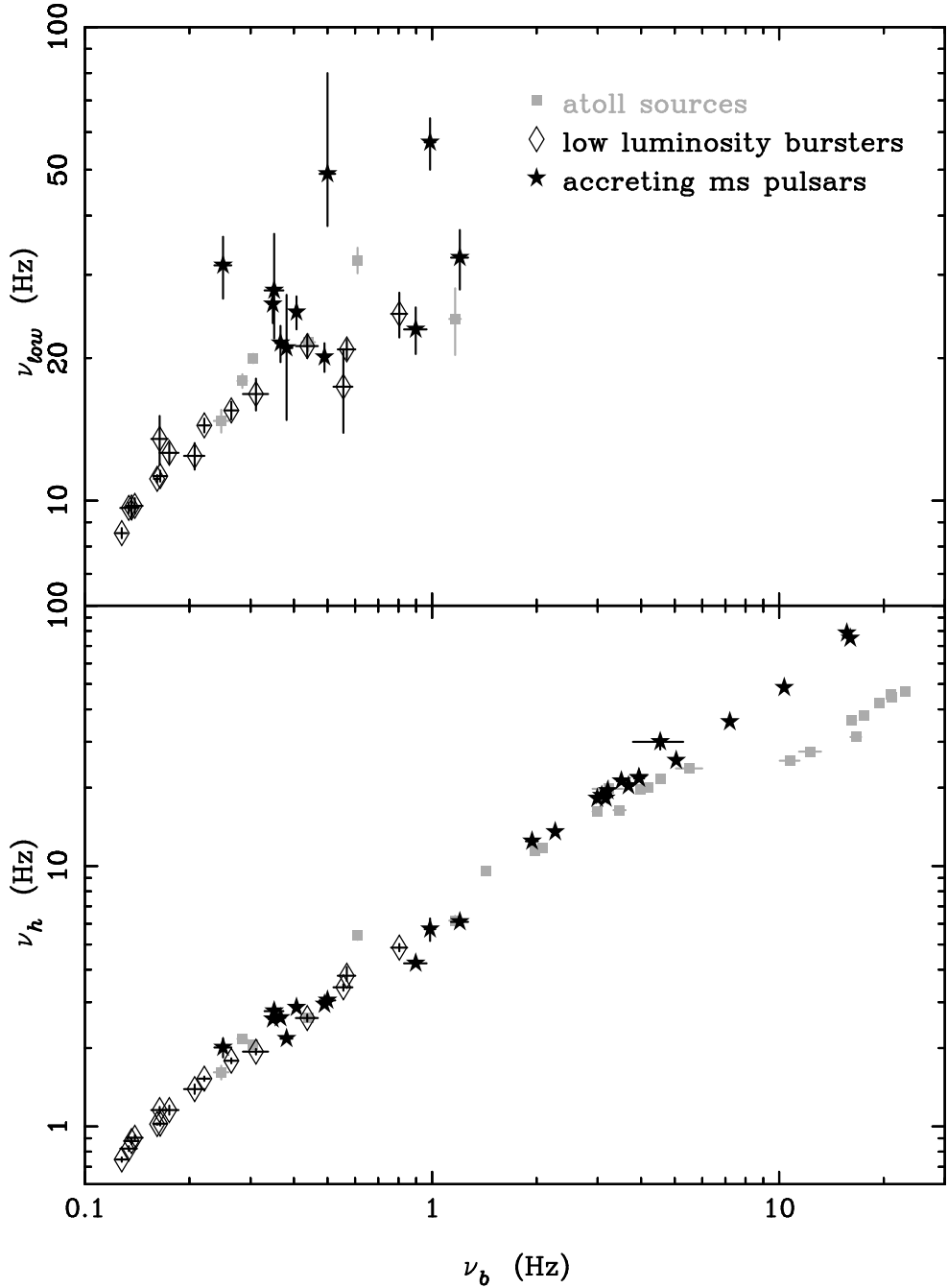}
\caption{
The characteristic frequencies $\nu_{\ell ow}$ (top panel) and $\nu_{h}$ (bottom panel) plotted versus $\nu_{b}$. 
The filled grey squares mark the atoll sources 4U 0614+09, 4U 1608-52, 4U 1728-34, and Aql X--1, 
the open black diamonds mark the low luminosity bursters 1E 1724--3045, GS 1826--24, and SLX 1735--269,
and the filled black stars mark the accreting millisecond pulsars SAX J1808.4--3658, 
XTE J1751--305, XTE J0929--314, and XTE J1814--338. 
}
\label{fig.nu_vs_nub}
\end{figure}

\begin{figure}
\figurenum{20}
\epsscale{0.8}
\plotone{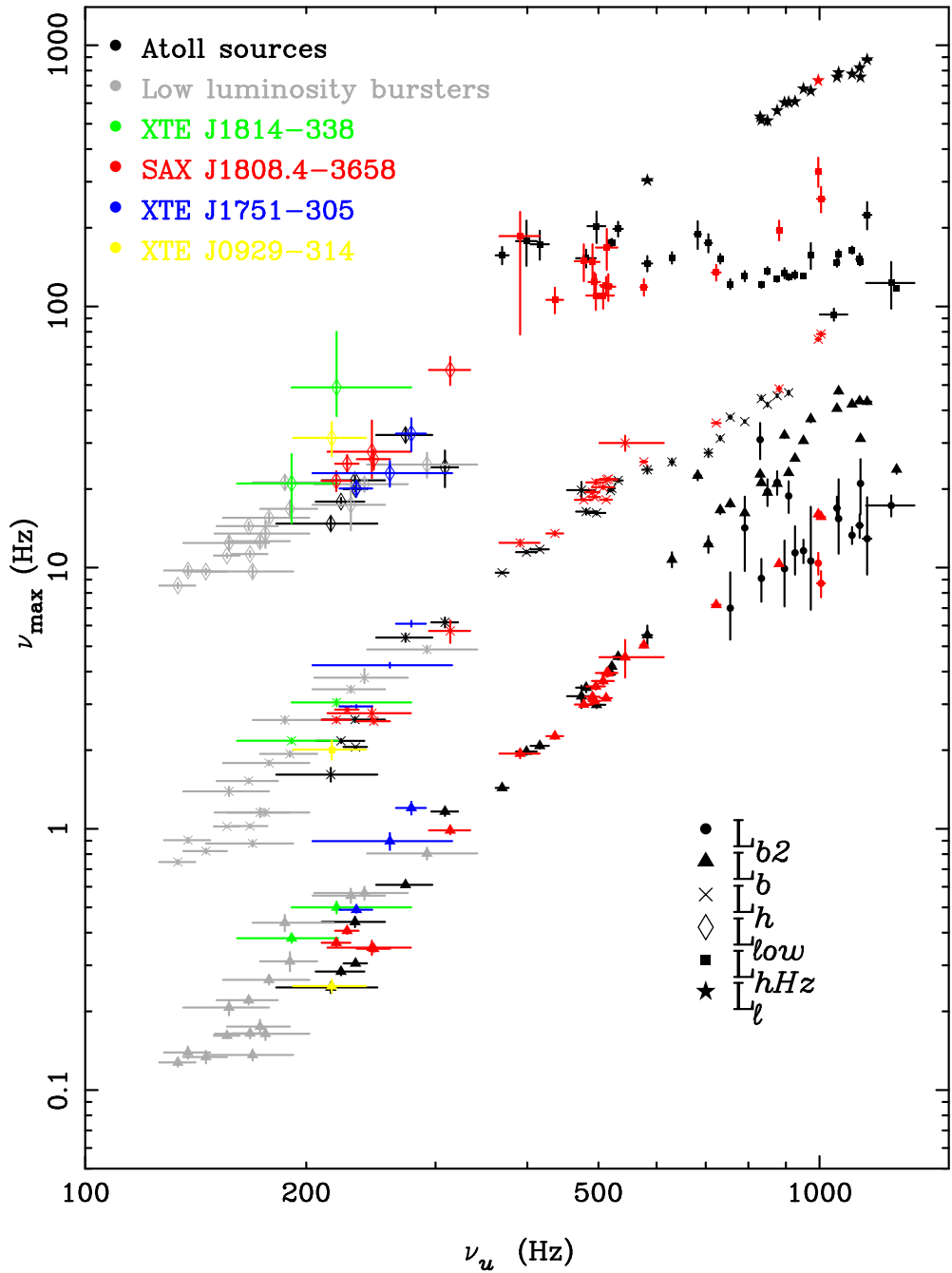}
\caption{As Figure \ref{fig.freq_freq}, but with $\nu_{u}$ and $\nu_\ell$ of SAX J1808.4--3658
and XTE J0929--314 multiplied by a factor 1.454.}
\label{fig.freq_freq_shifted}
\end{figure}

\begin{figure}
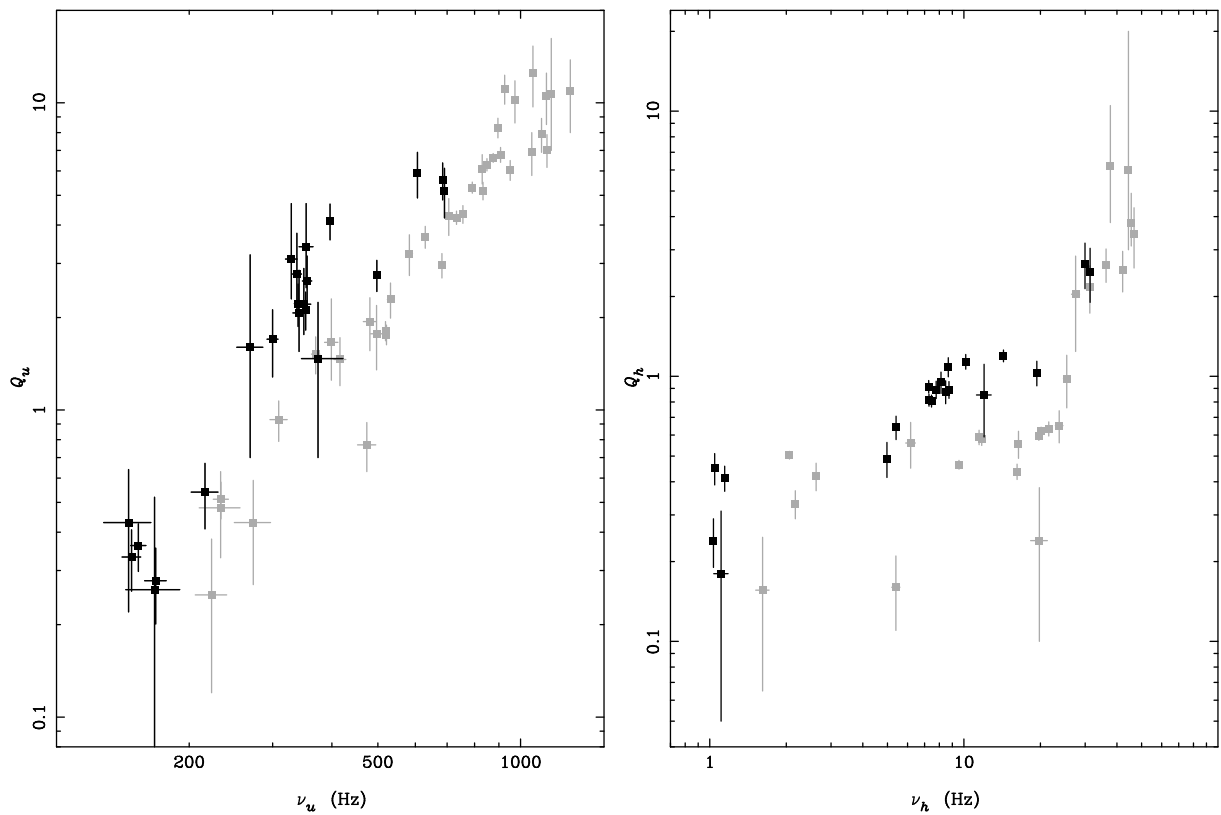

\figurenum{21}
\epsscale{0.48}
\plotone{f21a.eps}
\plotone{f21b.eps}
\caption{$Q$ versus $\nu_{\rm max}$ for L$_{h}$ and L$_{u}$. The black points mark SAX J1808.4--3658,
the gray points the atoll sources.}
\label{fig.nu_vs_q}
\end{figure}

\begin{figure}
\figurenum{22}
\epsscale{0.8}
\plotone{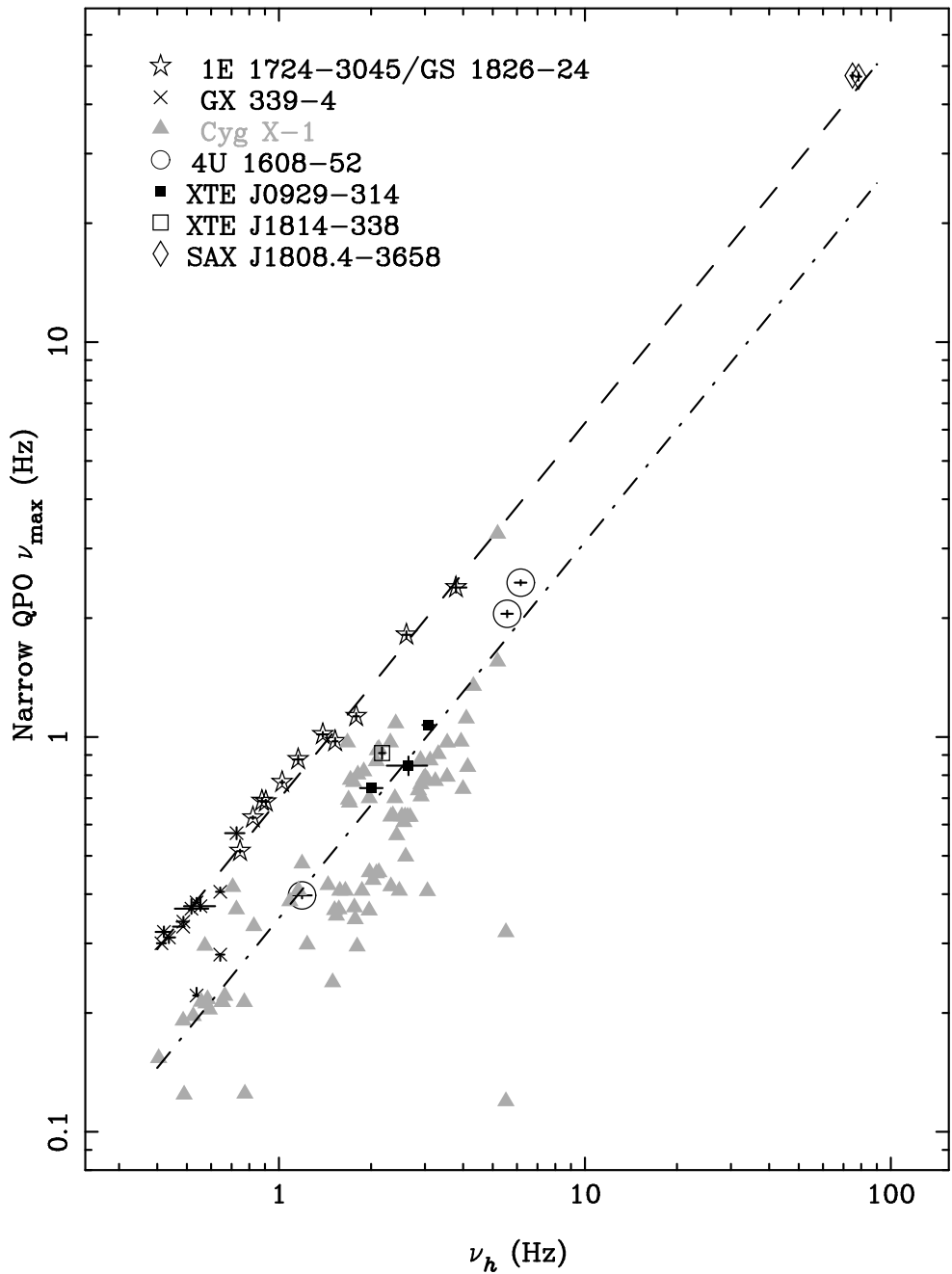}
\caption{Characteristic frequency of L$_{h}$ versus that of the narrow low-frequency QPOs that have a 
$\nu_{\rm max}$ between $\nu_{b}$ and $\nu_{h}$. The sources are indicated in the plot. 
Note that we use the results of this paper for the low luminosity bursters 1E 1724--3045 and GS 1826--24, 
which for the low frequencies are in agreement with the results of \citet{belloni02:apj572} that were used in Figure 
13 of \citet{vstr03:apj596}. The dashed line 
indicates a power law fit to the $\nu_{LF}$ vs. $\nu_{h}$ relation of the low-luminosity bursters 1E 1724--3045 
and GS 1826--24, and the BHC GX 339--4. 
The dash-dotted line is a power law with a normalization half of that of the dashed line.}
\label{fig.narrowqpo}
\end{figure}

\begin{figure}
\figurenum{23}
\epsscale{0.8}
\plotone{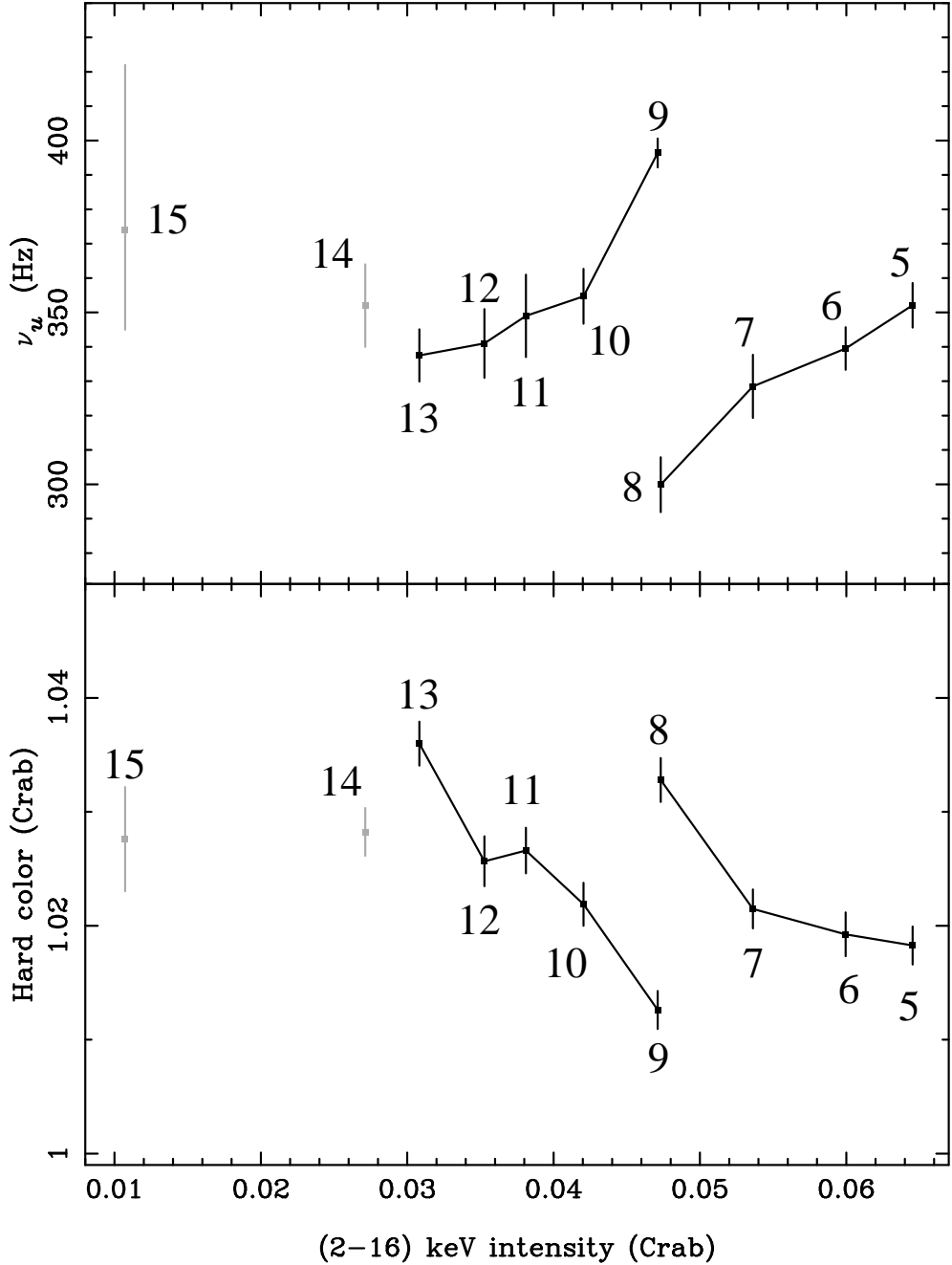}
\caption{Parallel lines during an island state of SAX J1808.4--3658. The top panel shows the $\nu_u$ vs. intensity diagram and
the bottom panel the hard color vs. intensity diagram. The group numbers are indicated. The points that span up the 
parallel tracks are connected by a line.}
\label{fig.parallel}
\end{figure}

Currently, to our knowledge, there is no physical model \citep[see][for a discussion of models]{vstr03:apj596} that explains the 
mutual relations 
between the characteristic frequencies of the power spectral features (e.g. Figures 
\ref{fig.freq_freq} and \ref{fig.nu_vs_nub}). Here we point out some clues that might help to construct such a model.
As described above in \S \ref{sec.identify}, in the lower left banana of the atoll sources (among which we now also 
classify SAX J1808.4--3658) 
the behavior of the low frequency components is complex; several timing features, additional to the ones observed in the 
island states, appear (see e.g. groups 2 and 3 of SAX J1808.4--3658). In Figure \ref{fig.freq_freq_shifted} this 
happens at $\nu_u \approx 600$ Hz.
The Z sources also show a complex low frequency behavior \citep[see Figure 14 in][]{vstr03:apj596}, also at high $\nu_u$.
If $\nu_u$ indeed represents the Keplerian frequency at the inner disk radius, a high $\nu_u$ means that then the inner disk is 
close to the neutron star surface, so the more complex behavior might be caused by 
relativistic or magnetic effects that are stronger close to the neutron star.  
Note that the power spectra in the extreme island state of some sources also show extra timing features, but these are in the form of 
weak narrow QPOs that do not dominate the power spectra.

Because of the complex behavior of the low-frequency noise above $\nu_u \approx 600$ Hz in Figure \ref{fig.freq_freq_shifted}
we concentrate now on the more simple frequency relations below $\nu_u \approx 600$ Hz 
in Figure \ref{fig.freq_freq_shifted}.  The low-frequency relations below $\nu_u \approx 600$ Hz can be fitted 
with power laws (see Table \ref{tbl.plfits}). For these fits we use the same method
as in \S \ref{sec.compare_timing}, which uses the FITEXY routine of Press \citet{press92:comp274}.
Note that although the $\chi^2$/dof are relatively high, this is not due to 
systematic deviations from a power law but to scatter around it.
The power law indexes of the $\nu_{b}$, $\nu_{h}$, and $\nu_{\ell ow}$ vs. $\nu_u$
relations fall in the range 1.8--2.7, progressively steeper towards lower frequency components.
The errors on the power law indexes of the L$_{LF}$ and L$_{LF/2}$ relations are large as these relations are only based on a few 
points, but also these power law indexes are consistent with this picture.

Here we use $\nu_{\rm max}$ as the characteristic frequency. Although $\nu_{\rm max}$ has proven to be of great practical value
and may represent a real underlying physical frequency \citep[e.g., the inner frequency of a disk annulus, see][]{belloni02:apj572}, this is
not assured. Two other frequencies, that have physical interpretations in terms of a damped harmonic oscillator, are $\nu_0$, the 
centroid frequency of
the Lorentzian and $\nu_{\rm damped}$ \citep[for a definition, see][]{titarchuk02:apj578}. These frequencies can be 
calculated from our fit results $\nu_{\rm max}$ and $Q$ \citep[see][]{vstr03:apj596}.
Note that using $\nu_0$ instead of $\nu_{\rm max}$ does not give further insights in the shift between SAX J1808.4--3658, 
and the atoll sources (see \S \ref{sec.compare_timing}). For the narrow Lorentzians the shift factor will be the same as 
$\nu_{\rm max}$ is approximately equal to $\nu_0$, for the broad features $\nu_0$ is badly constrained and a shift cannot be 
determined.
For the $\nu_{h,0}$ vs. $\nu_{u,0}$ relation of the atoll sources 4U 0614+09, 4U 1608-52, and 4U 1728-34 \citet{vstr03:apj596} found a 
quadratic relation. A quadratic relation was also found for the $\nu_{h,0}$ vs. $\nu_{u,0}$ relations of several Z sources 
\citep{stella98:apj492,psaltisetal99:apj520,jonker98:apj499,homan02:apj568}.
The relativistic precession model \citep{stella98:apj492} predicts a quadratic relation between the Lense-Thirring precession 
frequency and the Keplerian frequency of the inner disk; $\nu_{LT} = 4.4\times10^{-8} I_{45} M^{-1} \nu_s \nu_K^2$ Hz, where 
$I_{45}$ is the moment of inertia in units of $10^{45}$ g cm$^2$, $M$ is the mass of the neutron star in $M_{\odot}$, and $\nu_s$ 
is the neutron star spin frequency. However, the required $I_{45}/M$ values were too large for L$_{h}$ to 
be the fundamental Lense-Thirring precession frequency \citep{psaltisetal99:apj520}. Also the $\nu_{h,0}$ vs. $\nu_{u,0}$ 
relations coincided for 4U 1728--34 and 4U 1608--52, which is not expected in this interpretation in view of their different spin frequencies (363 Hz and 620 Hz, respectively, assuming burst oscillations happen at the spin frequency).
The relation of the frequency difference between the 410 Hz QPO and the $\sim401$ Hz pulse spike of SAX J1808.4--3658
was found to be consistent with a quadratic relation by \citet{wijnands03:nat424}.
We note, that this frequency difference is always about half of $\nu_{h,0}$ (see Table \ref{tbl.410hz}). 

Although the relations between the characteristic frequencies of the low frequency components and $\nu_u$ are 
different between SAX J1808.4--3658, XTE J0929--314 and the other sources (see \S \ref{sec.compare_timing}), 
the  mutual relations between the low frequency components are the same for all atoll 
sources (Fig. \ref{fig.nu_vs_nub}). The $\nu_{h}$ vs. $\nu_{b}$ relation is even the same for black hole candidates and 
atoll sources \citep[][]{wijnands98:apj507,vstr02:apj568}. So, the $\nu_{h}$ vs. $\nu_{b}$ relation 
does not depend on spin, mass, magnetic field, or the presence of a surface. Note, that the frequency range of 
$\nu_{h}$ and $\nu_{b}$ is about a factor 10 lower for black hole candidates than for neutron star low mass X-ray binaries.
Therefore L$_{b}$ and L$_{h}$ might be properties of the accretion disk that scale inversely with mass \citep{wijnands99:apj514}. 

The frequency correlations of SAX J1808.4--3658 and XTE J0929--314 are shifted compared to those of the other 
sources in a way that is most easily described as a shift in upper and lower kilohertz QPO frequency by about 
a factor 1.5 (see \S \ref{sec.compare_timing}).
In most models $\nu_{u}$ reflects the Keplerian frequency at the inner edge 
of the accretion disk (e.g. \cite{miller98:apj508}). If this is indeed the correct interpretation, then for similar characteristic 
frequencies of the low frequency 
components, this inner radius of the disk can somehow get closer to the neutron star surface in the atoll sources, 
low luminosity bursters, and the accreting millisecond pulsars XTE J1751--305 and XTE J1814--338 
than in the accreting millisecond pulsars SAX J1808.4--3658 and XTE J0929--314. It is unclear which physical parameter 
is the cause of this. In Table \ref{tbl.msp} we list several characteristic parameters for the accreting millisecond 
pulsars. There is no obvious relation between the shift and any of the parameters. 
In particular, there is no relation to spin frequency: SAX J1808.4--3658, and XTE J0929--314 which share one set 
of frequency relations have spin frequencies of respectively 401 and 185 Hz, and the other sources which share 
another set of relations have spin frequencies that range from 314 to 620 Hz. 
Also the magnetic field strength $B$,
which may be stronger for the accreting millisecond pulsars than for the atoll sources 
\citep[see e.g.][]{wijnands03:nat424,strohmayer03:apj596}, is not a good candidate as not all accreting millisecond pulsars show
the shift. 
The fact that the shift occurs by a factor close to $\sim3/2$ suggests a possible relation with recently proposed 
parametric resonance models for kilohertz QPOs \citep[e.g.][]{abramowicz03:pasj55}. In these models 2:3 frequency resonances
between general relativistic orbital/epicyclic frequencies play a central role. However, similarly to the remarks 
about inner disk radius above, it is not clear what would set SAX J1808.4--3658 and XTE J0929--314 apart from the other sources.

This work was supported by NWO SPINOZA grant 08--0 to E.P.J. van den Heuvel, 
by the Netherlands Organization for Scientific Research (NWO), and by 
the Netherlands Research School for Astronomy (NOVA). 
This research has made use of data obtained through
the High Energy Astrophysics Science Archive Research Center Online Service, 
provided by the NASA/Goddard Space Flight Center.
We would also like to thank the referee, whose comments helped us to improve the paper.

\clearpage


\begin{center}
\begin{deluxetable}{lccccccccc}
\tabletypesize{\tiny}
\tablewidth{0pt}
\tablecaption{Characteristic frequencies of the multi-Lorentzian fit for the 
accreting millisecond pulsars and low luminosity bursters\label{tbl.numax}} 
\tablehead{ 
\colhead{Group} & \colhead{L$_{b}$} & 
\colhead{L$_{b2}$} & \colhead{L$_{h}$} &
\colhead{L$_{LF}$} & \colhead{L$_{LF/2}$} &
\colhead{L$_{\ell ow}$} & \colhead{L$_{hHz}$} &
\colhead{L$_\ell$} &  \colhead{L$_u$} \\
\colhead{Number} & \colhead{$\nu_{\rm max}$ (Hz)} & 
\colhead{$\nu_{\rm max}$ (Hz)} & \colhead{$\nu_{\rm max}$ (Hz)} & 
\colhead{$\nu_{\rm max}$ (Hz)} & \colhead{$\nu_{\rm max}$ (Hz)} & 
\colhead{$\nu_{\rm max}$ (Hz)} & \colhead{$\nu_{\rm max}$ (Hz)} &
\colhead{$\nu_{\rm max}$ (Hz)} & \colhead{$\nu_{\rm max}$ (Hz)} 
}

\startdata
\multicolumn{10}{c}{the accreting millisecond pulsars}\\
\hline
\multicolumn{10}{c}{SAX J1808.4--3658 (the 2002 outburst)}\\
\hline
1 &      10.34$\pm$0.17  & --	     	& 48.4$\pm$1.2    & --		   & -- & --	& 196$\pm$17	& --		& 606.0$\pm$4.4\\
2 &      15.66$\pm$0.28  & 8.7$\pm$1.0  & 78.4$\pm$1.7	  & 47.0$\pm$1.1   & -- & --	& 258$\pm$29    & --    	& 691.2$\pm$8.7\\
3 &      16.02$\pm$0.25  & 10.4$\pm$1.0 & 74.9$\pm$1.2    & 47.28$\pm$0.81 & -- & --	& 329$\pm$42	& 503.6$\pm$5.3 & 685.1$\pm$5.6\\
4 &      7.204$\pm$0.097 & --   	& 35.78$\pm$0.37  & --		   & -- & --	& 134.9$\pm$9.3 & -- 		& 497.6$\pm$6.9\\
5 &      3.165$\pm$0.064 & --  	     	& 18.21$\pm$0.19  & --		   & -- & --	& 120.6$\pm$9.1 & -- 		& 352.1$\pm$6.5\\
6 &      3.083$\pm$0.058 & --  	     	& 18.63$\pm$0.20  & --		   & -- & --	& 124$\pm$11    & --    	& 339.5$\pm$6.2\\
7 &      2.991$\pm$0.049 & --  	     	& 18.17$\pm$0.19  & --		   & -- & --	& 149$\pm$24	& --      	& 328.5$\pm$9.2\\
8 &      2.264$\pm$0.056 & --   	& 13.52$\pm$0.28  & --		   & -- & --	& 106$\pm$12	& --      	& 299.9$\pm$8.0\\
9 &      5.05$\pm$0.11   & --    	& 25.41$\pm$0.28  & --		   & -- & --	& 118.6$\pm$8.7 & -- 		& 396.4$\pm$4.2\\
10 &     3.94$\pm$0.11   & --    	& 21.84$\pm$0.32  & --		   & -- & --	& 119$\pm$14	& --      	& 354.7$\pm$8.0\\
11 &     3.68$\pm$0.11   & --   	& 20.30$\pm$0.34  & --		   & -- & --	& 110$\pm$12	& --       	& 349$\pm$12\\
12 &     3.51$\pm$0.10   & --   	& 21.19$\pm$0.44  & --		   & -- & --	& 110$\pm$23	& --      	& 341$\pm$10\\   
13 &     3.205$\pm$0.077 & --   	& 19.44$\pm$0.30  & --		   & -- & --	& 148$\pm$25	& --        	& 337.5$\pm$7.6\\
14 &     3.93$\pm$0.13   & --     	& 21.85$\pm$0.34  & --		   & -- & --	& 173$\pm$30	& --        	& 352$\pm$12\\   
15 &     4.54$\pm$0.75   & --    	& 30.0$\pm$2.0    & --		   & -- & --	& --	        & --		& 374$^{+48}_{-29}$\\
\hline
\multicolumn{10}{c}{SAX J1808.4--3658 (the 1998 outburst)}\\
\hline
16 & 1.940$\pm$0.077   & -- & 12.44$\pm$0.37  & -- & -- & --           & 186$^{+44}_{-108}$\tablenotemark{1} &  -- & 269$\pm$17\tablenotemark{1}\\               
17 & 0.407$\pm$0.011   & -- & 2.861$\pm$0.056     & -- & -- & 25.0$\pm$2.0        & --		    &  -- & 156.3$\pm$5.8\\
18 & 0.366$\pm$0.012   & -- & 2.612$\pm$0.064     & -- & -- & 21.5$\pm$1.9        & --		    &  -- & 151.2$\pm$6.8\\
19 & 0.3476$\pm$0.0093 & -- & 2.583$\pm$0.065     & -- & -- & 26.0$\pm$2.3        & --		    &  -- & 170.0$\pm$8.9\\ 
20 & 0.351$\pm$0.022   & -- & 2.77$\pm$0.18       & -- & -- & 27.8$\pm$8.8        & --		    &  -- & 169$\pm$22\\     
21 & 0.987$\pm$0.035   & -- & 5.72$\pm$0.57       & -- & -- & 57.1$\pm$7.1        & --		    &  -- & 216$\pm$14 \\     
22 & 0.969$\pm$0.054   & -- & --\tablenotemark{2} & -- & -- & --\tablenotemark{2} & --		    &  -- & --\tablenotemark{2}\\ 
\hline
\multicolumn{10}{c}{XTE J1751--305}\\
\hline
23 & 1.202$\pm$0.068 & -- & 6.10$\pm$0.16   & -- & -- & 32.6$\pm$4.7 & -- & -- & 278$\pm$13\\     
24 & 0.897$\pm$0.067 & -- & 4.229$\pm$0.092 & -- & -- &	23.0$\pm$2.6 & -- & -- & 260$\pm$56\\
25 & 0.490$\pm$0.013 & -- & 2.942$\pm$0.034 & -- & -- & 20.1$\pm$1.4 & -- & -- & 234$\pm$12\\
\hline
\multicolumn{10}{c}{XTE J0929--314}\\
\hline
26 & 0.250$\pm$0.014 & -- & 2.01$\pm$0.17   & -- & 0.742$\pm$0.013  & 31.4$\pm$4.7                   & -- & -- & 149$\pm$17\\
27 & 0.463$\pm$0.048 & -- & 3.08$\pm$0.14   & -- & 1.072$\pm$0.022  & --\tablenotemark{3}            & -- & -- & --\tablenotemark{3}\\ 
28 & 0.42$\pm$0.10   & -- & 2.65$\pm$0.40   & -- & 0.846$\pm$0.046  & --\tablenotemark{3}            & -- & -- & --\tablenotemark{3}\\ 
\hline
\multicolumn{10}{c}{XTE J1814--338}\\
\hline
29 & 0.381$\pm$0.015 & -- & 2.173$\pm$0.050 & -- & 0.910$\pm$0.031     & 21.0$\pm$6.2           & -- & -- & 191$\pm$30\\
30 & 0.500$\pm$0.026 & -- & 3.048$\pm$0.099 & -- & --                  & 49$^{+31}_{-11}$       & -- & -- & 220$^{+58}_{-29}$\\
\hline
\hline
\multicolumn{10}{c}{the low luminosity bursters}\\
\hline
\multicolumn{10}{c}{1E 1724--3045}\\
\hline
C & 0.1614$\pm$0.0034	& -- & 1.0206$\pm$0.0064 & --			     & -- & 11.12$\pm$0.21  & -- & -- & 156.0$\pm$6.3  \\       
D & 0.555$\pm$0.035  	& -- & 3.421$\pm$0.100   & --			     & -- & 17.4$\pm$3.5    & -- & -- & 230$\pm$26  \\     
E & 0.311$\pm$0.026  	& -- & 1.936$\pm$0.048   & --			     & -- & 16.8$\pm$1.3    & -- & -- & 190$\pm$17 \\      
F & 0.1338$\pm$0.0074	& -- & 0.822$\pm$0.018   & 0.6246$\pm$0.0049         & -- & 9.65$\pm$0.25   & -- & -- & 146$\pm$10 \\       
G & 0.1392$\pm$0.0075	& -- & 0.905$\pm$0.021   & 0.6854$\pm$0.0082         & -- & 9.75$\pm$0.37   & -- & -- & 138$\pm$10\\        
H & 0.1276$\pm$0.0048 	& -- & 0.746$\pm$0.015   & 0.5137$\pm$0.0048         & -- & 8.54$\pm$0.21   & -- & -- & 133.7$\pm$7.6 \\   
I & 0.1362$\pm$0.0066 	& -- & 0.879$\pm$0.025   & 0.688$^{+0.013}_{-0.007}$ & -- & 9.67$\pm$0.45   & -- & -- & 169$\pm$23   \\    
J & 0.1647$\pm$0.0073 	& -- & 1.024$\pm$0.017   & 0.7676$\pm$0.0091         & -- & 11.27$\pm$0.30  & -- & -- & 167.7$\pm$9.4 \\   
K & 0.207$\pm$0.014  	& -- & 1.391$\pm$0.058   & 1.015$\pm$0.015           & -- & 12.43$\pm$0.80  & -- & -- & 157$\pm$21  \\      
L & 0.175$\pm$0.011  	& -- & 1.155$\pm$0.046   & 0.877$\pm$0.021           & -- & 12.62$\pm$0.61  & -- & -- & 173$\pm$17 \\       
\hline
\multicolumn{10}{c}{SLX 1735--269}\\
\hline
M & 0.1642$\pm$0.0090	& -- & 1.155$\pm$0.033   & --			     & -- & 13.5$\pm$1.6    & -- & -- & 176$\pm$26\\
\hline
\multicolumn{10}{c}{GS 1826--24}\\
\hline
N & 0.264$\pm$0.011 	& -- & 1.787$\pm$0.040   & 1.128$\pm$0.022           & -- & 15.51$\pm$0.67  & -- & -- & 178$\pm$24\\
O & 0.2208$\pm$0.0081	& -- & 1.525$\pm$0.031   & 0.975$\pm$0.020           & -- & 14.42$\pm$0.49  & -- & -- & 167$\pm$16\\
P & 0.437$\pm$0.032  	& -- & 2.609$\pm$0.084   & 1.814$\pm$0.039     	     & -- & 21.2$\pm$1.2    & -- & -- & 187$\pm$31\\
Q & 0.805$\pm$0.043  	& -- & 4.86$\pm$0.16     & --			     & -- & 24.8$\pm$2.7    & -- & -- & 292$\pm$50\\
R & 0.568$\pm$0.033  	& -- & 3.79$\pm$0.31     & 2.39$^{+0.11}_{-0.07}$    & -- & 20.87$\pm$0.98  & -- & -- & 240$\pm$35\\
\enddata
\tablenotetext{1}{the power of this Lorentzian could only be measured to an accuracy of better 
than 50\% (and not 33\%, see \S \ref{sec.1808_1998}).}
\tablenotetext{2}{due to bad statistics the power above 4 Hz was fitted with only one broad Lorentzian. 
L$_{h}$, L$_{\ell ow}$, and L$_{u}$ could not be identified (see \S \ref{sec.1808_1998}).}
\tablenotetext{3}{the high frequency part of the power spectra is fitted with one broad Lorentzian that probably 
fits both L$_{\ell ow}$ and L$_u$ (see \S \ref{sec.0929}).}
\tablecomments{Listed are the characteristic frequencies ($\equiv 
\nu_{\rm max}$) of the different Lorentzians used to fit 
the power spectra of the accreting millisecond pulsars (described 
in \S 3) and the low luminosity bursters (see \S \ref{sec.compare_timing}). 
The groups of the low luminosity bursters are the same as in \citet{belloni02:apj572}.
The quoted errors in $\nu_{\rm max}$ use $\Delta\chi^{2}$ = 1.0 (corresponding to a 68\% confidence level).\\
}

\end{deluxetable}
\end{center}

\begin{center}
\begin{deluxetable}{lccccccccc}
\tabletypesize{\tiny}
\tablewidth{0pt}
\tablecaption{$Q$ values of the multi-Lorentzian fit for the 
accreting millisecond pulsars and low luminosity bursters\label{tbl.qvalues}}
\tablehead{ 
\colhead{Group} & \colhead{L$_{b}$} & 
\colhead{L$_{b2}$} & \colhead{L$_{h}$} &
\colhead{L$_{LF}$} & \colhead{L$_{LF/2}$} &
\colhead{L$_{\ell ow}$} & \colhead{L$_{hHz}$} &
\colhead{L$_\ell$} &  \colhead{L$_u$} \\
\colhead{Number} & \colhead{$Q$} & 
\colhead{$Q$} & \colhead{$Q$} & 
\colhead{$Q$} & \colhead{$Q$} & 
\colhead{$Q$} & \colhead{$Q$} &
\colhead{$Q$} & \colhead{$Q$} 
}

\startdata
\multicolumn{10}{c}{the accreting millisecond pulsars}\\
\hline
\multicolumn{10}{c}{SAX J1808.4--3658 (the 2002 outburst)}\\
\hline
1  & 0.322$\pm$0.013        & --	     	& 1.03$\pm$0.11   & --             & -- & -- & 0.60$\pm$0.26   & --		&  5.9$\pm$1.0  \\   
2  & 1.36$\pm$0.20          & 0.337$\pm$0.034	& 2.47$\pm$0.57   & 2.25$\pm$0.37  & -- & -- & 0.83$\pm$0.38   & --		& 5.17$\pm$0.95 \\  
3  & 1.35$^{+0.28}_{-0.18}$ & 0.272$\pm$0.023   & 2.66$\pm$0.52   & 2.38$\pm$0.30  & -- & -- & 0.35$\pm$0.18   & 14.26 (fixed) 	& 5.60$\pm$0.77 \\
4  & 0.312$\pm$0.012        & --           	& 1.195$\pm$0.061 & --             & -- & -- & 0.43$\pm$0.12   & --		&  2.75$\pm$0.32 \\   
5  & 0.238$\pm$0.015        & --           	& 0.813$\pm$0.042 & --             & -- & -- & 0.281$\pm$0.094 & --		&  2.12$\pm$0.30\\   
6  & 0.257$\pm$0.014        & --           	& 0.807$\pm$0.042 & --             & -- & -- & 0.28$\pm$0.11   & --		&  2.22$\pm$0.35 \\   
7  & 0.286$\pm$0.014        & --           	& 0.914$\pm$0.050 & --             & -- & -- & 0.03$\pm$0.17   & --		&  3.1$^{+1.6}_{-0.8}$ \\
8  & 0.211$\pm$0.017        & --           	& 0.642$\pm$0.065 & --             & -- & -- & 0.18$\pm$0.10   & --		&  1.70$\pm$0.42 \\    
9  & 0.255$\pm$0.016        & --           	& 1.136$\pm$0.073 & --             & -- & -- & 0.32$\pm$0.13   & --		&  4.13$\pm$0.55 \\    
10 & 0.234$\pm$0.019        & --           	& 0.890$\pm$0.065 & --             & -- & -- & 0.30$\pm$0.17   & --		&  2.63$\pm$0.54\\    
11 & 0.245$\pm$0.021        & --           	& 0.949$\pm$0.088 & --             & -- & -- & 0.34$\pm$0.19   & --		&  2.21$^{+0.68}_{-0.45}$\\
12 & 0.217$\pm$0.021        & --           	& 0.875$\pm$0.087 & --             & -- & -- & 0.35$\pm$0.30   & --		&  2.07$^{+0.78}_{-0.52}$\\
13 & 0.237$\pm$0.017        & --           	& 0.891$\pm$0.067 & --             & -- & -- & 0.22$\pm$0.15   & --		&  2.77$^{+0.99}_{-0.69}$\\
14 & 0.208$\pm$0.022        & --           	& 1.085$\pm$0.090 & --             & -- & -- & 0 (fixed)       & --		&  3.4$^{+1.3}_{-0.9}$\\ 
15 & 0.13$\pm$0.10          & --           	& 0.85$\pm$0.26   & --             & -- & -- & --              & --		&  1.47$\pm$0.77 \\   
\hline
\multicolumn{10}{c}{SAX J1808.4--3658 (the 1998 outburst)}\\
\hline
16 & 0.342$\pm$0.035 & -- & 0.489$\pm$0.074  & -- & -- & --            & 0 (fixed)\tablenotemark{1}     &  -- & 1.6$^{+1.6}_{-0.9}$~\tablenotemark{1}\\          
17 & 0.186$\pm$0.022 & -- & 0.413$\pm$0.045  & -- & -- & 0 (fixed)     & --            &  -- & 0.362$\pm$0.064 \\
18 & 0.167$\pm$0.027 & -- & 0.450$\pm$0.061  & -- & -- & 0 (fixed)     & --            &  -- & 0.332$\pm$0.075 \\
19 & 0.222$\pm$0.021 & -- & 0.240$\pm$0.050  & -- & -- & 0 (fixed)     & --            &  -- & 0.278$\pm$0.077\\ 
20 & 0.309$\pm$0.060 & -- & 0.18$\pm$0.13    & -- & -- & 0 (fixed)     & --            &  -- & 0.26$\pm$0.26  \\   
21 & 0.186$\pm$0.020 & -- & 0 (fixed)  	     & -- & -- & 0.28$\pm$0.16 & --            &  -- & 0.54$\pm$0.13 \\   
22 & 0.020$\pm$0.045 & -- & --\tablenotemark{2}                & -- & -- & --\tablenotemark{2}            & --            &  -- & --\tablenotemark{2}            \\ 
\hline
\multicolumn{10}{c}{XTE J1751--305}\\
\hline
23 & 0.167$\pm$0.033 & -- & 0.70$\pm$0.10    & -- & -- & 0.13$^{+0.23}_{-0.93}$ & -- & -- & 1.01$\pm$0.28\\     
24 & 0.106$\pm$0.043 & -- & 0.787$\pm$0.091  & -- & -- & 0.39$\pm$0.20  	& -- & -- & 0.08$^{+0.22}_{-0.46}$\\
25 & 0.057$\pm$0.018 & -- & 0.724$\pm$0.038  & -- & -- & 0 (fixed)          	& -- & -- & 0.70$\pm$0.12\\  
\hline
\multicolumn{10}{c}{XTE J0929--314}\\
\hline
26 & 0.76$\pm$0.15          & -- & 0 (fixed)       & -- & 2.67$\pm$0.54          & 0.47$\pm$0.20  & -- & -- & 0.43$\pm$0.21\\
27 & 0.382$\pm$0.058        & -- & 0.367$\pm$0.097 & -- & 2.35$^{+0.77}_{-0.51}$ &  --\tablenotemark{3}            & -- & -- & --\tablenotemark{3} \\ 
28 & 0.49$^{+0.20}_{-0.12}$ & -- & 0.11$\pm$0.21   & -- & 2.0$^{+1.4}_{-0.7}$    &  -- \tablenotemark{3}           & -- & -- & --\tablenotemark{3} \\
\hline
\multicolumn{10}{c}{XTE J1814--338}\\
\hline
29 & 0.251$\pm$0.026 & -- & 0.581$\pm$0.071 & -- & 3.5$^{+1.5}_{-1.0}$ & 0 (fixed)              & -- & -- & 0.09$^{+0.19}_{-0.33}$\\
30 & 0.335$\pm$0.037 & -- & 0.290$\pm$0.065 & -- & --                  & 0.13$^{+0.20}_{-0.31}$ & -- & -- & 0.41$^{+0.33}_{-0.15}$\\
\hline
\hline
\multicolumn{10}{c}{the low luminosity bursters}\\
\hline
\multicolumn{10}{c}{1E 1724--3045}\\
\hline
C & 0.122$\pm$0.019 & -- & 0.570$\pm$0.016 & --                  &  -- & 0 (fixed)       & -- & -- & 0.268$\pm$0.051  \\    
D & 0.139$\pm$0.044 & -- & 0.653$\pm$0.093 & --                  &  -- & 0 (fixed)       & -- & -- & 0.38$\pm$0.15  \\    
E & 0.117$\pm$0.059 & -- & 0.600$\pm$0.083 & --                  &  -- & 0.10$\pm$0.13   & -- & -- & 0.72$\pm$0.20  \\   
F & 0.135$\pm$0.043 & -- & 0.470$\pm$0.037 & 7.5$^{+2.2}_{-1.5}$ &  -- & 0.081$\pm$0.042 & -- & -- & 0.208$\pm$0.085 \\   
G & 0.148$\pm$0.044 & -- & 0.512$\pm$0.044 & 6.6$^{+8.9}_{-2.2}$ &  -- & 0.036$\pm$0.068 & -- & -- & 0.30$\pm$0.12   \\
H & 0 (fixed)       & -- & 0.535$\pm$0.032 & 6.21$\pm$0.82       &  -- & 0 (fixed)       & -- & -- & 0.229$\pm$0.078 \\        
I & 0.210$\pm$0.046 & -- & 0.479$\pm$0.057 & 7$^{+17}_{-3}$      &  -- & 0.033$\pm$0.085 & -- & -- & 0.11$^{+0.15}_{-0.25}$ \\   
J & 0.083$\pm$0.036 & -- & 0.553$\pm$0.031 & 7.1$\pm$1.3         &  -- & 0 (fixed)       & -- & -- & 0.375$\pm$0.078  \\
K & 0.122$\pm$0.049 & -- & 0.505$\pm$0.064 & 4.4$^{+1.6}_{-1.1}$ &  -- & 0.07$\pm$0.12   & -- & -- & 0.25$\pm$0.16  \\   
L & 0.111$\pm$0.045 & -- & 0.491$\pm$0.050 & 3.2$^{+1.6}_{-0.9}$ &  -- & 0 (fixed)       & -- & -- & 0.30$\pm$0.13 \\    
\hline
\multicolumn{10}{c}{SLX 1735--269}\\
\hline
M & 0.040$\pm$0.042 & -- & 0.457$\pm$0.061 & --                  &  -- & 0 (fixed)       & -- & -- & 0.33$\pm$0.16 \\ 
\hline
\multicolumn{10}{c}{GS 1826--24}\\
\hline
N & 0.050$\pm$0.029 & -- & 0.562$\pm$0.050 & 7.1$\pm$2.9         & -- & 0.082$\pm$0.075 & -- & -- & 0.38$\pm$0.23\\ 
O & 0.139$\pm$0.028 & -- & 0.486$\pm$0.037 & 6.2$^{+2.5}_{-1.5}$ & -- & 0 (fixed)       & -- & -- & 0.47$\pm$0.14 \\          
P & 0.049$\pm$0.045 & -- & 0.69$\pm$0.10   & 5.6$^{+2.6}_{-1.3}$ & -- & 0.05$\pm$0.10 	& -- & -- & 1.4 $^{+1.3}_{-0.6}$\\ 
Q & 0.133$\pm$0.040 & -- & 0.76$\pm$0.12   & --			 & -- & 0.29$\pm$0.20 	& -- & -- & 0.82$\pm$0.44\\ 
R & 0.017$\pm$0.043 & -- & 0.684$\pm$0.095 & 2.9$\pm$1.1	 & -- & 0.43$\pm$0.11 	& -- & -- & 0.46$\pm$0.20\\ 
\enddata
\tablenotetext{1}{the power of this Lorentzian could only be measured to an accuracy of better 
than 50\% (and not 33\%, see \S \ref{sec.1808_1998}).}
\tablenotetext{2}{due to bad statistics the power above 4 Hz was fitted with only one broad Lorentzian. 
L$_{h}$, L$_{\ell ow}$, and L$_{u}$ could not be identified (see \S \ref{sec.1808_1998}).}
\tablenotetext{3}{the high frequency part of the power spectra is fitted with one broad Lorentzian that probably 
fits both L$_{\ell ow}$ and L$_u$ (see \S \ref{sec.0929}).}
\tablecomments{Listed are the $Q$ values ($\equiv \nu_{\rm 0}/2\Delta$) of 
the different Lorentzians used to fit 
the power spectra of the accreting millisecond pulsars (described 
in \S 3) and the low luminosity bursters (see \S \ref{sec.compare_timing}). 
The groups of the low luminosity bursters are the same as in \citet{belloni02:apj572}.
The quoted errors in $Q$ use $\Delta\chi^{2}$ = 1.0 (corresponding to a 68\% confidence level).\\
}

\end{deluxetable}
\end{center}

\begin{center}
\begin{deluxetable}{lccccccccc}
\tabletypesize{\tiny}
\tablewidth{0pt}
\tablecaption{Integrated fractional rms of the multi-Lorentzian fit for the 
accreting millisecond pulsars and low luminosity bursters\label{tbl.rms}}
\tablehead{ 
\colhead{Group} & \colhead{L$_{b}$} & 
\colhead{L$_{b2}$} & \colhead{L$_{h}$} &
\colhead{L$_{LF}$} & \colhead{L$_{LF/2}$} &
\colhead{L$_{\ell ow}$} & \colhead{L$_{hHz}$} &
\colhead{L$_\ell$} &  \colhead{L$_u$} \\
 & \colhead{rms (\%)} & 
\colhead{rms (\%)} & \colhead{rms (\%)} & 
\colhead{rms (\%)} & \colhead{rms (\%)} & 
\colhead{rms (\%)} & \colhead{rms (\%)} &
\colhead{rms (\%)} & \colhead{rms (\%)} 
}

\startdata
\multicolumn{10}{c}{the accreting millisecond pulsars}\\
\hline
\multicolumn{10}{c}{SAX J1808.4--3658 (the 2002 outburst)}\\
\hline
1  & 17.05$\pm$0.14 & --             & 11.78$\pm$0.65 & --             & -- & -- & 10.2$\pm$1.1   & --             & 8.92$\pm$0.50\\  
2  & 8.68$\pm$0.91  & 10.13$\pm$0.85 & 6.96$\pm$0.75  & 6.87$\pm$0.57  & -- & -- & 7.98$\pm$0.98  & --             & 8.32$\pm$0.57\\  
3  & 8.7$\pm$1.2    & 12.55$\pm$0.96 & 6.79$\pm$0.66  & 7.17$\pm$0.49  & -- & -- & 11.66$\pm$0.95 & 2.94$\pm$0.48  & 8.78$\pm$0.58\\  
4  & 16.61$\pm$0.11 & --             & 13.94$\pm$0.38 & --             & -- & -- & 12.43$\pm$0.72 & --             & 10.87$\pm$0.49\\  
5  & 15.40$\pm$0.15 & --             & 15.71$\pm$0.35 & --             & -- & -- & 14.99$\pm$0.78 & --             & 10.84$\pm$0.74\\  
6  & 15.11$\pm$0.14 & --             & 15.73$\pm$0.35 & --             & -- & -- & 14.71$\pm$0.87 & --             & 10.76$\pm$0.82\\  
7  & 15.23$\pm$0.12 & --             & 15.25$\pm$0.40 & --             & -- & -- & 17.3$\pm$1.3   & --             & 8.8$\pm$1.5\\     
8  & 16.23$\pm$0.20 & --             & 14.21$\pm$0.59 & --             & -- & -- & 17.0$\pm$1.1   & --             & 11.4$\pm$1.3\\    
9  & 16.03$\pm$0.16 & --             & 14.58$\pm$0.48 & --             & -- & -- & 14.39$\pm$0.87 & --             & 10.29$\pm$0.57\\  
10 & 15.30$\pm$0.21 & --             & 16.11$\pm$0.56 & --             & -- & -- & 14.1$\pm$1.2   & --             & 10.67$\pm$0.99\\  
11 & 15.56$\pm$0.22 & --             & 15.58$\pm$0.67 & --             & -- & -- & 14.5$\pm$1.4   & --             & 10.9$\pm$1.2\\    
12 & 15.67$\pm$0.22 & --             & 16.03$\pm$0.84 & --             & -- & -- & 13.9$\pm$2.2   & --             & 11.9$\pm$1.6\\    
13 & 16.37$\pm$0.18 & --             & 15.60$\pm$0.51 & --             & -- & -- & 16.5$\pm$1.5   & --             & 10.7$\pm$1.5\\    
14 & 16.14$\pm$0.23 & --             & 14.28$\pm$0.48 & --             & -- & -- & 19.22$\pm$0.93 & --             & 9.0$\pm$1.4\\     
15 & 15.4$\pm$1.1   & --             & 17.1$\pm$1.5   & --             & -- & -- & $<16.0$\tablenotemark{1} & --   & 19.5$\pm$2.8 \\  
\hline
\multicolumn{10}{c}{SAX J1808.4--3658 (the 1998 outburst)}\\
\hline
16 & 15.32$\pm$0.38 & -- & 16.09$\pm$0.61 & -- & -- & --              & 18.7$^{+2.1}_{-5.0}$~\tablenotemark{2} 	&  -- & 10.1$^{+9.5}_{-2.6}$~\tablenotemark{2}\\ 
17 & 19.76$\pm$0.25 & -- & 15.85$\pm$0.49      & -- & -- & 18.14$\pm$0.46  & --            		&  -- & 19.85$\pm$0.68\\ 
18 & 20.85$\pm$0.31 & -- & 15.34$\pm$0.64      & -- & -- & 19.04$\pm$0.51  & --            		&  -- & 20.58$\pm$0.79\\ 
19 & 20.26$\pm$0.31 & -- & 17.33$\pm$0.60      & -- & -- & 19.04$\pm$0.55  & --            		&  -- & 19.65$\pm$0.85\\ 
20 & 18.83$\pm$0.82 & -- & 18.0$\pm$1.5        & -- & -- & 18.1$\pm$2.0    & --            		&  -- & 20.8$\pm$2.6 \\  
21 & 16.33$\pm$0.39 & -- & 16.53$\pm$0.35      & -- & -- & 15.8$\pm$2.0    & --            		&  -- & 17.3$\pm$1.6 \\  
22 & 25.08$\pm$0.48 & -- & --\tablenotemark{3} & -- & -- & --\tablenotemark{3} & --            		&  -- & --\tablenotemark{3} \\  
\hline
\multicolumn{10}{c}{XTE J1751--305}\\
\hline
23 & 10.70$\pm$0.29  & -- & 12.0$^{+0.8}_{-1.2}$  & -- & -- & 11.2$\pm$2.5    & -- & -- & 12.2$\pm$1.1 \\  
24 & 11.89$\pm$0.39  & -- & 12.37$\pm$0.65	  & -- & -- & 10.9$\pm$1.5    & -- & -- & 15.3$\pm$1.5 \\  
25 & 12.53$\pm$0.13  & -- & 12.05$\pm$0.26	  & -- & -- & 14.57$\pm$0.23  & -- & -- & 13.49$\pm$0.57 \\ 
\hline
\multicolumn{10}{c}{XTE J0929--314}\\
\hline
26 & 5.36$\pm$0.44   & -- & 15.99$\pm$0.25  & -- & 5.00$\pm$0.49        & 13.3$\pm$2.1  & -- & -- & 18.8$\pm$2.0\\
27 & 8.13$\pm$0.52   & -- & 12.66$\pm$0.61  & -- & 5.23$\pm$0.73        & --\tablenotemark{4} & -- & -- & --\tablenotemark{4}\\ 
28 & 7.1$\pm$1.6     & -- & 12.6$\pm$1.2    & -- & 5.0$^{+1.8}_{-1.2}$  & --\tablenotemark{4} & -- & -- & --\tablenotemark{4}\\ 
\hline
\multicolumn{10}{c}{XTE J1814--338}\\
\hline
29 & 11.82$\pm$0.28   & -- & 13.92$\pm$0.65  & -- & 2.89$\pm$0.58       & 15.4$\pm$1.9           & -- & -- & 22.0$\pm$2.1\\
30 & 9.83$\pm$0.46    & -- & 15.81$\pm$0.705 & -- & --                  & 16.5$^{+5.7}_{-2.8}$   & -- & -- & 19.2$^{+2.4}_{-4.5}$\\
\hline
\hline
\multicolumn{10}{c}{the low luminosity bursters}\\
\hline
\multicolumn{10}{c}{1E 1724--3045}\\
\hline
C & 14.16$\pm$0.14  & -- & 16.06$\pm$0.14 & --                     & -- & 18.201$\pm$0.098        & -- & -- & 13.87$\pm$0.30 \\
D & 12.00$\pm$0.33  & -- & 12.20$\pm$0.74 & --                     & -- & 12.24$\pm$0.52   	  & -- & -- & 14.37$\pm$0.93 \\
E & 11.87$\pm$0.42  & -- & 13.57$\pm$0.69 & --                     & -- & 14.14$\pm$0.79   	  & -- & -- & 11.73$\pm$0.92 \\
F & 14.37$\pm$0.39  & -- & 16.46$\pm$0.46 & 3.25$\pm$0.31          & -- & 18.50$\pm$0.40   	  & -- & -- & 14.75$\pm$0.56 \\
G & 14.32$\pm$0.36  & -- & 16.27$\pm$0.49 & 2.51$^{+0.66}_{-0.32}$ & -- & 18.20$\pm$0.57   	  & -- & -- & 14.13$\pm$0.76 \\
H & 16.07$\pm$0.24  & -- & 15.15$\pm$0.35 & 4.21$\pm$0.31          & -- & 19.33$\pm$0.13   	  & -- & -- & 13.70$\pm$0.41 \\
I & 14.12$\pm$0.38  & -- & 16.00$\pm$0.55 & 3.29$^{+0.74}_{-0.46}$ & -- & 18.47$^{+0.65}_{-0.93}$ & -- & -- & 14.3$\pm$1.0   \\
J & 14.71$\pm$0.29  & -- & 15.42$\pm$0.34 & 3.32$\pm$0.34          & -- & 18.39$\pm$0.13	  & -- & -- & 13.65$\pm$0.45 \\
K & 14.29$\pm$0.44  & -- & 15.36$\pm$0.74 & 4.47$^{+0.88}_{-0.61}$ & -- & 16.71$\pm$0.99	  & -- & -- & 14.0$\pm$1.1   \\
L & 14.15$\pm$0.39  & -- & 15.38$\pm$0.59 & 4.7$^{+1.3}_{-0.8}$    & -- & 18.03$\pm$0.24	  & -- & -- & 13.39$\pm$0.71 \\
\hline
\multicolumn{10}{c}{SLX 1735--269}\\
\hline
M & 9.95$\pm$0.24   & -- & 9.96$\pm$0.38 & --                      & -- & 12.37$\pm$0.41	  & -- & -- & 12.78$\pm$0.94 \\
\hline
\multicolumn{10}{c}{GS 1826--24}\\
\hline
N & 12.66$\pm$0.21  & -- & 11.90$\pm$0.45 & 1.92$\pm$0.52	   & -- & 15.28$\pm$0.61	  & -- & -- & 9.01$\pm$0.99  \\
O & 14.28$\pm$0.24  & -- & 14.20$\pm$0.36 & 2.38$\pm$0.39	   & -- & 18.60$\pm$0.16	  & -- & -- & 10.27$\pm$0.63 \\
P & 14.14$\pm$0.41  & -- & 12.73$\pm$0.88 & 3.15$\pm$0.57	   & -- & 17.24$\pm$0.66	  & -- & -- & 7.5$\pm$1.4    \\
Q & 13.61$\pm$0.31  & -- & 13.36$\pm$1.00 & --                     & -- & 13.2$^{+1.5}_{-1.1}$ 	  & -- & -- & 10.0$\pm$1.5   \\
R & 14.25$\pm$0.34  & -- & 13.02$\pm$0.84 & 4.6$^{+1.8}_{-0.7}$    & -- & 13.75$\pm$0.78	  & -- & -- & 12.6$\pm$1.1   \\
\enddata
\tablenotetext{1}{95 \% confidence upper limits (see \S \ref{sec.1808_2002}).}
\tablenotetext{2}{the power of this Lorentzian could only be measured to an accuracy of better 
than 50\% (and not 33\%, see \S \ref{sec.1808_1998}).}
\tablenotetext{3}{due to bad statistics the power above 4 Hz was fitted with only one broad Lorentzian. 
L$_{h}$, L$_{\ell ow}$, and L$_{u}$ could not be identified (see \S \ref{sec.1808_1998}).}
\tablenotetext{4}{the high frequency part of the power spectra is fitted with one broad Lorentzian that probably 
fits both L$_{\ell ow}$ and L$_u$ (see \S \ref{sec.0929}).}

\tablecomments{Listed are the values of the integrated fractional rms
(over the full PCA energy band) of the different Lorentzians used to fit 
the power spectra of the accreting millisecond pulsars (described 
in \S 3) and the low luminosity bursters (see \S \ref{sec.compare_timing}). 
The groups of the low luminosity bursters are the same as in \citet{belloni02:apj572}.
The quoted errors in the rms use $\Delta\chi^{2}$ = 1.0 (corresponding to a 68\% confidence level).\\
}
\end{deluxetable}
\end{center}

\begin{center}
\begin{deluxetable}{ll}
\tabletypesize{\footnotesize}
\tablewidth{0pt}
\tablecaption{{\it RXTE} observation IDs of the fit groups\label{tbl.obsid}}
\tablehead{ 
\colhead{Group} & \colhead{{\it RXTE} observation IDs} 
}
\startdata
1  & C-04-00, A-01-000\\
2  & A-01-00, A-01-03, A-01-04\\
3  & A-01-01\\
4  & A-01-020, A-01-02\\
5  & A-02-01, A-02-000, A-02-00\\
6  & A-02-02, A-02-03, A-02-04\\
7  & A-02-05, A-02-06, A-02-07\\
8  & A-02-10, A-02-08\\
9  & A-02-20, A-02-09, A-02-19, A-02-23\\
10 & A-02-21, A-02-11, A-02-18, A-02-12, A-02-22\\
11 & A-02-13, A-02-15, A-02-14, A-02-16\\
12 & A-02-17, A-03-000, A-03-00, A-03-04, A-03-05\\
13 & A-03-06, A-03-07, A-03-010, A-03-01, A-03-08, A-03-09, A-03-10, A-03-11\\
14 & A-03-12, A-03-13, A-03-02, A-03-14, A-03-03, C-05-00, B-01-04\\
15 & B-01-000, B-01-00\\
16 & D-01-03S\\
17 & D-03-00, D-04-00\\
18 & D-05-00\\
19 & D-06-01, D-06-000, D-06-00\\
20 & D-07-00\\
21 & D-08-00, D-09-01, D-09-02, D-09-03, D-09-04\\
22 & D-09-00, D-10-02, D-10-01, D-10-00\\
23 & E-01-00, F-01-00, E-02-00	\\
24 & F-02-00 \\
25 & F-03-01, F-03-00, F-04-00 F-05-03, F-05-02, F-05-01, F-05-04, F-05-000, F-05-00, F-06-00,\\
   & F-06-01, F-07-00, F-07-01, F-08-000, F-08-00, F-09-000	\\
26 & G-03-01, G-03-00, G-04-00, G-04-01, G-05-01, G-07-01\\
27 & G-02-00, G-02-01, G-06-01, G-06-00, G-05-00, G-07-00 \\	
28 & G-08-00, G-08-01, G-08-02, G-08-03, G-08-04, G-08-05, G-08-06, G-09-00, G-09-01, G-09-02, \\
   & G-10-00, G-10-01, G-10-02, G-10-03, G-10-04, G-11-00, G-12-00, G-12-01, G-12-02\\
29 & H-01-01, H-01-02, H-01-03, H-01-04, H-01-05, H-01-06, H-02-04, H-02-05, H-03-12, H-03-13,\\ 
   & H-03-000, H-03-00, H-03-01, H-03-03, H-03-04, H-03-05, H-03-02, H-03-06, H-03-07\\
30 & H-01-07, H-01-08, H-01-09, H-02-01, H-02-00, H-02-03, H-02-02, H-02-06, H-02-07, H-02-09, \\
   & H-02-08, H-03-08, H-03-09, H-03-10, H-03-11, H-04-07, H-04-00, H-04-01, H-04-02, H-04-08, \\
   & H-04-03, H-04-04, H-04-05, H-04-09, H-04-06, H-05-00, H-05-01\\
\enddata

\tablecomments{\scriptsize Listed are {\it RXTE} observation IDs of each fit group. The letters stand for part of the observation ID;
A = 70080-01, B = 70080-02, C = 70080-03, D = 30411-01, E = 70134-03, F = 70131-01, G = 70096-03, and H = 80418-01.}

\end{deluxetable}
\end{center}

\newpage

\begin{center}
\begin{deluxetable}{lcccc}
\tabletypesize{\footnotesize}
\tablewidth{0pt}
\tablecaption{Characteristic frequencies of the 410 Hz QPO\label{tbl.410hz}}
\tablehead{ 
\colhead{Groups}   & \colhead{410 Hz QPO} & \colhead{$\nu_{\rm 410}-\nu_{\rm pulse}$} &
\multicolumn{2}{c}{L$_{h}$} \\
 & \colhead{$\nu_{\rm max}$ (Hz)} & \colhead{(Hz)} & 
\colhead{$\nu_{\rm max}$ (Hz)} & \colhead{$\nu_{0}$ (Hz)}  
}
\startdata
5--6   & 409.2$^{+0.6}_{-1.0}$ & 8.3$^{+0.6}_{-1.0}$ & 18.67$\pm$0.15 & 16.81$\pm$0.13  \\  
9      & 413.41$\pm$0.30       & 12.47$\pm$0.30      & 25.55$\pm$0.29 & 23.41$\pm$0.26  \\  
10--11 & 409.3$\pm$1.7         & 8.36$\pm$1.7        & 20.95$\pm$0.24 & 18.54$\pm$0.21  \\
\enddata

\tablecomments{Listed are the characteristic frequency ($\nu_{\rm max}$) of the 410 Hz QPO, 
the frequency difference between the $\sim401$ Hz pulse spike ($\nu_{\rm pulse}$) and the 410 Hz 
QPO ($\nu_{\rm 410}-\nu_{\rm pulse}$). 
Also listed are the $\nu_{\rm max}$ and $\nu_0$ of L$_{h}$. $\nu_0$ is the 
centroid frequency of the Lorentzian and can be obtained from our fit parameters $\nu_{\rm max}$ and $Q$ as
$\nu_0 = 2 \nu_{\rm max} Q/\sqrt{4 Q^2 + 1}$. For the  410 Hz QPO $\nu_{\rm max} = \nu_0$ as $Q$ is 
between 50 and 100.}

\end{deluxetable}
\end{center}

\begin{center}
\begin{deluxetable}{lccc}
\tabletypesize{\footnotesize}
\tablewidth{0pt}
\tablecaption{Shift factors\label{tbl.shftfactors}}
\tablehead{ 
\colhead{$\nu_i$} & \colhead{$\nu_u$ shift factor}  & \colhead{$\nu_i$ shift factor} & \colhead{$\chi^2$/dof}\tablenotemark{1}\\
                  & \colhead{($\nu_u \times$ factor)} & \colhead{($\nu_i \times$ factor)} & 
}
\startdata
\multicolumn{4}{c}{SAX J1808.4--3658}\\
\hline
$\nu_b$		& 1.420$\pm$0.013	& 0.355$\pm$0.013	& 35.4/31\\
$\nu_h$		& 1.481$\pm$0.013	& 0.375$\pm$0.011	& 54.5/31\\
$\nu_{\ell ow}$	& 1.653$\pm$0.073	& 0.334$\pm$0.061	& 7.8/9\\
$\nu_\ell$\tablenotemark{2}	& --			& 1.359$\pm$0.021	& 183.8/15\\
$\nu_\ell$	& --			& 0.804$\pm$0.015	& 183.8/15\\
\hline
\multicolumn{4}{c}{XTE J0929--314}\\
\hline
$\nu_b$         & 1.40$\pm$0.16		& 0.35$\pm$0.13		& 15.8/15\\
$\nu_h$         & 1.41$\pm$0.16		& 0.42$\pm$0.13 	& 30.1/15\\
$\nu_{\ell ow}$ & 2.03$\pm$0.30		& 0.283$\pm$0.090	& 5.5/5\\
\hline
\multicolumn{4}{c}{XTE J1751--305}\\
\hline
$\nu_b$         & 1.188$\pm$0.045	& 0.585$\pm$0.073	& 18.6/17\\
$\nu_h$         & 1.112$\pm$0.042	& 0.763$\pm$0.076	& 32.9/17\\
$\nu_{\ell ow}$ & 1.036$\pm$0.058	& 0.93$\pm$0.10		& 6.3/7\\
\hline
\multicolumn{4}{c}{XTE J1814--338}\\
\hline
$\nu_b$         & 1.23$\pm$0.14		& 0.53$\pm$0.19		& 15.8/16\\
$\nu_h$         & 1.13$\pm$0.13		& 0.73$\pm$0.22		& 30.1/16\\
$\nu_{\ell ow}$ & 1.42$\pm$0.24		& 0.52$\pm$0.16		& 6.5/6\\
\hline
\multicolumn{4}{c}{1E 1724--3045}\\
\hline
$\nu_b$         & 1.167$\pm$0.043	& 0.622$\pm$0.076	& 22.2/24\\
$\nu_h$         & 1.037$\pm$0.040	& 0.915$\pm$0.091	& 36.7/24\\
$\nu_{\ell ow}$ & 1.051$\pm$0.077	& 0.92$\pm$0.12		& 9.7/14 \\
\hline
\multicolumn{4}{c}{SLX 1735--269}\\
\hline
$\nu_b$         & 1.04$\pm$0.15		& 0.90$\pm$0.40		& 15.8/15\\
$\nu_h$         & 0.96$\pm$0.14		& 1.11$\pm$0.41		& 30.1/15\\
$\nu_{\ell ow}$ & 1.07$\pm$0.18		& 0.88$\pm$0.27		& 5.5/5\\
\hline
\multicolumn{4}{c}{GS 1826--24}\\
\hline
$\nu_b$         & 1.195$\pm$0.071	& 0.58$\pm$0.11		& 17.4/19\\
$\nu_h$         & 1.134$\pm$0.067 	& 0.73$\pm$0.11		& 31.1/19\\
$\nu_{\ell ow}$ & 1.117$\pm$0.075	& 0.83$\pm$0.11		& 9.1/9\\
\enddata
\tablenotetext{1}{the quoted $\chi^2$/dof are independent of whether the shift was in $\nu_i$ or $\nu_u$.}
\tablenotetext{2}{after $\nu_u$ was multiplied with 1.454 (see \S \ref{sec.compare_timing}).}

\tablecomments{Shift factors between the frequency relations of the listed sources and the atoll 
sources (see \S \ref{sec.compare_timing}). The errors in the shift factor use $\Delta\chi^{2}$ = 1.0.}

\end{deluxetable}
\end{center}

\begin{center}
\begin{deluxetable}{lccc}
\tabletypesize{\footnotesize}
\tablewidth{0pt}
\tablecaption{Power law fits to the frequency relations\label{tbl.plfits}}
\tablehead{ 
\colhead{Component} & \colhead{Index} & \colhead{Normalization} & 
\colhead{$\chi^2$/dof}
}
\startdata
L$_{b}$       & 2.724$\pm$0.038 & $1.49 (\pm0.34) \times 10^{-7}$ & 152/53\\
L$_{h}$       & 2.439$\pm$0.030 & $4.62 (\pm0.85) \times 10^{-6}$ & 89/53\\
L$_{\ell ow}$ & 1.81$\pm$0.11   & $1.15 (\pm0.66) \times 10^{-3}$ & 34/31\\
L$_{LF}$      & 3.00$\pm$0.63   & $2.0 (\pm6.9) \times 10^{-7}$ & 6/9\\
L$_{LF/2}$    & 3.13$\pm$0.84   & $4 (\pm22) \times 10^{-8}$ & 1.7/1\\
\enddata
\tablecomments{
Listed are the index, the normalization and the $\chi^2$/dof of power law fits to 
the $\nu_{\rm max}$ versus $\nu_u$ relations in Figure \ref{fig.freq_freq_shifted}. Only the 
data below about $\nu_u \approx 600$ Hz is included in the fit.}
\end{deluxetable}
\end{center}

\begin{center}
\begin{deluxetable}{lccccc}
\tabletypesize{\footnotesize}
\tablewidth{0pt}
\tablecaption{Characteristic parameters for the accreting millisecond 
pulsars\label{tbl.msp}}
\tablehead{ 
\colhead{Source} & \colhead{$\nu_{spin}$} & \colhead{P$_{orb}$} & 
\colhead{Bursts} & \colhead{Shift factor} & \colhead{References\tablenotemark{1}}
}
\startdata
SAX J1808.4--3658 & 401 Hz & 2   hr  & Y & $\sim1.5$      & a, b, c, d\\
XTE J0929--314   & 185 Hz & 44  min & N & $\sim1.5$      & e \\
XTE J1751--305   & 435 Hz & 42  min & N\tablenotemark{2} & $\sim1$   & f \\
XTE J1814--338   & 314 Hz & 4.3 hr  & Y & $\sim1$        & g, h, i\\
\enddata
\tablenotetext{1}{a: \cite{wijnands98:nat394}; b: \cite{chakrabarty98:nat394}; 
c: \cite{intzand01:aa372}; d: \cite{chakrabarty03:nat424};
e: \cite{galloway02:apj576}; f: \cite{markwardt02:apj575}; g: \cite{markwardt03:iauc8144}; h: \cite{markwardt03:atel164};
i: \cite{strohmayer03:apj596}}
\tablenotetext{2}{the {\it RXTE} data of XTE J1751--305 shows 1 type I X-ray bursts which 
originated from from GRS 1747--312, see \citet{intzand03:aa409}}
\tablecomments{Listed are the source name, the spin frequency, $\nu_{spin}$, the orbital 
period, P$_{orb}$, whether or not the source has shown type I X-ray bursts, and 
the shift factor that has to be applied to $\nu_{u}$ to make the relations coincide with 
those of the atoll sources (see \S \ref{sec.compare_timing}).
}

\end{deluxetable}
\end{center}
\onecolumn

\clearpage

\clearpage 

\clearpage 

\begin{thebibliography}{57}
\expandafter\ifx\csname natexlab\endcsname\relax\def\natexlab#1{#1}\fi

\bibitem[{{Abramowicz} {et~al.}(2003){Abramowicz}, {Karas}, {Kluzniak}, {Lee},
  \& {Rebusco}}]{abramowicz03:pasj55}
{Abramowicz}, M.~A., {Karas}, V., {Kluzniak}, W., {Lee}, W.~H., \& {Rebusco},
  P. 2003, \pasj, 55, 467

\bibitem[{{Barret} \& {Olive}(2002)}]{barret02:apj576}
{Barret}, D. \& {Olive}, J. 2002, \apj, 576, 391

\bibitem[{{Barret} {et~al.}(2000){Barret}, {Olive}, {Boirin}, {Done},
  {Skinner}, \& {Grindlay}}]{barret00:apj533}
{Barret}, D., {Olive}, J.~F., {Boirin}, L., {et~al.} 2000, \apj, 533, 329

\bibitem[{{Belloni} {et~al.}(2002){Belloni}, {Psaltis}, \& {van der
  Klis}}]{belloni02:apj572}
{Belloni}, T., {Psaltis}, D., \& {van der Klis}, M. 2002, \apj, 572, 392

\bibitem[{{Chakrabarty} \& {Morgan}(1998)}]{chakrabarty98:nat394}
{Chakrabarty}, D. \& {Morgan}, E.~H. 1998, \nat, 394, 346

\bibitem[{{Chakrabarty} {et~al.}(2003){Chakrabarty}, {Morgan}, {Muno},
  {Galloway}, {Wijnands}, {van der Klis}, \&
  {Markwardt}}]{chakrabarty03:nat424}
{Chakrabarty}, D., {Morgan}, E.~H., {Muno}, M.~P., {et~al.} 2003, \nat, 424, 42

\bibitem[{{Di Salvo} {et~al.}(2001){Di Salvo}, {M{\' e}ndez}, {van der Klis},
  {Ford}, \& {Robba}}]{disalvo01:546}
{Di Salvo}, T., {M{\' e}ndez}, M., {van der Klis}, M., {Ford}, E., \& {Robba},
  N.~R. 2001, \apj, 546, 1107

\bibitem[{{Ford} {et~al.}(2000){Ford}, {van der Klis}, {M{\' e}ndez},
  {Wijnands}, {Homan}, {Jonker}, \& {van Paradijs}}]{ford00:apj537}
{Ford}, E.~C., {van der Klis}, M., {M{\' e}ndez}, M., {et~al.} 2000, \apj, 537,
  368

\bibitem[{{Galloway} {et~al.}(2002){Galloway}, {Chakrabarty}, {Morgan}, \&
  {Remillard}}]{galloway02:apj576}
{Galloway}, D.~K., {Chakrabarty}, D., {Morgan}, E.~H., \& {Remillard}, R.~A.
  2002, \apjl, 576, L137

\bibitem[{{Gierli{\' n}ski} \&
  {Done}(2002{\natexlab{a}})}]{gierlinski02:mnras331}
{Gierli{\' n}ski}, M. \& {Done}, C. 2002{\natexlab{a}}, \mnras, 331, L47

\bibitem[{{Gierli{\' n}ski} \&
  {Done}(2002{\natexlab{b}})}]{gierlinski02:mnras337}
---. 2002{\natexlab{b}}, \mnras, 337, 1373

\bibitem[{{Gierli{\' n}ski} {et~al.}(2002){Gierli{\' n}ski}, {Done}, \&
  {Barret}}]{gierdonebarret02:mnras331}
{Gierli{\' n}ski}, M., {Done}, C., \& {Barret}, D. 2002, \mnras, 331, 141

\bibitem[{{Gilfanov} {et~al.}(1998){Gilfanov}, {Revnivtsev}, {Sunyaev}, \&
  {Churazov}}]{gilfanov98:aa338}
{Gilfanov}, M., {Revnivtsev}, M., {Sunyaev}, R., \& {Churazov}, E. 1998, \aap,
  338, L83

\bibitem[{{Hasinger} \& {van der Klis}(1989)}]{hasinger89:aa225}
{Hasinger}, G. \& {van der Klis}, M. 1989, \aap, 225, 79

\bibitem[{{Heindl} \& {Smith}(1998)}]{heindl98:apj506}
{Heindl}, W.~A. \& {Smith}, D.~M. 1998, \apjl, 506, L35

\bibitem[{{Homan} {et~al.}(2002){Homan}, {van der Klis}, {Jonker}, {Wijnands},
  {Kuulkers}, {M{\' e}ndez}, \& {Lewin}}]{homan02:apj568}
{Homan}, J., {van der Klis}, M., {Jonker}, P.~G., {et~al.} 2002, \apj, 568, 878

\bibitem[{{in't Zand} {et~al.}(2001){in't Zand}, {Cornelisse}, {Kuulkers},
  {Heise}, {Kuiper}, {Bazzano}, {Cocchi}, {Muller}, {Natalucci}, {Smith}, \&
  {Ubertini}}]{intzand01:aa372}
{in't Zand}, J.~J.~M., {Cornelisse}, R., {Kuulkers}, E., {et~al.} 2001, \aap,
  372, 916

\bibitem[{{in't Zand} {et~al.}(2003){in't Zand}, {Strohmayer}, {Markwardt}, \&
  {Swank}}]{intzand03:aa409}
{in't Zand}, J.~J.~M., {Strohmayer}, T.~E., {Markwardt}, C.~B., \& {Swank}, J.
  2003, \aap, 409, 659

\bibitem[{{Jonker} {et~al.}(1998){Jonker}, {Wijnands}, {van der Klis},
  {Psaltis}, {Kuulkers}, \& {Lamb}}]{jonker98:apj499}
{Jonker}, P.~G., {Wijnands}, R., {van der Klis}, M., {et~al.} 1998, \apjl, 499,
  L191

\bibitem[{{Klein-Wolt} {et~al.}(2004){Klein-Wolt}, {Homan}, \& {van der
  Klis}}]{klein04:prep}
{Klein-Wolt}, M., {Homan}, J., \& {van der Klis}, M. 2004, in preparation

\bibitem[{{Leahy} {et~al.}(1983){Leahy}, {Darbro}, {Elsner}, {Weisskopf},
  {Kahn}, {Sutherland}, \& {Grindlay}}]{leahy83:apj266}
{Leahy}, D.~A., {Darbro}, W., {Elsner}, R.~F., {et~al.} 1983, \apj, 266, 160

\bibitem[{{M{\' e}ndez}(2000)}]{mendez00:texas}
{M{\' e}ndez}, M. 2000, in Proc. 19th Texas Symposium on Relativistic
  Astrophysics and Cosmology, ed. J. Paul, T. Montmerle, \& E. Aubourg
  (Amsterdam: Elsevier), 15

\bibitem[{{M{\' e}ndez} {et~al.}(2001){M{\' e}ndez}, {van der Klis}, \&
  {Ford}}]{mendez01:apj561}
{M{\' e}ndez}, M., {van der Klis}, M., \& {Ford}, E.~C. 2001, \apj, 561, 1016

\bibitem[{{M{\' e}ndez} {et~al.}(1999){M{\' e}ndez}, {van der Klis}, {Ford},
  {Wijnands}, \& {van Paradijs}}]{mendez99:apj511}
{M{\' e}ndez}, M., {van der Klis}, M., {Ford}, E.~C., {Wijnands}, R., \& {van
  Paradijs}, J. 1999, \apjl, 511, L49

\bibitem[{{Markwardt} {et~al.}(2003){Markwardt}, {Strohmayer}, \&
  {Swank}}]{markwardt03:atel164}
{Markwardt}, C.~B., {Strohmayer}, T.~E., \& {Swank}, J.~H. 2003, The
  Astronomer's Telegram, 164, 1

\bibitem[{{Markwardt} \& {Swank}(2003)}]{markwardt03:iauc8144}
{Markwardt}, C.~B. \& {Swank}, J.~H. 2003, in IAU Circ., 8144, 1

\bibitem[{{Markwardt} {et~al.}(2002){Markwardt}, {Swank}, {Strohmayer}, {Zand},
  \& {Marshall}}]{markwardt02:apj575}
{Markwardt}, C.~B., {Swank}, J.~H., {Strohmayer}, T.~E., {Zand}, J.~J.~M.~i.,
  \& {Marshall}, F.~E. 2002, \apjl, 575, L21

\bibitem[{{Migliari} {et~al.}(2003){Migliari}, {van der Klis}, \&
  {Fender}}]{migliari03:mnras345}
{Migliari}, S., {van der Klis}, M., \& {Fender}, R.~P. 2003, \mnras, 345, L35

\bibitem[{{Miller} {et~al.}(1998){Miller}, {Lamb}, \&
  {Psaltis}}]{miller98:apj508}
{Miller}, M.~C., {Lamb}, F.~K., \& {Psaltis}, D. 1998, \apj, 508, 791

\bibitem[{{Muno}(2004)}]{muno04:rossi}
{Muno}, M.~P. 2004, in "X-Ray Timing 2003: Rossi and Beyond", ed. P. Kaaret, F.
  K. Lamb, \& J. H. Swank (Melville, NY: American Institute of Physics).
  (astro-ph/0403394)

\bibitem[{{Muno} {et~al.}(2002){Muno}, {Remillard}, \&
  {Chakrabarty}}]{muno02:apj568}
{Muno}, M.~P., {Remillard}, R.~A., \& {Chakrabarty}, D. 2002, \apjl, 568, L35

\bibitem[{{Olive} {et~al.}(2003){Olive}, {Barret}, \& {Gierli{\'
  n}ski}}]{olive03:apj583}
{Olive}, J., {Barret}, D., \& {Gierli{\' n}ski}, M. 2003, \apj, 583, 416

\bibitem[{{Olive} {et~al.}(1998){Olive}, {Barret}, {Boirin}, {Grindlay},
  {Swank}, \& {Smale}}]{olive98:aa333}
{Olive}, J.~F., {Barret}, D., {Boirin}, L., {et~al.} 1998, \aap, 333, 942

\bibitem[{{Pottschmidt} {et~al.}(2003){Pottschmidt}, {Wilms}, {Nowak},
  {Pooley}, {Gleissner}, {Heindl}, {Smith}, {Remillard}, \&
  {Staubert}}]{pottschmidt03:aa407}
{Pottschmidt}, K., {Wilms}, J., {Nowak}, M.~A., {et~al.} 2003, \aap, 407, 1039

\bibitem[{{Press} \& {Teukolsky}(1992)}]{press92:comp274}
{Press}, W.~H. \& {Teukolsky}, S.~A. 1992, Computers in Physics, 6, 274

\bibitem[{{Prins} \& {van der Klis}(1997)}]{prins97:aa319}
{Prins}, S. \& {van der Klis}, M. 1997, \aap, 319, 498

\bibitem[{{Psaltis} {et~al.}(1999{\natexlab{a}}){Psaltis}, {Belloni}, \& {van
  der Klis}}]{psaltis99:apj520}
{Psaltis}, D., {Belloni}, T., \& {van der Klis}, M. 1999{\natexlab{a}}, \apj,
  520, 262

\bibitem[{{Psaltis} {et~al.}(1999{\natexlab{b}}){Psaltis}, {Wijnands}, {Homan},
  {Jonker}, {van der Klis}, {Miller}, {Lamb}, {Kuulkers}, {van Paradijs}, \&
  {Lewin}}]{psaltisetal99:apj520}
{Psaltis}, D., {Wijnands}, R., {Homan}, J., {et~al.} 1999{\natexlab{b}}, \apj,
  520, 763

\bibitem[{{Reig} {et~al.}(2000){Reig}, {M{\' e}ndez}, {van der Klis}, \&
  {Ford}}]{reig00:apj530}
{Reig}, P., {M{\' e}ndez}, M., {van der Klis}, M., \& {Ford}, E.~C. 2000, \apj,
  530, 916

\bibitem[{{Reig} {et~al.}(2004){Reig}, {van Straaten}, \& {van der
  Klis}}]{reig04:apj}
{Reig}, P., {van Straaten}, S., \& {van der Klis}, M. 2004, \apj, 602, 918

\bibitem[{{Stella} \& {Vietri}(1998)}]{stella98:apj492}
{Stella}, L. \& {Vietri}, M. 1998, \apjl, 492, L59

\bibitem[{{Strohmayer} {et~al.}(2003){Strohmayer}, {Markwardt}, {Swank}, \&
  {in't Zand}}]{strohmayer03:apj596}
{Strohmayer}, T.~E., {Markwardt}, C.~B., {Swank}, J.~H., \& {in't Zand}, J.
  2003, \apjl, 596, L67

\bibitem[{{Titarchuk}(2002)}]{titarchuk02:apj578}
{Titarchuk}, L. 2002, \apjl, 578, L71

\bibitem[{{van der Klis}(1995)}]{vdk95:xrb252}
{van der Klis}, M. 1995, in X-ray binaries, editors W. H. G. Lewin, J. van
  Paradijs, \& E. P. J. van den Heuvel, Cambridge Univ. Press, Great Britain,
  252

\bibitem[{{van der Klis}(2001)}]{vdk01:apj561}
{van der Klis}, M. 2001, \apj, 561, 943

\bibitem[{{van Straaten} {et~al.}(2000){van Straaten}, {Ford}, {van der Klis},
  {M{\' e}ndez}, \& {Kaaret}}]{vstr00:apj540}
{van Straaten}, S., {Ford}, E.~C., {van der Klis}, M., {M{\' e}ndez}, M., \&
  {Kaaret}, P. 2000, \apj, 540, 1049

\bibitem[{{van Straaten} {et~al.}(2002){van Straaten}, {van der Klis}, {di
  Salvo}, \& {Belloni}}]{vstr02:apj568}
{van Straaten}, S., {van der Klis}, M., {di Salvo}, T., \& {Belloni}, T. 2002,
  \apj, 568, 912

\bibitem[{{van Straaten} {et~al.}(2003){van Straaten}, {van der Klis}, \& {M{\'
  e}ndez}}]{vstr03:apj596}
{van Straaten}, S., {van der Klis}, M., \& {M{\' e}ndez}, M. 2003, \apj, 596,
  1155

\bibitem[{{Wijnands}(2004)}]{wijnands04:bepposax}
{Wijnands}, R. 2004, in To appear in the Proceedings of the Symposium ``The
  Restless High-Energy Universe'', 5--8 May 2003, Amsterdam, The Netherlands,
  E. P. J. van den Heuvel, J. J. M. in 't Zand, and R. A. M. J. Wijers Eds
  (astro-ph/0309347)

\bibitem[{{Wijnands} {et~al.}(2001){Wijnands}, {M{\' e}ndez}, {Markwardt}, {van
  der Klis}, {Chakrabarty}, \& {Morgan}}]{wijnands01:apj560}
{Wijnands}, R., {M{\' e}ndez}, M., {Markwardt}, C., {et~al.} 2001, \apj, 560,
  892

\bibitem[{{Wijnands} \& {van der Klis}(1998{\natexlab{a}})}]{wijnands98:nat394}
{Wijnands}, R. \& {van der Klis}, M. 1998{\natexlab{a}}, \nat, 394, 344

\bibitem[{{Wijnands} \& {van der Klis}(1998{\natexlab{b}})}]{wijnands98:apj507}
---. 1998{\natexlab{b}}, \apjl, 507, L63

\bibitem[{{Wijnands} \& {van der Klis}(1999)}]{wijnands99:apj514}
---. 1999, \apj, 514, 939

\bibitem[{{Wijnands} {et~al.}(2003){Wijnands}, {van der Klis}, {Homan},
  {Chakrabarty}, {Markwardt}, \& {Morgan}}]{wijnands03:nat424}
{Wijnands}, R., {van der Klis}, M., {Homan}, J., {et~al.} 2003, \nat, 424, 44

\bibitem[{{Yoshida} {et~al.}(1993){Yoshida}, {Mitsuda}, {Ebisawa}, {Ueda},
  {Fujimoto}, {Yaqoob}, \& {Done}}]{yoshida93:pasj45}
{Yoshida}, K., {Mitsuda}, K., {Ebisawa}, K., {et~al.} 1993, \pasj, 45, 605

\bibitem[{{Zhang} {et~al.}(1993){Zhang}, {Giles}, {Jahoda}, {Soong}, {Swank},
  \& {Morgan}}]{zhang93:spie2006}
{Zhang}, W., {Giles}, A.~B., {Jahoda}, K., {et~al.} 1993, in Proc. SPIE Vol.
  2006, p. 324-333, EUV, X-Ray, and Gamma-Ray Instrumentation for Astronomy IV,
  Oswald H. Siegmund; Ed., 324--333

\bibitem[{{Zhang} {et~al.}(1995){Zhang}, {Jahoda}, {Swank}, {Morgan}, \&
  {Giles}}]{zhang95:apj449}
{Zhang}, W., {Jahoda}, K., {Swank}, J.~H., {Morgan}, E.~H., \& {Giles}, A.~B.
  1995, \apj, 449, 930

\end{thebibliography}
\end{document}